\documentclass{aa}
\usepackage{graphicx}
\usepackage{txfonts}
\usepackage[utf8]{inputenc}
\usepackage{amssymb,amsmath}
\usepackage{lscape}
\usepackage{longtable}
\usepackage{float}
\usepackage{natbib}
\usepackage{multicol}
\usepackage[bookmarks=false,colorlinks,allcolors=blue]{hyperref}
\usepackage[para,multiple]{footmisc}

\begin{document}

%
%
\title{
Distribution of star formation in galactic bars \\ as seen with H$\alpha$ and stacked GALEX UV imaging
}
\titlerunning{The distribution of star formation in galactic bars}
\author{S. D\'iaz-Garc\'ia\inst{1,2,3}
          \and        
          F. D. Moyano\inst{1,2,4}
          \and
          S. Comer\'on\inst{1,2,5}
          \and
          J. H. Knapen\inst{1,2}
          \and
          H. Salo\inst{5}
          \and
          A. Y. K. Bouquin\inst{1,2}      
          }
\institute{Instituto de Astrof\'isica de Canarias, E-38205, La Laguna, Tenerife, Spain \\
              \email{simondiazgar@gmail.com}
         \and
             Departamento de Astrof\'isica, Universidad de La Laguna, E-38205, La Laguna, Tenerife, Spain
         \and
             Department for Physics, Engineering Physics and Astrophysics, Queen's University, Kingston, ON K7L 3N6, Canada
         \and
             Geneva Observatory, University of Geneva, Chemin des Maillettes 51, 1290 Sauverny, Switzerland       
         \and 
             Space Physics and Astronomy Research Unit, University of Oulu, FI-90014 Finland
             }
\date{Received 11 August 2020; accepted 2 September 2020}
\abstract
{
\emph{Context.} 
Stellar bars are known to gradually funnel gas to the central parts of disk galaxies. 
It remains a matter of debate why the distribution of ionized gas along bars and in the circumnuclear regions varies among galaxies.
\\
\emph{Aims.} 
Our goal is to investigate the spatial distribution of star formation (SF) within bars of nearby low-inclination disk galaxies 
($i < 65^{\circ}$) from the S$^4$G survey. We aim to link the loci of SF to global properties of the hosts 
(morphological type, stellar mass, gas fraction, and bar-induced gravitational torques), 
providing constraints for the conditions that regulate SF in bars.
\\
\emph{Methods.} 
We use archival GALEX far- and near-UV imaging for 772 barred galaxies, and for a control sample of 423 non-barred galaxies. 
We also assemble a compilation of continuum-subtracted H$\alpha$ images for 433 barred galaxies, 
70 of which we produced from ancillary photometry and MUSE and CALIFA integral field unit data cubes. 
We employ two complementary approaches: i) the analysis of bar (2D) and disk (1D) stacks built from co-added UV images (oriented and 
scaled with respect to the stellar bars and the extent of disks) 
of hundreds of galaxies that are binned based on their Hubble stage ($T$) and bar family; 
and ii) the visual classification of the morphology of ionized regions (traced from H$\alpha$ and UV data) in individual galaxies 
into three main SF classes: A) only circumnuclear SF; B) SF at the bar ends, but not along the bar; and C) SF along the bar. 
Barred galaxies with active and passive inner rings are likewise classified.
\\
\emph{Results.} 
Massive, gas-poor, lenticular galaxies typically belong to SF class A; this is probably related to bar-induced quenching of SF in the disk. 
The distribution of SF class B peaks for early- and intermediate-type spirals; 
this most likely results from the interplay of gas flow, shocks, and enhanced shear in massive 
centrally concentrated galaxies with large bar amplitudes 
(the latter is supported by the lack of a dip in the radial distribution of SF in  non-barred galaxies). 
Late-type gas-rich galaxies with high gravitational torques are mainly assigned to SF class C; we argue that this is a consequence of low shear 
among the faintest galaxies. In bar stacks of spiral galaxies the UV emission traces the stellar bars and dominates on their leading side, 
as witnessed in simulations. Among early-type spirals the central UV emission is $\sim$0.5 mag brighter in strongly barred galaxies, 
relative to their weakly barred counterparts; this is probably related to the efficiency of strong bars 
sweeping the disk gas and triggering central starbursts. On the contrary, in later types 
the UV emission is stronger at all radii in strongly barred galaxies than in weakly barred and non-barred ones. 
We also show that the distributions of SF in inner-ringed galaxies are broadly the 
same in barred and non-barred galaxies, including a UV and H$\alpha$ deficit in the middle part of the bar; 
this hints at the effect of resonance rings trapping gas that is no longer funneled inwards.
\\
\emph{Conclusions.}
Distinct distributions of SF within bars are reported in galaxies of different morphological types. 
Star-forming bars are most common among late-type gas-rich galaxies. Bars are important agents in the regulation of SF in disks.
}
\maketitle
%
%
\section{Introduction}\label{introduction}
%
%
Stellar bars are common in the local Universe, with well over half of disk galaxies having a bar visible on 
optical and near-infrared images 
\citep[e.g.,][]{1963ApJS....8...31D,1993RPPh...56..173S,2000ApJ...529...93K,
2002MNRAS.336.1281W,2004ApJ...607..103L,2007ApJ...659.1176M,2007ApJ...657..790M,2009A&A...495..491A,2012ApJ...761L...6M,
2015ApJS..217...32B,2016A&A...587A.160D,2019A&A...631A..94D}. 
Due to the non-axisymmetric mass distribution in bars, they stimulate angular momentum transfer and 
gas inflow in galaxy disks \citep[][]{1979MNRAS.187..101L,1980ApJ...237..404S,1989Natur.338...45S,1992MNRAS.259..345A}, 
and are thus an important agent in the secular evolution of galaxies \citep[see the review by][and references therein]{2013seg..book....1K}.

The distribution of massive  star formation (SF) in galaxy disks is conditioned by localized zones where gas 
clouds are both stable and dense enough to form stars, which \citet[][]{1989ApJ...344..685K}, 
following \citet[][]{1964ApJ...139.1217T}, parameterized to depend on gas surface density and velocity dispersion. 
Velocity shear can limit SF, however,  acting against the condensation of massive clouds \citep[e.g.,][]{1998A&A...337..671R,2005MNRAS.361L..20S}. 
\citet[][]{2004A&A...413...73Z} nicely illustrated how shear in a strong bar can locally inhibit the massive SF from a 
clear systematic offset they observed in their Fabry-P\'erot H$\alpha$ data between regions of high non-circular motions 
and active SF in the bar of NGC~1530. Although this effect is hard to observe, and has not been seen in many other galaxies, 
it does illustrate graphically the relation between bar dynamics and SF morphology. In general, 
the occurrence of massive SF zones is governed by the location of the spiral arms and dynamical resonances, 
with SF often concentrated in spiral arms and the rings that can form near the resonances. 

In galactic bars, there is no uniform picture of where the SF occurs. Often there are regions of SF near the ends of the bar, 
and these can form parts of inner rings, as in NGC~5850 (Fig.~\ref{Fig_classB}, lower panel), 
or highlight the start of grand-design spiral arms, as in NGC~1300 (Fig.~\ref{Fig_classB}, upper panel). 
The sets of symmetric enhancements of stellar emission near the ends of the bar known as ansae are typically not star forming 
and have a stellar dynamical origin \citep[][]{2007AJ....134.1863M}.

Bars stimulate gas inflow \citep[][]{1984MNRAS.209...93S,1985A&A...150..327C}, and where this inflow is slowed down 
in the vicinity of inner Lindblad resonances \citep[e.g.,][]{1994ApJ...424...84H,1995ApJ...454..623K,2010MNRAS.402.2462C} 
a nuclear ring can form and the gas accumulated within them can lead to important and visually striking star-forming nuclear rings, 
as in NGC~1097 (Fig.~\ref{Fig_classC}, upper panel). 
Bars have statistically been linked to enhanced gas concentration, 
and very clearly linked to enhanced SF in the central kpc region \citep[e.g.,][]{1980A&A....88..365H,1981A&A....93...93H,1986MNRAS.221P..41H,
1987ApJ...323...91D,1999ApJ...525..691S,2005ApJ...630..837J,2005ApJ...632..217S,2006ApJ...652.1112R,2017ApJ...838..105L} 
\citep[for a review of the early papers, see][]{2004ASSL..319..189K}, and often this manifests itself not as a ring but as a 
(circum)nuclear starburst, as in NGC~2712 or NGC~3185 \citep[Figs.~2~and~3 in][]{2016MNRAS.457..917J}. 
The SF can  be limited to the central region, as in NGC~0936 (Fig.~\ref{Fig_classA}).

Star formation can occur along the bar, but often does not. When it does, it can occur in a narrow linear or curved morphology 
either in the middle of the bar, as in NGC~7479 \citep[see Fig.~1 in][]{2001Ap&SS.276..491Z},
or along one of its edges, as in NGC~1365 (Fig.~\ref{Fig_classC}, middle panel). 
Many bars are devoid of SF, as in the case of NGC~5850 (Fig.~\ref{Fig_classB}, lower panel), showing only a central SF peak. 
Finally, the bar sweeping up gaseous material often leads to a dearth of gas, and thus SF, 
in symmetric regions on either side of the bar \citep[][]{2009A&A...501..207J}; 
a good example is NGC~3351, shown in Fig.~4 of \citet[][]{2016MNRAS.457..917J}. 
This was referred to as the ``SF desert'' by \citet[][]{2016MNRAS.457..917J,2018MNRAS.474.3101J} 
and the desert was confirmed from numerical modeling to consist of older stars by \citet[][]{2019MNRAS.489.4992D}.

In this paper we use ultraviolet (UV) and H$\alpha$ imaging to study the distribution of SF in bars in a statistical 
manner rather than by considering the detailed morphology of individual galaxies, 
for a sample of more than 800 barred galaxies from the \emph{Spitzer} Survey of 
Stellar Structure in Galaxies \citep[S$^4$G;][]{2010PASP..122.1397S}. 
As a result, we do not use all the possible categories described earlier in this Introduction, but concentrate on whether SF 
occurs at the inner or outer ends of a bar, and/or within it, as described in Sect.~\ref{individual}. 

Our investigation builds on a small but very interesting body of past work. \citet[][]{2007A&A...474...43V} characterized 
the H$\alpha$ morphology of 45 suitable isolated galaxies from their AMIGA sample \citep[see][]{2007A&A...472..121V}, 
classifying them into three main groups depending on whether or not emission is present from the central and bar regions of a galaxy 
\citep[see also work by][]{1997A&A...326..449M,2019A&A...627A..26N}. 
Recently, \citet[][]{2020MNRAS.495.4158F} used 684 relatively face-on galaxies from 
the Mapping Nearby Galaxies at APO \citep[MaNGA;][]{2015ApJ...798....7B} survey, 
which have a high probability of being barred following the classification by volunteer citizens in 
a Galaxy Zoo 2 project \citep[][]{2013MNRAS.435.2835W}. 
They then classified their H$\alpha$ images according to whether a galaxy shows SF in the center,  inner ring, ends of the bar, 
or within the bar, concluding that only low-mass galaxies host SF along their bars, and that both the physical and SF properties of 
bars are mostly governed by the galaxy stellar mass. 

We improve on several aspects of previous work, for example sample size (the samples 
of \citealt[][]{2007A&A...474...43V} or \citealt[][]{2019A&A...627A..26N} are small 
and did not probe the plentiful galaxies at the end of the Hubble sequence), 
the set of explored morphological and physical disk and bar parameters, and the quality of the multiwavelength imaging data. 
\citet[][]{2020MNRAS.495.4158F} use H$\alpha$ images derived from MaNGA that have limited physical resolution 
across their sample, and depend on criteria for bar classification that are hard to quantify but can introduce important biases 
(e.g., towards the most prominent bars, judging from their too small overall bar fractions). 

In addition, in Sect.~\ref{stacks} we introduce the stacking of UV bars  (2D), 
based on the techniques developed by \citet[][]{2016A&A...596A..84D} at 3.6~$\mu$m, 
and significantly improve on the averaging of SF radial profiles (in 1D) 
pioneered by \citet[][]{2009A&A...501..207J} (in H$\alpha$), using hundreds of images per sample bin. 
These techniques   probe with unprecedented statistical significance the spatial distribution of SF in disks, 
whose dependence on global galaxy properties is discussed in Sect~\ref{discussion_chapter}, 
as well as the possible effect of stellar bars enhancing or inhibiting SF. 
Finally, in Sect.~\ref{summarysection} we summarize the main results 
of this paper and their interpretation in light of galaxy evolution.
%
%
\section{Stacking GALEX near- and far-UV images}\label{stacks}
%
%
\citet[][]{2016A&A...596A..84D} obtained average 1D disk profiles and 2D bar density maps by stacking 
\emph{Spitzer} Infrared Array Camera (IRAC) 3.6 $\mu$m images, which trace old stellar populations, 
in order to characterize the stellar mass distribution of more than a thousand disk galaxies and 
reveal signatures of bar-induced secular evolution. 

Here, these averaging techniques are applied to Galaxy Evolution Explorer (GALEX) near-UV 
(NUV; $\lambda_{\rm eff}=2267\,\AA$) and far-UV (FUV; $\lambda_{\rm eff}=1516\,\AA$) images, 
so that SF activity in bars is analyzed  with unprecedented 
statistical significance \citep[emission at these UV wavelengths trace recent SF, up to $\sim 100$ Myr; ][]{1998ARA&A..36..189K}. 
The UV stacks constitute a non-parametric characterization of the distribution of SF in bars, 
which may be useful for comparison with numerical models.

We use the sky-subtracted and masked images from 
the \emph{GALEX/S$^4$G UV-IR Catalog} by \citet[][]{2018ApJS..234...18B} \citep[see also][]{2015ApJ...800L..19B}, 
comprising 1931 galaxies of all morphological types 
that were gathered from the GALEX GR6/7 Data Release\footnote{\href{http://galex.stsci.edu/GR6/}{http://galex.stsci.edu/GR6/}}, 
cross-matched with those of the S$^4$G, 
and reduced following \citet[][]{2007ApJS..173..185G}. {Roughly one-half of the} data belong to the GALEX All-Sky Imaging Survey 
(AIS), whose images had exposure times of $\sim$ 100 seconds and allow the detection of point sources down to 
$\approx 20$ AB mag \citep[e.g.,][]{2017ApJS..230...24B}. 
The rest of the galaxies were imaged in deeper GALEX surveys and had exposure times of 1000 seconds or more. The pixel size is 1.5 arcsec.
%
%
\begin{figure*}
\centering
\includegraphics[width=0.99\textwidth]{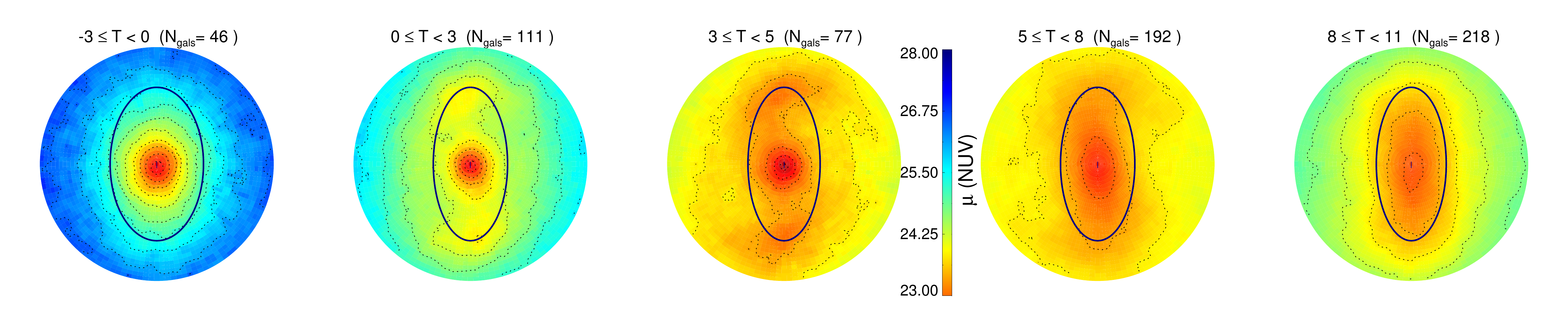}

\includegraphics[width=0.99\textwidth]{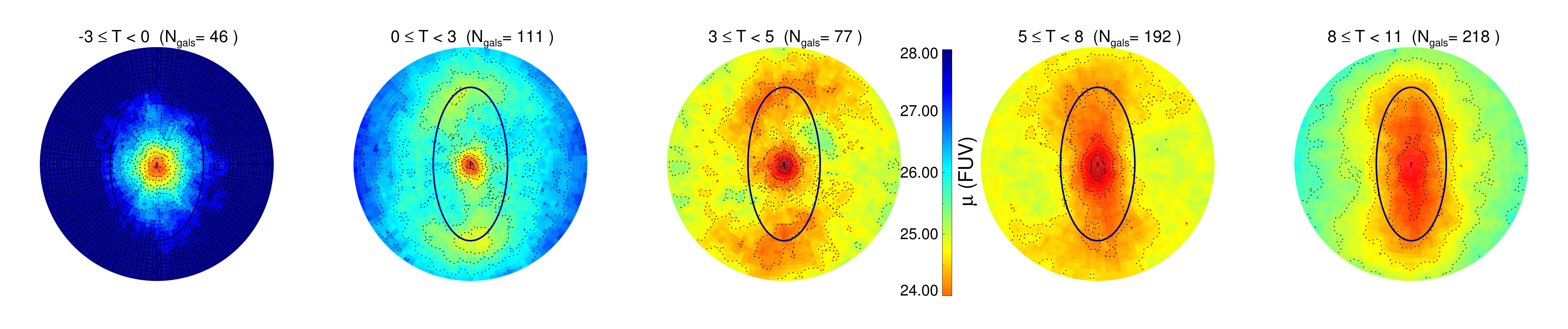}
\caption{
Two-dimensional synthetic stellar bars constructed from co-added NUV (top) and FUV (bottom) 
images of disk galaxies that were oriented and scaled with respect to the bars, 
flipped to make the spiral arms wind clockwise (if needed), and grouped based on revised Hubble stage ($T$, increasing from left to right). 
The number of galaxies in each  subsample is also indicated. 
Bar stacks are shown in units of mag arcsec$^{-2}$ (see vertical bar for thresholds and color-coding) and 
cropped to a radius $1.5 \cdot r_{\rm bar}$, so that all binned galaxies are covered radially. 
The dotted lines show isophotal contours with a step of 0.35 mag arcsec$^{-2}$. 
The ellipse represents the average ellipticity (3.6 $\mu$m) of the galaxies in the bin \citep[from][]{2015A&A...582A..86H,2016A&A...587A.160D}. 
The mean bar length is used as a unit, but the actual mean 3.6 $\mu$m bar lengths in kpc vary for each $T$-bin 
\citep[see Fig.~11 and Table~3 in][]{2016A&A...587A.160D} and are lowest among the faintest galaxies.
}
\label{Fig_ttype_bars_NUV}
\end{figure*}
%
%
\begin{figure*}
\centering
\includegraphics[width=0.49\textwidth]{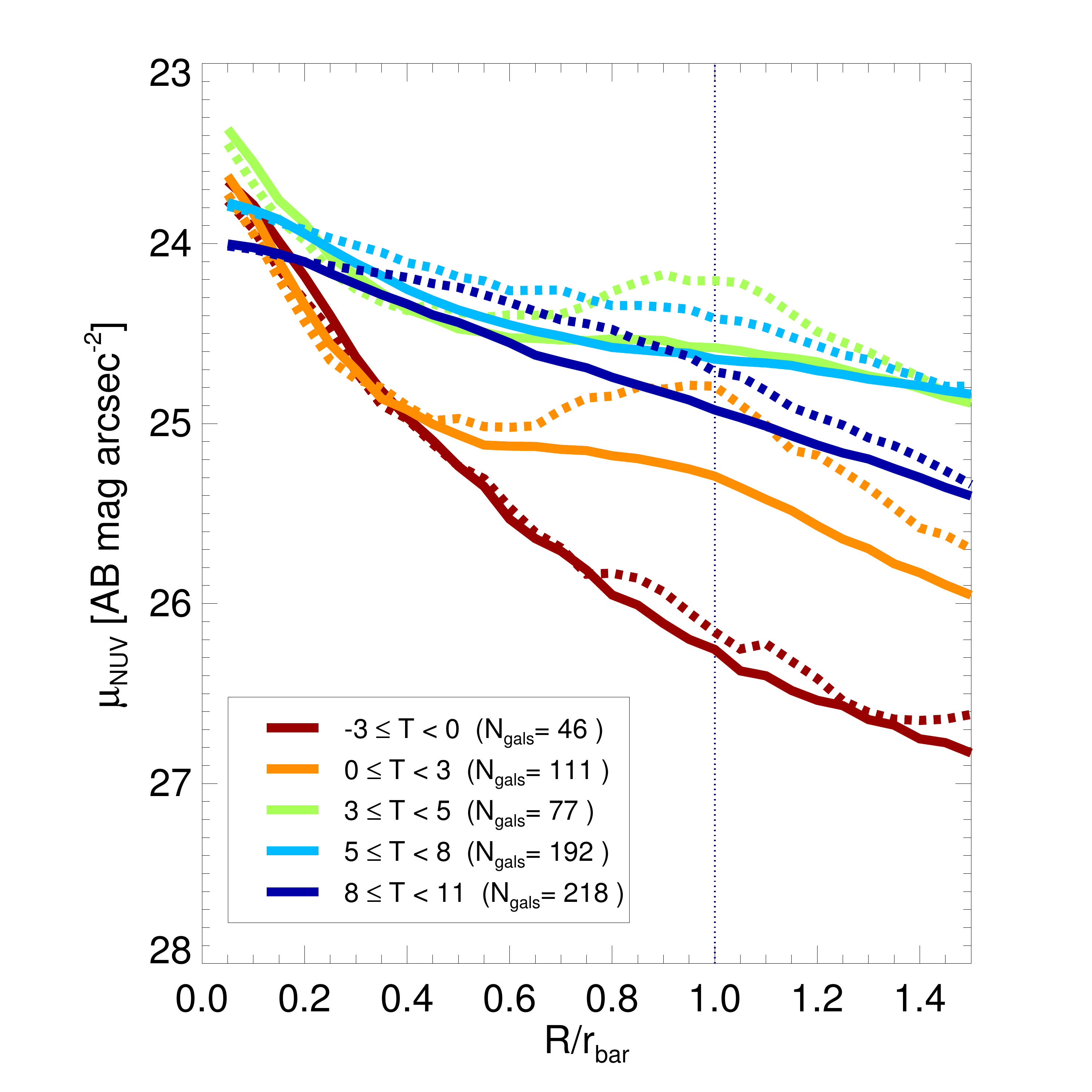}
\includegraphics[width=0.49\textwidth]{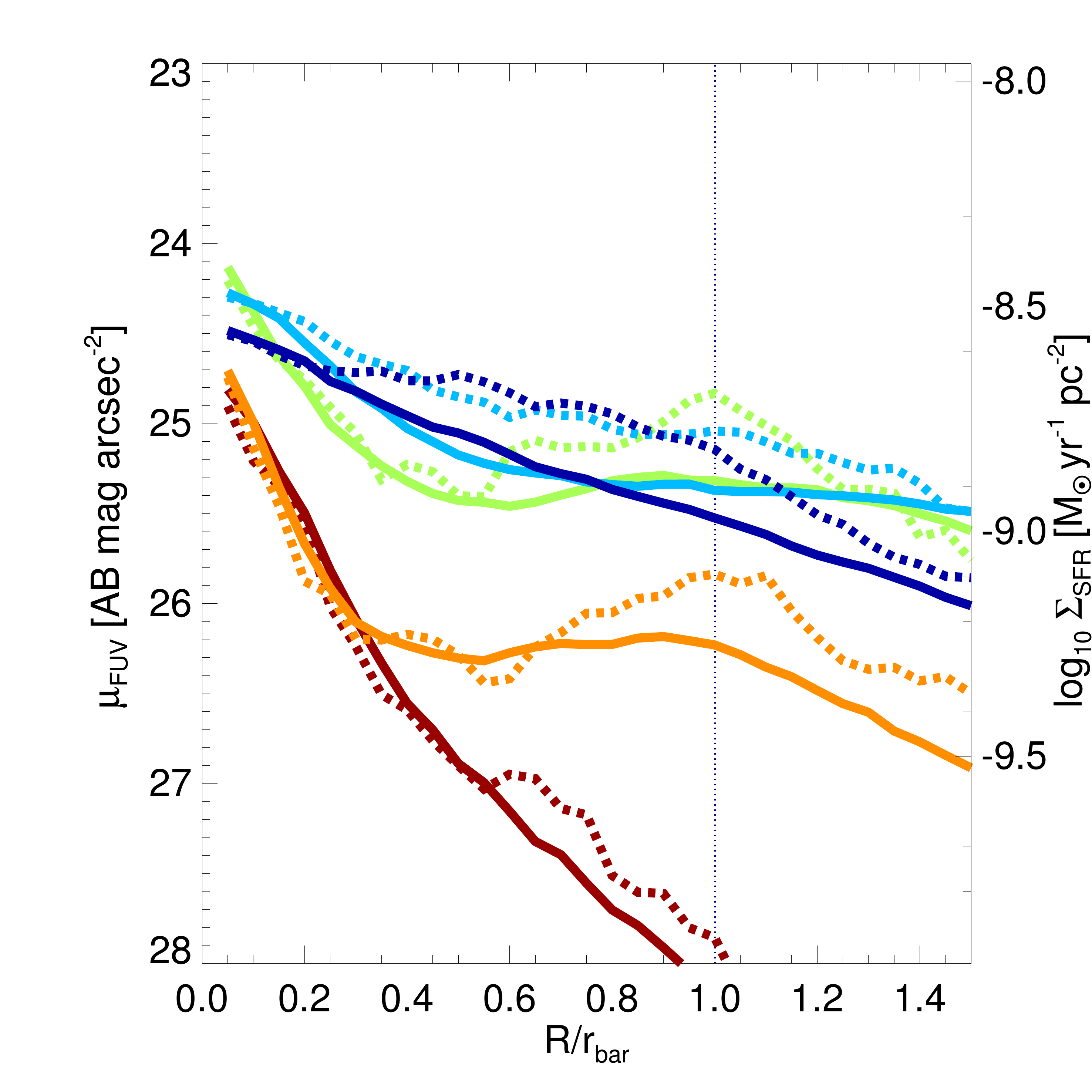}
\caption{
Azimuthally averaged mean NUV (left) and FUV (right) luminosity profiles (solid lines),  in bins of numerical Hubble type,  obtained from the 
2D bar stacks shown in Fig.~\ref{Fig_ttype_bars_NUV}. The dashed lines correspond to the 
surface brightness cut along the bar major axis. The vertical dotted line indicates the bar end. 
FUV luminosities are converted to $\Sigma_{\rm SFR}$ (right $y$-axis of the right panel) using Eq.~\ref{SFR_eq}.
}
\label{Fig_ttype_bars_NUV_1D}
\end{figure*}

Mean FUV surface brightnesses ($\mu_{\rm FUV}$) are converted to SF rate surface densities ($\Sigma_{\rm SFR}$) 
following the prescription by \citet[][]{2012ARA&A..50..531K} and \citet[][]{2014ARA&A..52..415M} \citep[see Appendix~B in][]{2018ApJS..234...18B},
%
%
\begin{equation}\label{SFR_eq}
{\rm log_{10}} (\Sigma_{\rm SFR}) [M_{\odot}\,{\rm yr}^{-1}\,{\rm pc}^{-2}] =1.239-0.4\cdot \mu_{\rm FUV} [{\rm AB \, {\rm mag} \, {\rm arcsec}^{-2}}],
\end{equation}
%
%
assuming a Kroupa initial mass function \citep[][]{2001MNRAS.322..231K}. These estimates are not corrected for extinction, 
and thus the values are lower boundaries of the true $\Sigma_{\rm SFR}$.

Our parent sample is made up of the 1345 disk galaxies with inclinations $< 65^{\circ}$ \citep[according to][]{2015ApJS..219....4S} in the S$^4$G. 
Of these, 860 are barred according to \citet[][]{2015ApJS..217...32B}, of which 760 ($\sim 88\%$) have available NUV and FUV imaging 
from \citet[][]{2018ApJS..234...18B}. We also use a control subsample of 423 non-barred and not highly inclined 
galaxies with available GALEX UV data.
%
%
\subsection{Average UV bars (2D)}\label{bar_uv_stack}
%
%
In order to study in detail the distribution of SF in bars, 
FUV and NUV images are scaled to a common frame determined by the sizes ($r_{\rm bar}$) and orientations of the bars, 
measured visually by \citet[][]{2015A&A...582A..86H} using 3.6 $\mu$m S$^4$G images. 
Here we present a  summary of the way the UV images are treated  \citep[for further details see][]{2016A&A...596A..84D}:
\begin{enumerate}
 \item Deprojection to face-on view using the orientation parameters for the outer disk from \citet[][]{2015ApJS..219....4S}. 
 To make sure that deprojections are reliable, we only use galaxies with ``ok'' quality flags for the orientations.
 \item Fourier decomposition of the UV light distribution of the 
 galaxy images \citep[up to 40 azimuthal modes, using the NIR-QB code;][]{1999AJ....117..792S,2002MNRAS.337.1118L}, 
 and reconstruction of the image in a polar grid with 128 bins in the azimuthal direction \citep[][]{1999AJ....117..792S}.
 \item Rotation of the image with respect to the bar major axis, imposing a bar position angle equal to zero.
 \item Geometric reflection across the bar major axis to make the spiral arms wind clockwise (S-shaped) 
 in case they wind counterclockwise (Z-shaped) in the 3.6 $\mu$m images. 
 The correction of the orientation of the spiral arms  (normally trailing, relative to the disk rotation) 
 is important for our analysis:  H{\sc\,ii} regions typically appear 
 on the leading side of the bar \citep[e.g.,][]{2002AJ....124.2581S,2010A&A...521A...8P}.
 \item Scaling of the reoriented image to a grid of radius $3 \cdot \rm r_{\rm bar}$, 
 and width of the radial bin of $0.05 \cdot r_{\rm bar}$. This ensures a good sampling of the bar 
 (the median bar radii in our sample are $\sim$10 and $\sim$20 resolution elements in GALEX and \emph{IRAC} images, respectively) 
 and also of the spiral arms slightly beyond the bar region. 
 \item Having uniformly scaled all the images of barred galaxies to a common physical framework, 
 we are in the position to take subsamples and perform the bar stacks: 
 the mean FUV and NUV surface brightness (weighted in mag arcsec$^{-2}$) is obtained within each of the bins of the polar grid. 
 Our stacking techniques yield roughly the same results (within uncertainties) regardless of the employed weighting when 
 co-adding the light (flux or magnitudes) or the used measure of central tendency 
 (mean or median) \citep[for further details see Fig.~2 and explanations in][]{2016A&A...596A..84D}.
\end{enumerate}
%
%
\begin{figure}
\centering
\includegraphics[width=0.49\textwidth]{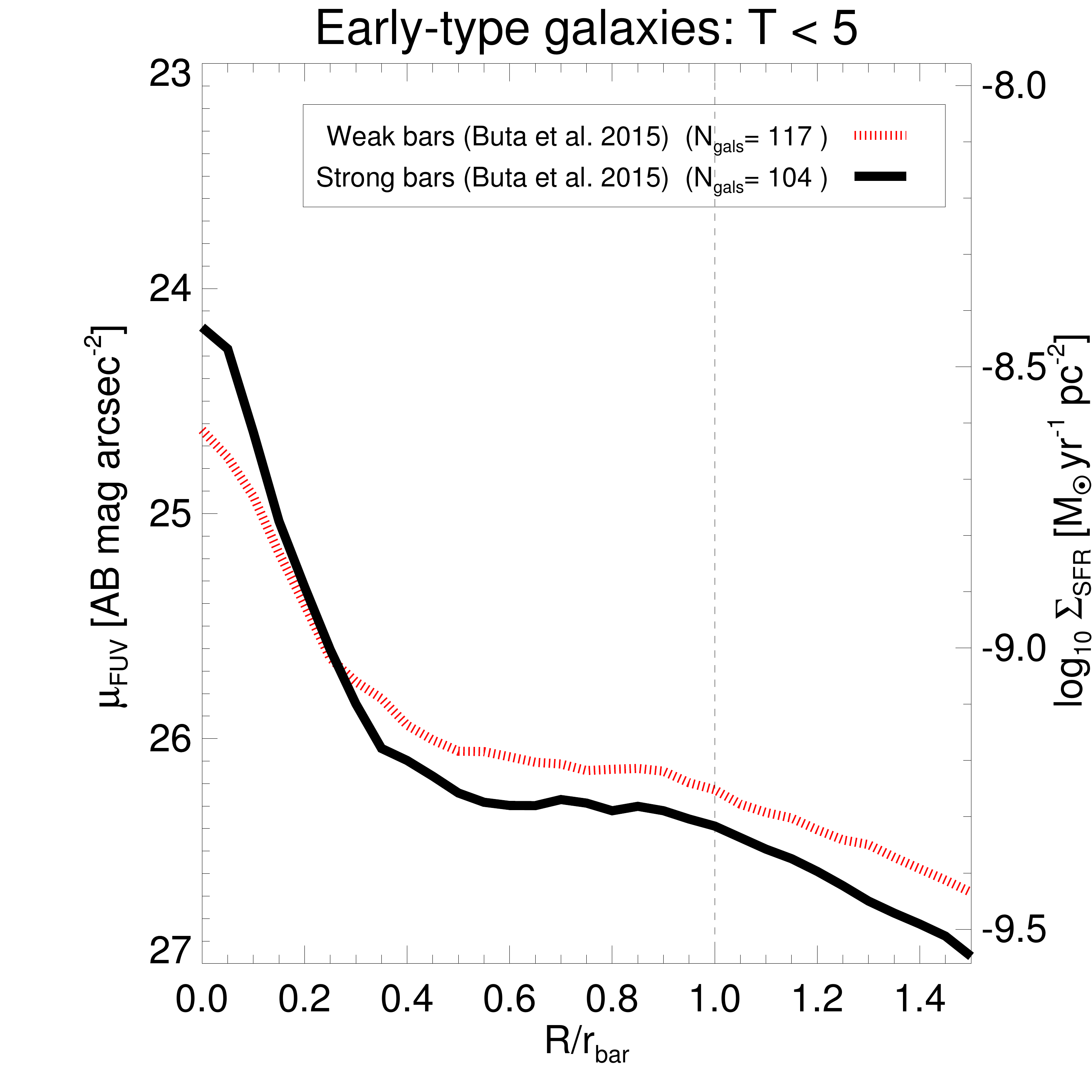}

\includegraphics[width=0.49\textwidth]{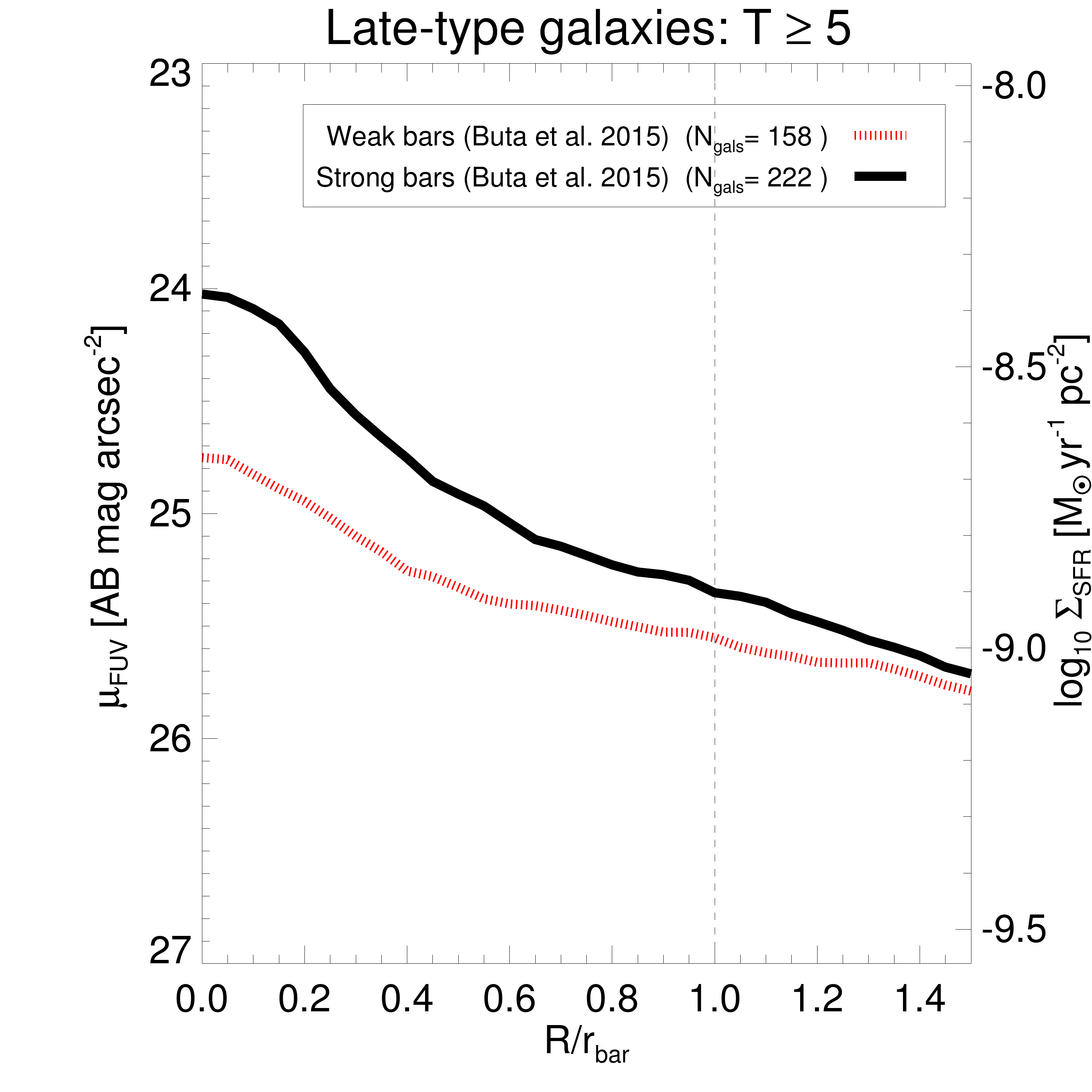}
\caption{
Azimuthally averaged mean FUV luminosity profiles obtained from the bar stacks of weakly and strongly barred 
galaxies \citep[based on the family classification from][]{2015ApJS..217...32B} with 
total stellar masses $10^{8.5}M_{\odot} < M_{\ast} < 10^{11}M_{\odot}$, 
considering separately early-type ($T<5$, \emph{upper panel}) and late-type galaxies ($T\ge 5$, \emph{lower panel}). 
The same plots using NUV are shown in Fig.~\ref{Fig_family_bars_NUV_1D}.
}
\label{Fig_family_bars_UV_1D}
\end{figure}
%
%
\begin{figure*}
\centering
\includegraphics[width=0.49\textwidth]{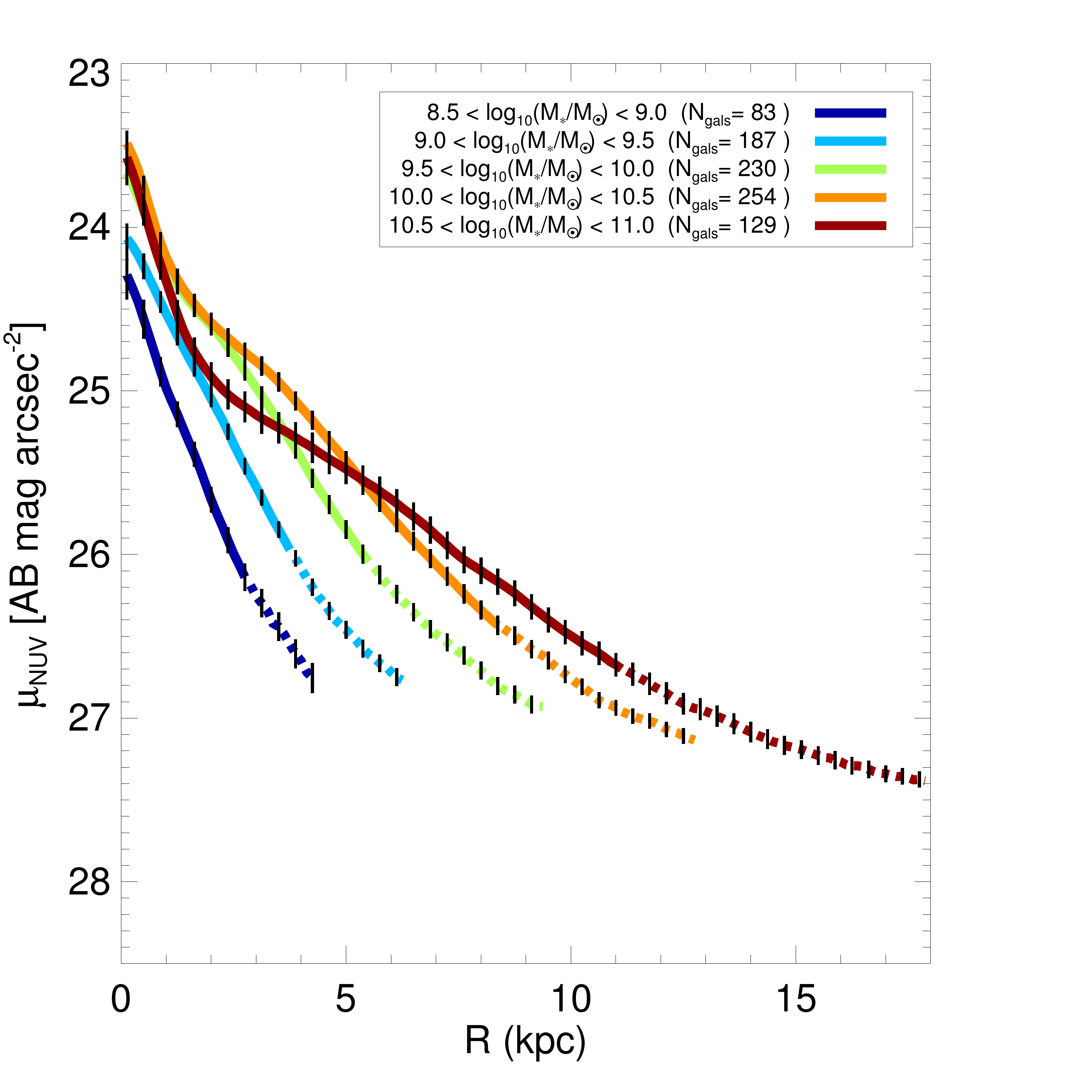}
\includegraphics[width=0.49\textwidth]{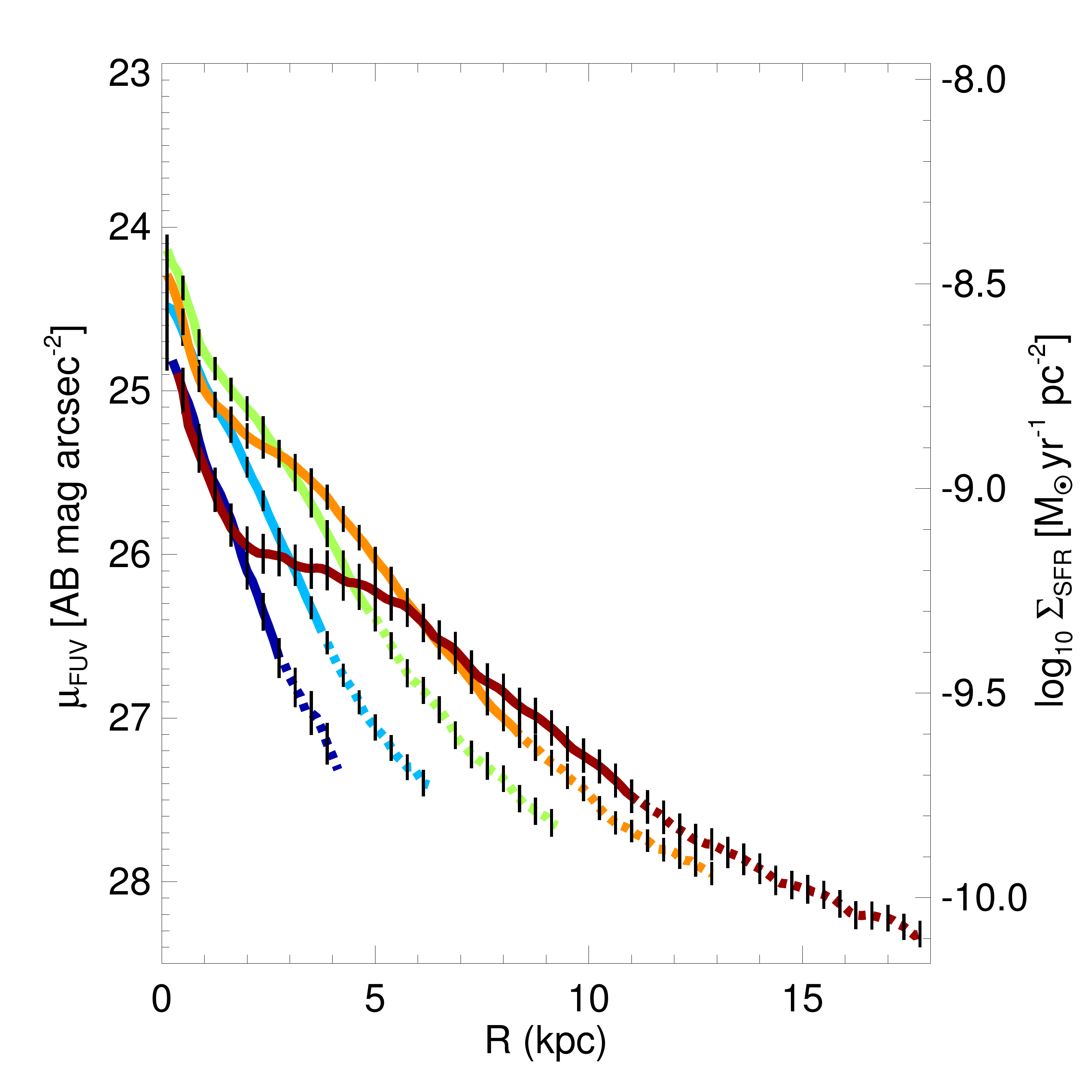}
\caption{
Mean $\mu_{\rm NUV}$ (left) and $\mu_{\rm FUV}$ (right) 1D profiles as a function of galactocentric radius 
for different subsamples defined as a function of the total stellar mass (in bins of 0.5 dex; see legend) 
\citep[see also][]{2018ApJS..234...18B}. Error bars correspond to the standard deviation of the mean ($\sigma/\sqrt{N_{\rm gals}}$). 
The dashed lines show the average luminosity profiles where the radial sample coverage is greater than 75$\%$ and lower than 100$\%$, 
and thus where uncertainties are larger (e.g., artificially created up-bending sections due to dominance of more extended UV disks with 
fainter extrapolated central surface brightnesses). 
}
\label{Fig_mass_bars_NUV_FUV_1Dstack}
\end{figure*}
%
%
\begin{figure*}
\centering
\includegraphics[width=0.49\textwidth]{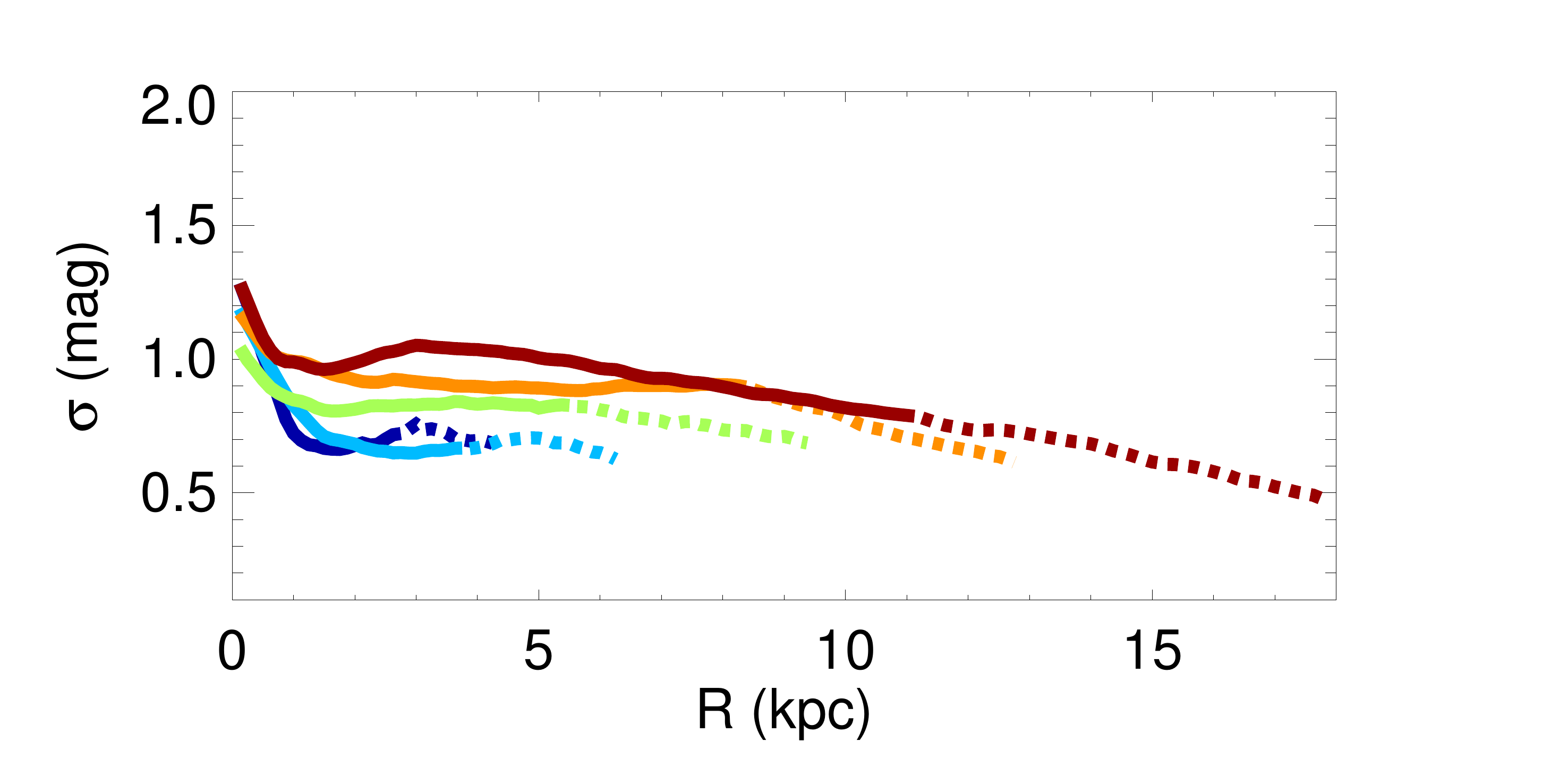}
\includegraphics[width=0.49\textwidth]{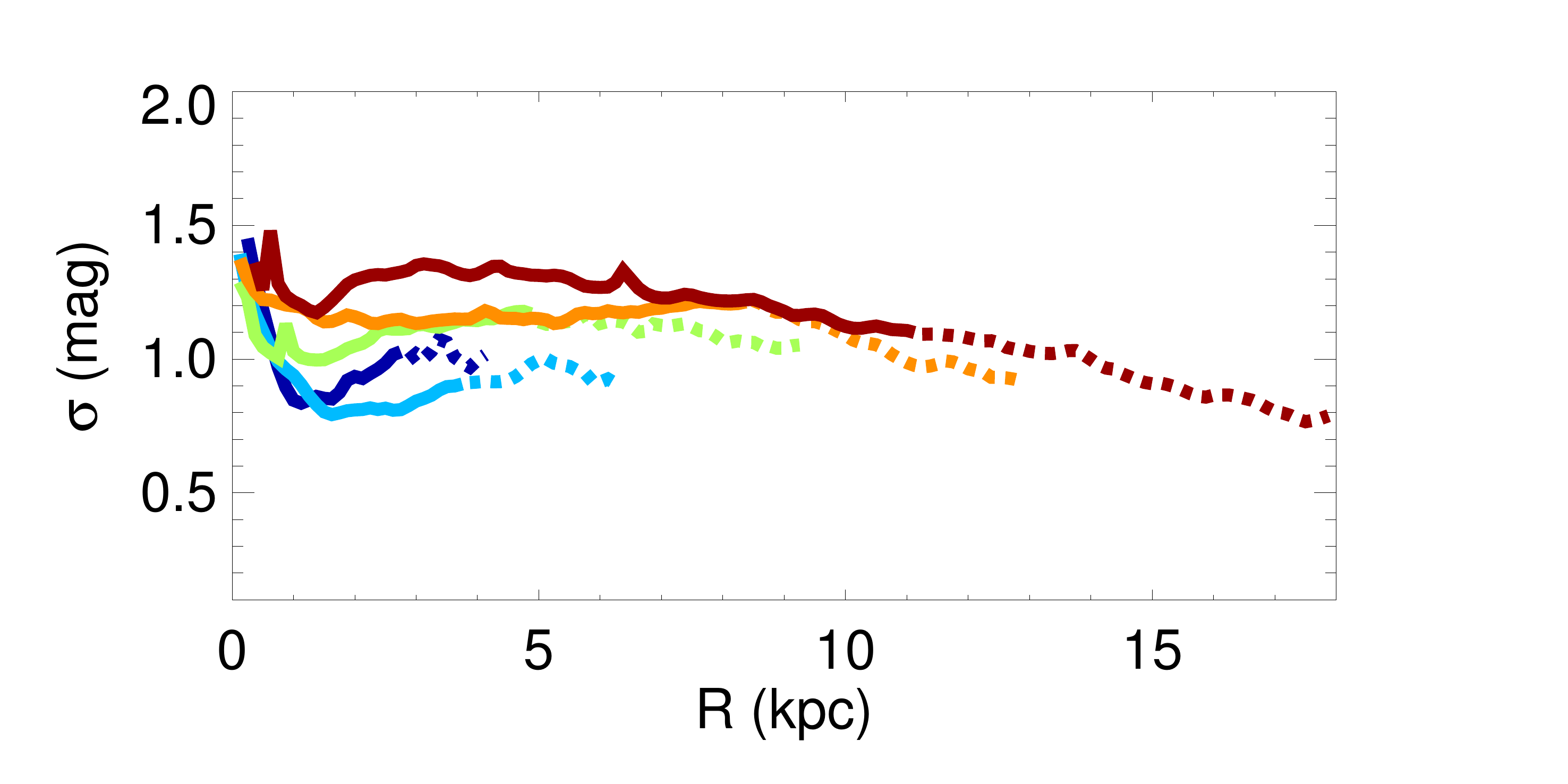}
\caption{
As in Fig.~\ref{Fig_mass_bars_NUV_FUV_1Dstack}, but for the dispersion of the $\mu_{\rm NUV}$ (left) 
and $\mu_{\rm FUV}$ (right) luminosity profiles.
}
\label{Fig_mass_bars_FUV_1Dstack_disp}
\end{figure*}
%
%
Bar stacks resulting from the co-adding of FUV and NUV images and the 
binning of our sample in the Hubble sequence are shown in Fig.~\ref{Fig_ttype_bars_NUV}. 
Azimuthally averaged luminosity radial profiles and the surface brightness along the bar major axis 
are directly extracted from the bar stacks, and are shown in Fig.~\ref{Fig_ttype_bars_NUV_1D}. Uncertainties are estimated via the 
standard deviation of the mean ($\sigma/\sqrt{N_{\rm gals}}$), which is typically $\lesssim 0.2$ mag, as shown in Sect.~\ref{1-Dstacks}.

The subsamples were binned by morphological types, separating S0s ($-3 \le T < 0$), early-type spirals ($0 \le T < 3$), 
intermediate-type spirals ($3 \le T < 5$), late-type spirals ($5 \le T < 8$), and Magellanic and irregular galaxies ($8 \le T \le 10$). 
The average ellipticity of stellar bars from \citet[][]{2015A&A...582A..86H}, 
obtained via ellipse fitting \citep[][]{1987MNRAS.226..747J} from 3.6~$\mu$m imaging, is highlighted with a black ellipse. 
A similar characterization of bar stacks 
as a function of the total stellar mass of the binned galaxies can be found in  Appendix~\ref{mass_average_bar}  (Figs.~\ref{Fig_mass_bars_NUV} and \ref{Fig_mass_bars_NUV_1D}).
%
%
\subsubsection{Spatial distribution of UV emission}\label{UV_emission_spatial}
%
%
Among spirals ($0 \le T< 8$), the UV emission leads with respect to the stellar bar \citep[e.g.,][]{2002AJ....124.2581S}. 
This is not the case for the S0s ($T \le 0$), where the UV emission is circumnuclear and does not follow the bars. 
In addition, the leading and trailing sides of the bars cannot be identified when $T>8$ because no spiral pattern is present, 
and thus the UV emission does not occupy a preferential side in bars hosted by irregular galaxies. 
Within the outer half of the bar ellipse (semi-major axis distances $> 0.5\cdot r_{\rm bar}$), 
the FUV flux on the leading side of the bar stacks (averaged over the two quadrants) 
is $21\%$, $16\%$, and $11\%$ higher than on the trailing side for early-, intermediate-, and late-type spirals, respectively.

Clear differences stand out for early- and late-type spirals: 
when $0\le T<5$ (Cols. 2 and 3 of Fig.~\ref{Fig_ttype_bars_NUV}), the UV emission dominates in the circumnuclear regions and 
at the bar ends,  with a deficit of light in the middle part of the bar, 
whereas for $T\ge5$ the distribution of UV light is almost uniform across the bar. 
These trends are more clearly seen in Fig.~\ref{Fig_ttype_bars_NUV_1D}: a hump at the bar end is noticeable in 
the surface brightness profiles of early- and intermediate-type spirals (especially in the cut along the bar major axis), 
whereas late-type galaxies present an  exponential radial decay of the UV surface brightness. 
Late-type barred galaxies are brighter in UV wavelengths in general: this is not surprising, as these galaxies are known 
to be richer in gas and form stars more actively. We note that, in general, the trends are very similar in the  NUV and FUV passbands.
%
%
\subsubsection{Differences in UV emission between strongly and weakly barred galaxies}\label{diffs_strong_weak}
%
%
A relation between the strength of the bar and the presence of SF regions along the bar has been 
hypothesized \citep[see, e.g., discussion in][and references therein]{2002ApJ...570L..55J}, but whether such a connection exists remains unclear. 
To test this, we  derived FUV bar stacks after splitting our sample into weakly barred (I/S$\underline{\rm A}$B+I/SAB) 
and strongly barred (I/SA$\underline{\rm B}$+I/SB) galaxies (Fig.~\ref{Fig_family_bars_UV_1D}), 
based on the classification of galaxy families by \citet[][]{2015ApJS..217...32B}
\footnote{\citet[][]{2016A&A...596A..84D,2016A&A...587A.160D} showed a correspondence between visual 
\citep[\underline{A}B/AB/A\underline{B}/B, from][]{2015ApJS..217...32B} 
and quantitative estimates (tangential-to-radial forces, normalized $m = 2$ Fourier amplitudes, intrinsic ellipticity) of the bar strength.}. 
Early-type ($T<5$) and late-type ($T\ge5\equiv$ Sc) galaxies are studied separately;  not only are they characterized by remarkably 
different structural properties \citep[e.g.,][]{2016A&A...596A..84D,2016A&A...587A.160D}, 
but also by distinct distributions of SF, as seen from the UV stacks (see Sect.~\ref{UV_emission_spatial}).

Among early-type spirals, the central FUV emission is $\sim$0.5 mag brighter for strongly barred galaxies, on average, 
than for  their weakly barred counterparts. This translates into a difference in $\Sigma_{\rm SFR}$ larger than $50 \%$. 
On the other hand, weakly barred galaxies are characterized by a somewhat higher level of  FUV emission in the middle and 
end parts of the bar. 
As discussed in Sect.~\ref{discussion_chapter}, we interpret that such differences can be related to the subtle effect 
of strong bars sweeping the disk gas and inducing circumnuclear starbursts. 
A different picture is identified among late-type galaxies ($T\ge 5$): 
strongly barred galaxies present more intense UV emission than weakly barred galaxies at all bar radii, 
and by more than 0.5 mag in the central parts in particular. 
The same trends are identified using NUV imaging (see Fig.~\ref{Fig_family_bars_NUV_1D}).
%
%
\subsection{Average UV disks (1D)}\label{1-Dstacks}
%
%
\begin{figure}
\centering
\includegraphics[width=0.49\textwidth]{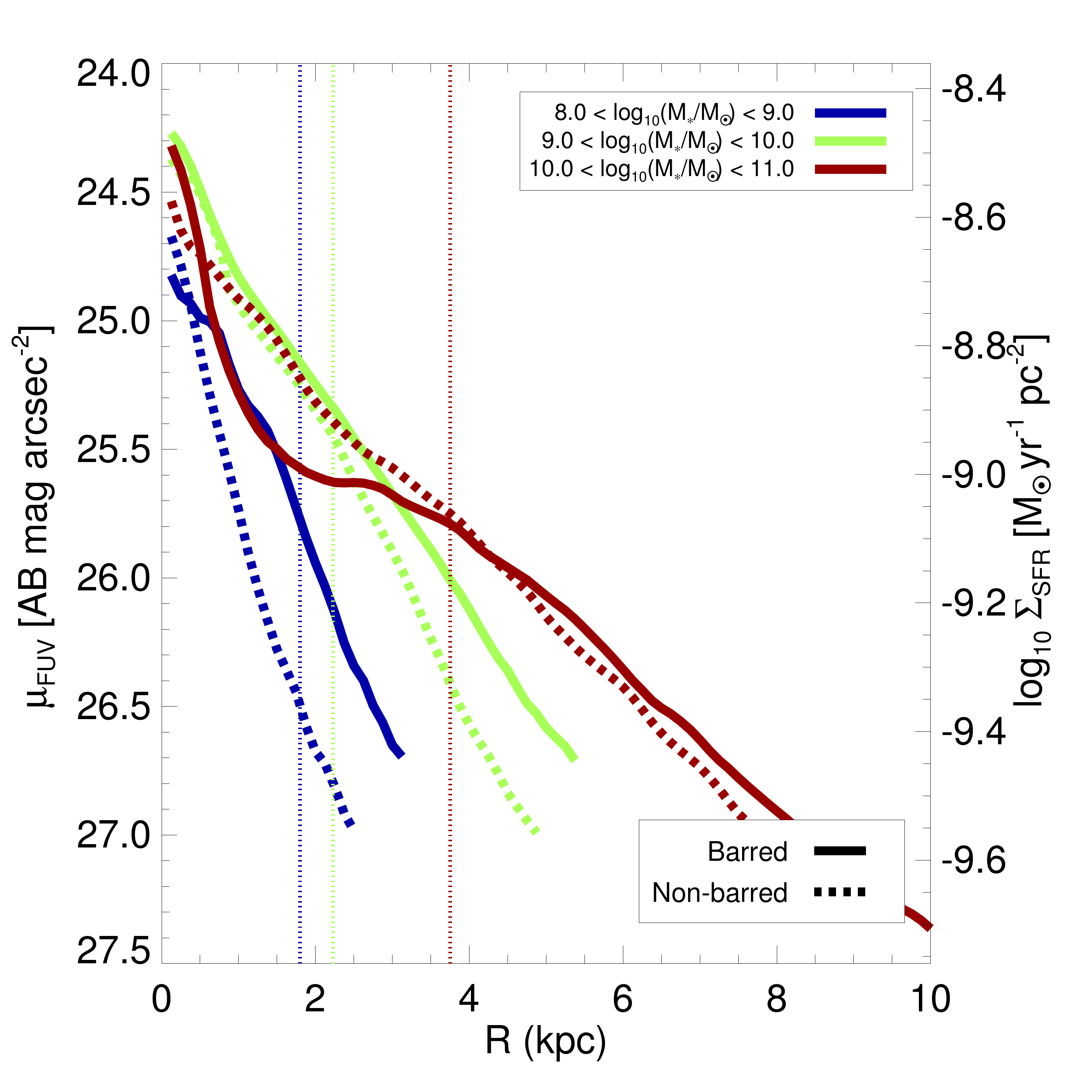}
\caption{
Mean $\mu_{\rm FUV}$ 1D radial profiles in bins of total stellar mass (1 dex in width), 
separating barred (solid line) and non-barred galaxies (dashed line) ($95\,\%$ radial coverage). 
The vertical dotted lines indicate the mean bar size of the barred galaxies in each of the $M\ast$-bins. 
The same plot for $\mu_{\rm NUV}$ can be found in Fig.~\ref{Fig_mass_bars_NUV_1Dstack_bars_separated}.
}
\label{Fig_mass_bars_NUV_FUV_1Dstack_bars_separated}
\end{figure}
%
%
\begin{figure}
\centering
\includegraphics[width=0.49\textwidth]{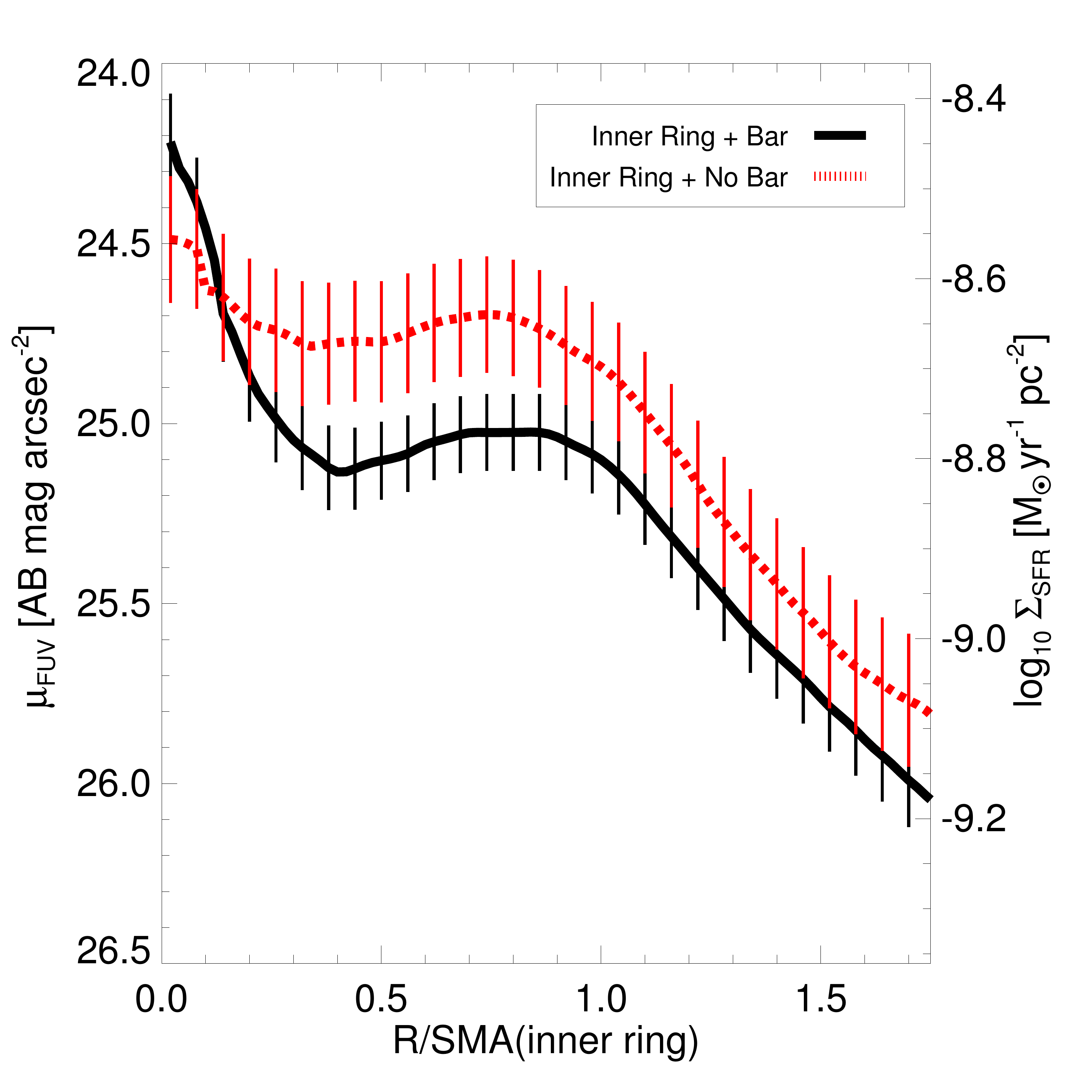}
\caption{
Mean FUV 1D average luminosity profiles for barred (black) and non-barred (red) galaxies hosting an inner ring, 
with total stellar masses $10^{9.5} < M_{\ast}/M_{\odot} < 10^{11}$. Vertical bars correspond to the standard deviation of the mean. 
The luminosity profiles were scaled with respect to the deprojected semi-major axis  of 
the inner ring, labeled SMA(inner ring), before they were co-added. 
}
\label{Fig_inner_rings_SMA}
\end{figure}
%
%
To perform a direct comparison of the mean UV luminosity profiles of barred and non-barred galaxies, 
and also to study those hosting inner rings, we apply 1D averaging techniques \citep[for a characterization of average profiles without 
separation into barred and non-barred galaxies, see also Fig.~4 in][]{2018ApJS..234...18B}. 
This also allows  a direct estimate of the dispersion and uncertainties in our stacks.

Prior to the co-adding, the $m=0$ Fourier intensity profiles are resized to a common frame determined by 
the extent of the disk in physical units (up to 25 kpc, using a 0.125 kpc wide radial bin), using spline interpolation 
\citep[see][]{2016A&A...596A..84D,2017ApJ...835..252S}. 
The radial extent of the grid is controlled from the rough estimate of the galaxy outer 
radius that \citet[][]{2015ApJS..219....4S} used to encompass the image region in the 2D photometric decompositions. 
In the construction of the average profile we must take into account that 
for some galaxies the extent of their profiles is limited by the image field of view. 
We therefore limit the averaging to those radii that are covered by at least $75 \%$ of the galaxies in the bin, unless stated otherwise. 
Uncertainties on the stacks are estimated from the standard deviation of the mean, $\sigma/\sqrt{N_{\rm gals}}$, 
where $N_{\rm gals}$ corresponds to the number of galaxies comprised in a certain bin.

We characterize the 1D radial profiles of UV surface brightness ($\mu$) after binning the sample based on 
the total stellar masses ($M_{\ast}$) of the host galaxies \citep[from][]{2015ApJS..219....3M}\footnote{$M_{\ast}$ was derived by 
\citet[][]{2015ApJS..219....3M} from 3.6~$\mu$m imaging using the calibration of the mass-to-light ratio by \citet[][]{2012AJ....143..139E}.}. 
In Fig.~\ref{Fig_mass_bars_NUV_FUV_1Dstack} we show the mean $\mu_{\rm NUV}$ and $\mu_{\rm FUV}$ obtained by 
scaling the density profiles to a common frame in physical units. The statistical dispersion of the luminosity profiles 
among the galaxies in each of the $M_{\ast}$-bins is shown in Fig.~\ref{Fig_mass_bars_FUV_1Dstack_disp}. 
It is larger in the FUV ($\sigma \le 1.3$ mag) than in NUV ($\sigma \le 1$ mag), 
while the standard deviation of the mean $\sigma/\sqrt{N_{\rm gals}}\lesssim 0.2$ mag 
at all radii due to the rich sampling, and hence the differences in the mean $\mu_{\rm FUV}$ and $\mu_{\rm NUV}$ for the different 
$M_{\ast}$ and bar family bins probed in this work are statistically significant.

The UV luminosity follows an exponential slope as a function of radius (Fig.~\ref{Fig_mass_bars_NUV_FUV_1Dstack}) 
\citep[see also Fig.~3 in][]{2018ApJS..234...18B}, with a  scale length that increases with increasing $M_{\ast}$. 
The luminosity profiles in the outskirts are brighter in more massive galaxies. 
When $M_{\ast}>10^{10}M_{\odot}$ a hump is detected in the inner regions, more clearly identified when $M_{\ast}>10^{10.5}M_{\odot}$. 
This feature is associated with the presence of bars. This is confirmed in Fig.~\ref{Fig_mass_bars_NUV_FUV_1Dstack_bars_separated}, 
where we study the mean $\mu_{\rm FUV}$ for barred and non-barred galaxies separately {(see also Fig.~\ref{Fig_mass_bars_NUV_1Dstack_bars_separated} 
for $\mu_{\rm NUV}$)}. 
When $M_{\ast}>10^{10}M_{\odot}$, barred galaxies have a deficit of FUV light within the bar region, 
which is not identified at the same radial distances in non-barred galaxies. 
Beyond that bar radius, the average FUV emission is again somewhat stronger in barred galaxies, 
hinting at a more active rate of SF in the spiral arms of barred galaxies, and 
possibly to the effect of bars redistributing gas across the disk. 
For the smaller $M_{\ast}$ bins, barred galaxies have stronger UV emission at all radii.

Finally, we test a possible causal connection between the detection of SF along the bar and the presence 
of inner rings \citep[e.g.,][and references therein]{2019A&A...627A..26N}, where gas can accumulate \citep[][]{1984MNRAS.209...93S} 
and no longer migrate inwards. We note that GALEX UV imaging has been used to study SF in 
rings in previous work in the literature \citep[e.g.,][]{2013A&A...555L...4C,2015BaltA..24..426K}. 
In Fig.~\ref{Fig_inner_rings_SMA} we show the mean FUV emission for the 
disk galaxies hosting inner rings and pseudorings \citep[according to][]{2015ApJS..217...32B}. 
The $\mu_{\rm FUV}$ profiles are scaled with respect to the deprojected ring semi-major axis \citep[from][]{2015A&A...582A..86H}. 
We find an FUV peak close to the ring radius, showing the intense SF taking place in rings. 
We note that the peak of mean SF is not located at the semi-major axis  (SMA) distance, but at $\sim 0.8\,$SMA, 
because inner rings are not intrinsically circular \citep[e.g.,][]{2014A&A...562A.121C}; 
in particular, the mean de-projected axis ratio of inner rings in the S$^4$G (inclinations lower than 65$^{\circ}$) is 
$0.76 \pm 0.01$ ($\sigma=0.12$), in the  range $0.4-1$ \citep[][]{2019A&A...625A.146D}. 

We study separately barred and non-barred galaxies, resulting in very similar radial FUV distributions. 
Non-barred inner-ringed  galaxies present slightly higher mean star formation rates (SFRs) along the disk than their barred counterparts. 
In Fig.~\ref{Fig_mass_bars_NUV_1Dstack_bars_separated} in Appendix~\ref{bars_NUV} we show similar profiles for the NUV, finding the same trends. 
We conclude that the spatial distribution of UV light in ringed galaxies is roughly the same for barred and non-barred galaxies; 
the implications are discussed in Sect~\ref{inner_ring_disc}.
%
%
\section{Visual classification of the distribution of SF within bars in individual galaxies}\label{individual}
%
%
In Sect.~\ref{stacks} we showed the statistical power of stacking techniques to characterize 
the SF activity in bars with a high signal-to-noise ratio and to detect low levels of SF. 
Nevertheless, by averaging hundreds of UV images we lose information on individual galaxies. 
In addition, the UV passbands are also not necessarily optimal for tracing the most recent SF bursts. 
Here, we compensate for these disadvantages by individually inspecting the distribution of  FUV (same dataset as in Sect.~\ref{stacks}, 
comprising 760 galaxies, with the inclusion of 12 additional images from the GALEX GR6/7 Data Release) 
and continuum-subtracted H$\alpha$ emission \citep[that traces SF in the last $\sim$20 Myr;][]{1998ARA&A..36..189K} 
in a large comprehensive sample with accurately determined disk and bar physical properties. 
%
%
\subsection{Compilation of H$\alpha$ images for S$^4$G barred galaxies and continuum subtraction}\label{compilations_halpha}
%
%
The sources of the H${\alpha}$ images used in this work are 
listed in Table~\ref{table_SF_class_sources} in Appendix~\ref{SF_class_sources}, 
and were mainly gathered from the NASA/IPAC Extragalactic Database (NED)\footnote{\href{http://ned.ipac.caltech.edu}{http://ned.ipac.caltech.edu}}. 
We started with the compilation of 281 continuum-subtracted images that were used in \citet[][]{2013A&A...555L...4C}, 
mostly from the \emph{Hubble} Space Telescope (HST) 
Archive\footnote{\href{http://archive.stsci.edu/hst/search.php}{http://archive.stsci.edu/hst/search.php}}. 
We  updated this compilation by adding 152 new images, making a final sample 433 S$^4$G galaxies 
with available H$\alpha$ continuum-subtracted imaging.

We produced additional H$\alpha$ continuum-subtracted images for 17 galaxies following 
\citet[][]{2004A&A...426.1135K} and \citet[][]{2006A&A...448..489K} 
\citep[see also][]{1999ApJS..124...95B}, namely four from the SPLUS survey \citep[][]{2019MNRAS.489..241M}, 
nine from JPLUS \citep[][]{2019A&A...622A.176C}, three from the ESO archive, and one (IC~1158) 
from the original compilation by \citet[][]{2013A&A...555L...4C}. 
We scaled $R$-band continuum images to match the intensity level of the continuum emission in the H$\alpha$ image. 
We measured the integrated intensity of at least six non-saturated foreground stars in the H$\alpha$ and 
$R$-band continuum images, and obtained the scaling factor from the average ratio of the intensities obtained for each star. 
If not enough foreground stars were available, 
we employed a second method in which the intensity of each pixel in the H$\alpha$ and $R$-band images were compared. 
If the  color is constant across the image, the relation is expected to be roughly linear, 
with deviations associated with strong emission-line regions. 
In order to reduce the scatter and avoid possible saturated pixels, we re-binned some of the images and removed $\sim 1-10 \%$ 
of the brightest pixels of each image. Finally, the scaling factor was obtained from the slope of the linear regression fit. 
For further details the reader is referred to \citet{2004A&A...426.1135K}, 
who showed that the two methods described above (scaling based on stars and on pixel-to-pixel matching) 
give similar values of the H$\alpha$ continuum-level. 

For 31 galaxies we also used state-of-the-art data gathered with the 
Multi-Unit Spectroscopic Explorer \citep[MUSE;][]{2010SPIE.7735E..08B} integral field unit at the Very Large Telescope (VLT). 
The mosaics have two sources. The first is the 
ESO Science portal \footnote{\href{http://archive.eso.org/scienceportal/home}{http://archive.eso.org/scienceportal/home}} 
where science-ready mosaics of several of our galaxies can be downloaded. 
We produced the others  by downloading the raw MUSE data from the ESO archive (see Table~\ref{table_SF_class_sources}). 
We reduced each exposure using the MUSE pipeline \citep{2012SPIE.8451E..0BW,2014ASPC..485..451W} 
under the \texttt{Reflex} environment \citep{2013A&A...559A..96F} using standard parameters. 
We manually aligned the exposures before combining them using the \texttt{muse\_exp\_combine} recipe to produce the final cube. 
For another 23 galaxies in our sample, fully reduced data cubes in the wavelength range 3750 - 7500 \AA\ exist from the 
CALIFA survey \citep{2012A&A...538A...8S,2014A&A...569A...1W,2016A&A...594A..36S} and are publicly available via 
the CALIFA DR3 website\footnote{\href{http://califa.caha.es/DR3}{http://califa.caha.es/DR3}}. 

We produced the continuum-subtracted H$\alpha$ maps from integral field unit (IFU) data 
by convolving with a filter of 20 \AA\ FWHM 
centered at the H$\alpha$ line (accounting for the Doppler shift using recession velocities taken from NED) and 
subtracting the adjacent continuum contribution within $\pm 50 \AA$, either manually (MUSE) or 
using PINGSoft\footnote{\href{https://www.inaoep.mx/~frosales/pings/html/software/}{https://www.inaoep.mx/$\sim$frosales/}} 
software \citep[][]{2011NewA...16..220R} (CALIFA). We checked the quality of the subtraction by comparing the resulting maps 
with archival H$\alpha$ continuum-subtracted images that existed for a few galaxies 
\citep[e.g., the image of IC$\,$0776  also appears in][]{2003A&A...400..451G}, in which case the CALIFA maps are not used (poorer resolution). 
Galaxies in which the bar was not fully covered by the IFU field of view were discarded.

In total, we produce continuum-subtracted H$\alpha$ images for 70 galaxies 
that are also used for the statistical analysis presented in this paper. 
All of the 433 continuum-subtracted H$\alpha$ images used here are 
publicly available at the CDS associated with this publication and at NED.
%
%
\begin{figure}
\centering
\includegraphics[width=0.5\textwidth]{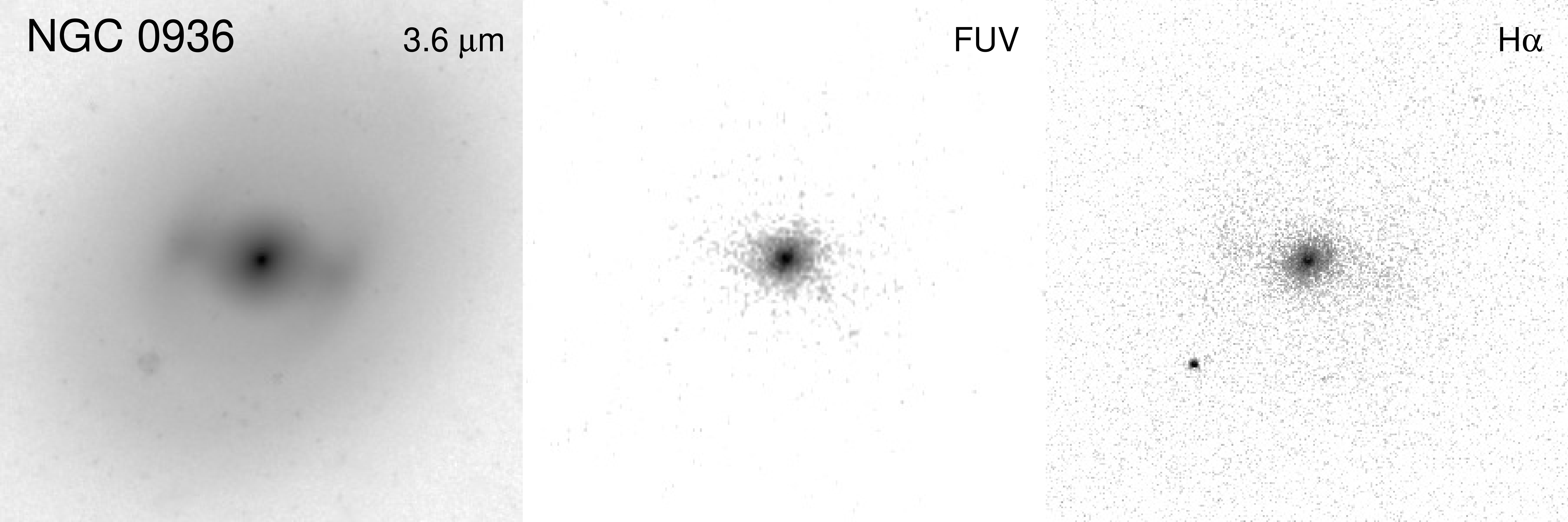}
\caption{
Illustrative example of SF class A (only circumnuclear star formation): NGC~0936. 
Shown are the \emph{Spitzer} 3.6~$\mu$m (S$^4$G) (left), GALEX FUV (center), 
and continuum-subtracted H$\alpha$ images \citep[right; from][but subtraction   performed by us]{2019MNRAS.489..241M}.
}
\label{Fig_classA}
\end{figure}
%
%
\begin{figure}
\centering
\includegraphics[width=0.5\textwidth]{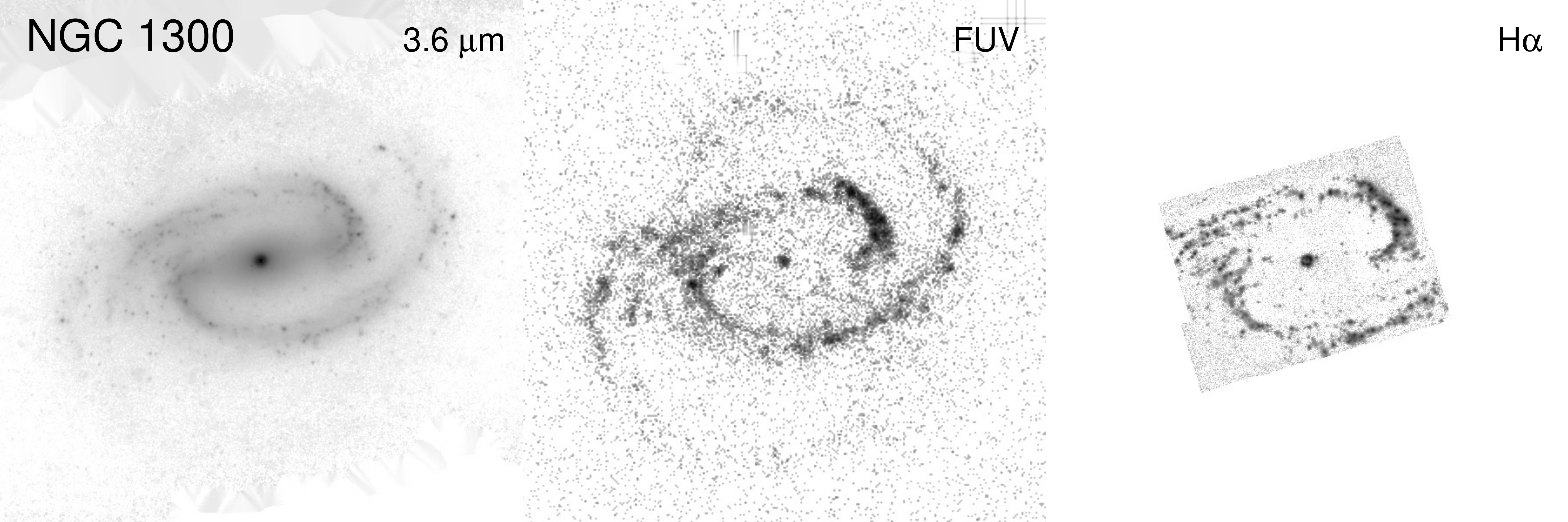}
\includegraphics[width=0.5\textwidth]{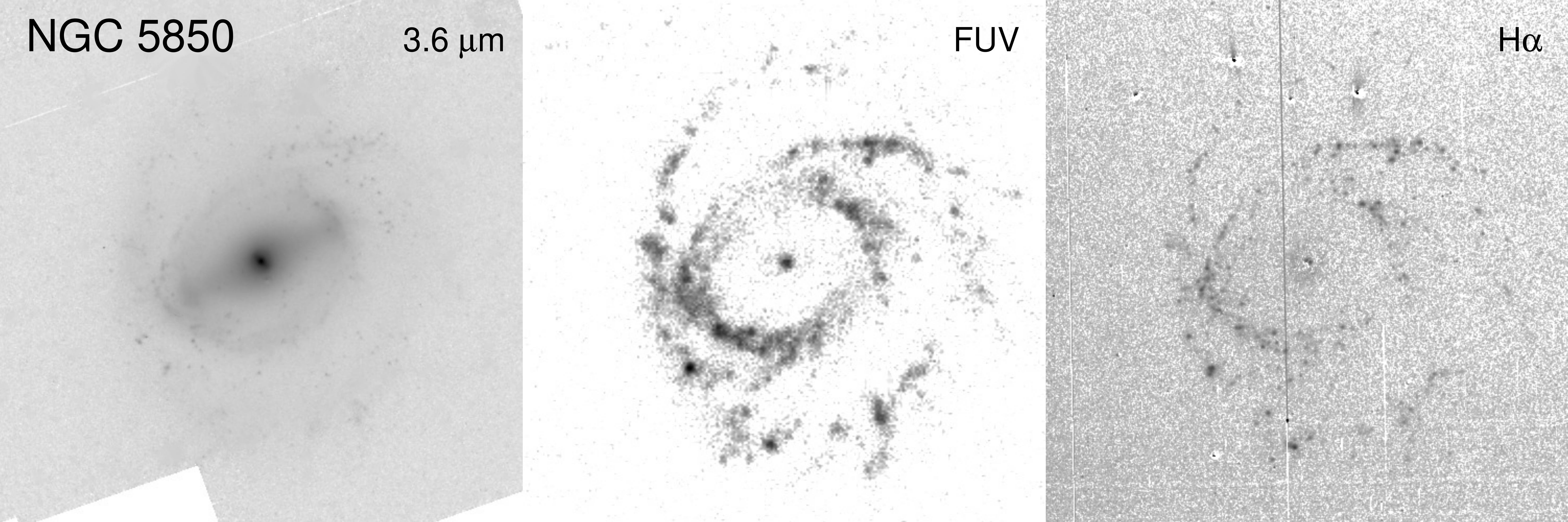}
\caption{
As in Fig.~\ref{Fig_classA}, but for SF class B 
(star formation at the bar ends, but not along the bar) 
and subclass ``a'' (circumnuclear SF): NGC~1300 (top) and NGC~5850 (bottom). 
H$\alpha$ images are from \citet[][]{2004A&A...426.1135K}.
}
\label{Fig_classB}
\end{figure}
%
%
\begin{figure}
\centering
\includegraphics[width=0.5\textwidth]{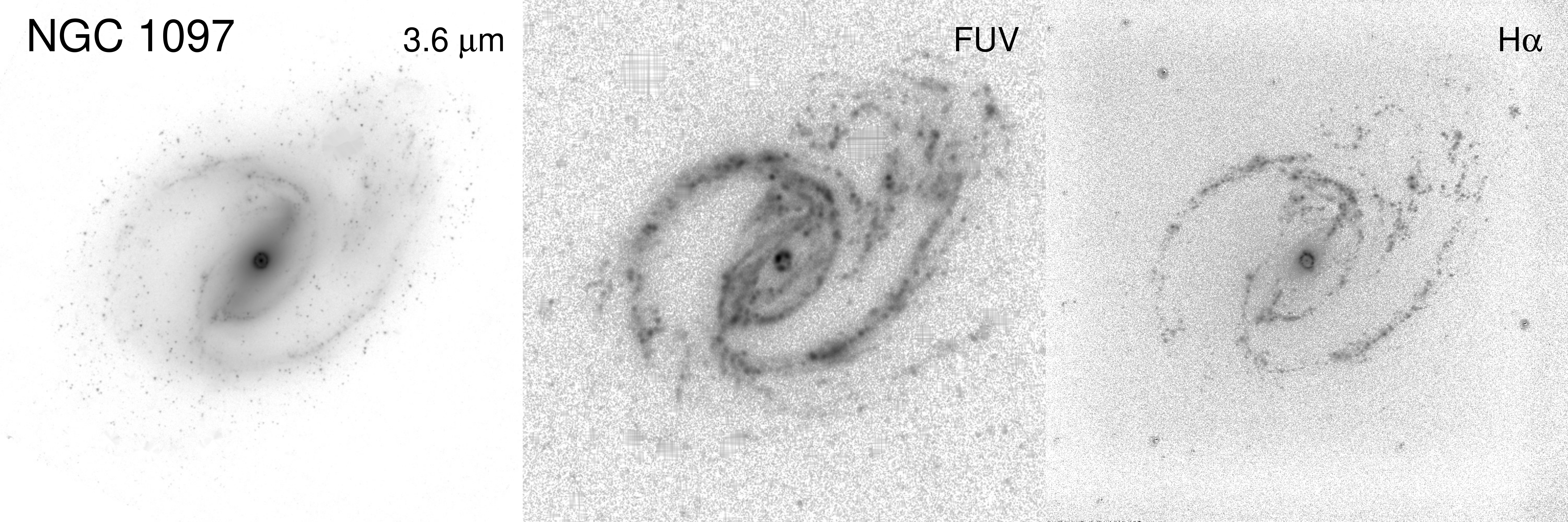}
\includegraphics[width=0.5\textwidth]{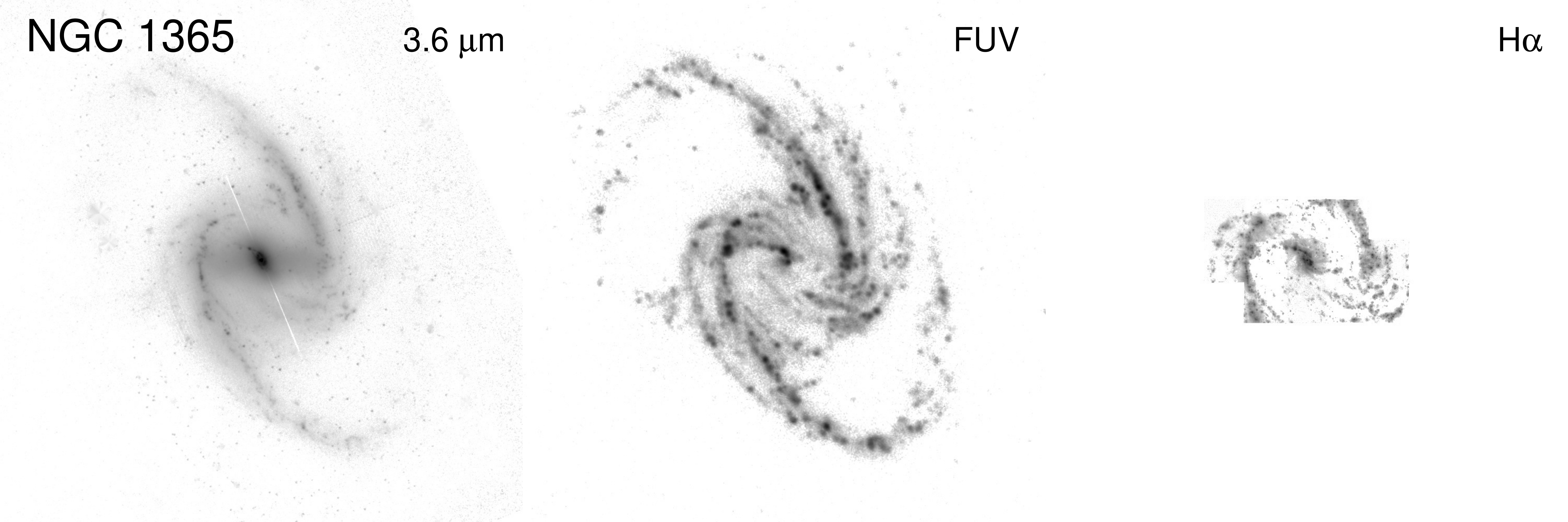}
\includegraphics[width=0.5\textwidth]{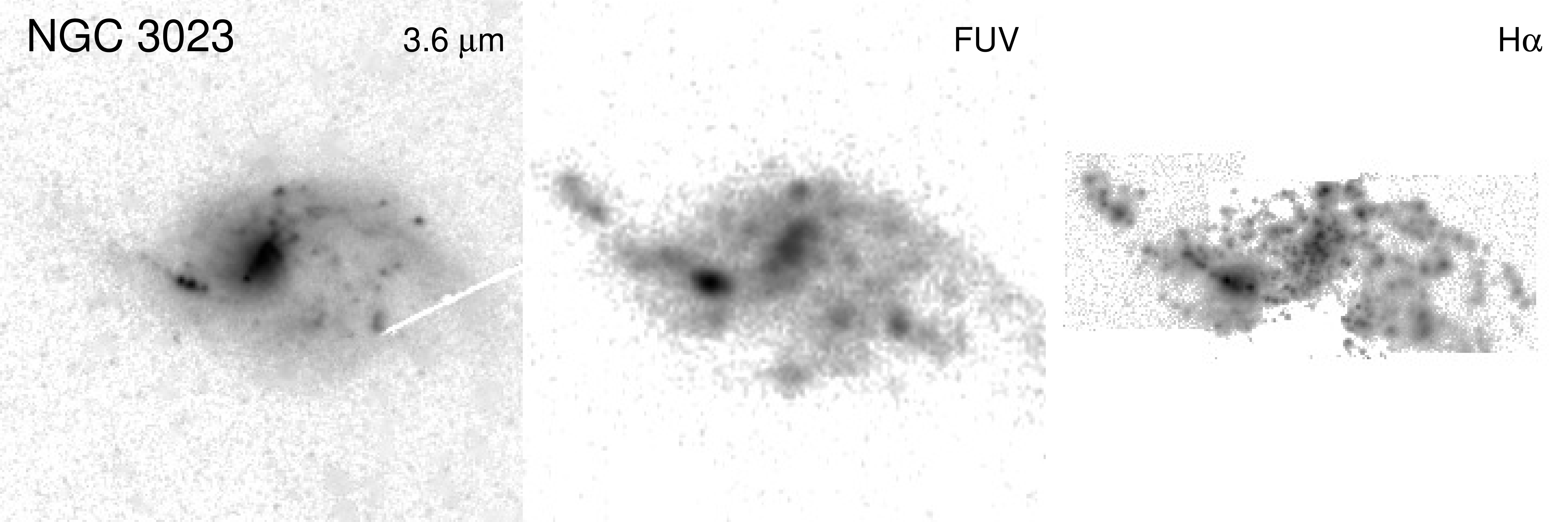}
\caption{
As in Fig.~\ref{Fig_classA}, but for SF class C (galaxies with SF along the bar) and subclass ``a'' (circumnuclear SF): 
NGC~1097 (top), NGC~1365 (middle), and NGC~3023 (bottom). 
H$\alpha$ images are respectively from \citet[][]{2003PASP..115..928K} and from the ESO archive (see Table~\ref{table_SF_class_sources}) (mosaics and continuum-subtraction performed by us from the MUSE archive, 
covering a smaller field of view than the FUV and 3.6~$\mu$m images).
}
\label{Fig_classC}
\end{figure}
%
%
\subsection{Classification method}\label{class_met}
%
%
\begin{table}
\centering
\small
\begin{tabular}{| c| c|}
\hline
\multicolumn{1}{|c}{\centering SF class} & \multicolumn{1}{|c|}{\centering Distribution of star formation} \tabularnewline
\hline
\hline
A & SF only in the bar central region.\\
\hline
B & SF at the ends of the bar, but not along the bar;\\
\hspace{0.75cm} Ba & with circumnuclear SF, \\
\hspace{0.75cm} Bb & without circumnuclear SF. \\
\hline
C & SF along the bar;\\
\hspace{0.75cm} Ca & with circumnuclear SF,\\
\hspace{0.75cm} Cb & without circumnuclear SF.\\
\hline
N & No flux detection.\\
\hline
U & Uncertain classification.\\
\hline
\end{tabular}
\caption{
Classification system of bar SF classes adopted in this work.
}
\label{sf_system}
\end{table}
%
%
\begin{table}
\centering
\small
\begin{tabular}{| c| c| c| c| c| c|}
\hline
\multicolumn{1}{|c}{\centering Sample} & \multicolumn{1}{|c}{\centering A} & \multicolumn{1}{|c}{\centering B} & \multicolumn{1}{|c}{\centering C} & \multicolumn{1}{|c}{\centering N} & \multicolumn{1}{|c|}{\centering U} \tabularnewline
\hline
\hline
FUV & 100 & 196 & 382 &  11 &  83 \\
H$\alpha$ & 54 & 129 & 215 &  19 &  16 \\
\hline
\end{tabular}
\caption{
Number of galaxies belonging to each SF class A-B-C (see Table~\ref{sf_system}) 
in the samples with available FUV (upper row) and H$\alpha$ (lower row), 
and number of cases without emission (N) or uncertain classifications (U).
}
\label{number_classified}
\end{table}
%
%
We devise a classification method in which the distribution of SF at the bar region is 
assigned to a class (hereafter SF class) using  
criteria similar to those used in   \citet[][]{2007A&A...474...43V} \citep[see also][]{1997A&A...326..449M,2019A&A...627A..26N,2020MNRAS.495.4158F}. 
Our classification system is given in Table~\ref{sf_system}. 

In Fig.~\ref{Fig_classA} we display an illustrative example (NGC~0936) of class A (only circumnuclear SF), 
showing the 3.6~$\mu$m, GALEX FUV, and continuum-subtracted H$\alpha$ images; in 
Fig.~\ref{Fig_classB} we show those of NGC~1300 (top) and NGC~5850 (bottom), of class B (SF at bar ends, but not along the bar); 
and in  Fig.~\ref{Fig_classC} we show the images of NGC~1097 (top), NGC~1365 (middle), and NGC~3023 (bottom), which belong to class C (SF along the bar). 
In Table~\ref{table_SF_class_sources} in Appendix~\ref{SF_class_sources} we list the SF class assigned to each galaxy in our sample. 
For classes B and C, subclasses are also considered depending on whether we detect circumnuclear SF (``a'') or not (``b''); 
these are also listed in Table~\ref{table_SF_class_sources}, but are not analyzed here. 
We also note that in a number of cases (94  in FUV and 35 in H$\alpha$) we 
either did not detect SF or could not reliably classify its distribution.

The assignment of a SF class to each galaxy in our sample was performed by F.D.M., 
who examined the whole sample twice (nine months time-spacing, allowing him to re-visit contradicting cases), 
consistently obtaining the same statistical trends (see next sections). 
The classifications of the FUV and H$\alpha$ samples were done independently, so that the visual analysis was unbiased. 
Images were navigated using \emph{SAOImage DS9} and the constrast was varied to make the H{\sc\,ii} knots, clumps, and filaments stand out. 

In  Table~\ref{number_classified} we indicate the number of galaxies classified in each category A, B, and C, as well as the 
number of    non-detections (N) and uncertain cases (class U). Of 772 barred galaxies with available FUV imaging, 
the percentages (and binomial errors) of SF classes A, B, and C are $13 \pm 1.2 \%$, $25.4 \pm 1.6 \%$, and $49.5 \pm 1.8 \%$, respectively, 
while $1.4 \pm 0.4 \%$ present no emission (N) and $10.8 \pm 1.1 \%$ are uncertain (U). 
For the 433 with available H$\alpha$ images, the percentages are consistently 
$12.5 \pm 1.6 \%$, $30 \pm 2.2 \%$, and $49.7 \pm 2.4 \%$ for SF classes A, B, and C, 
while $4.4 \pm 1 \%$ and $3.7 \pm 1 \%$ belong to classes N and U, respectively. 
We note that statistical trends of SF classes presented in the next sections are roughly the same regardless of the   passband used; however,  in some cases the classification in H$\alpha$ is not the same as in FUV (23 cases in class A, 29 in class B, 3 in class C) 
mainly as a consequence of differences in the traced SF timescales in the two passbands, 
image depth, resolution, size and irregularity of bars, or unavoidable subjectivity.

In addition, we reassess the number of inner rings that are active, expanding the work by \citet[][]{2013A&A...555L...4C} 
by enlarging his collection of H$\alpha$ images for the barred galaxies in the S$^4$G. 
We focus on the inner-ringed galaxies identified by \citet[][]{2015ApJS..217...32B} in \emph{IRAC} 3.6 $\mu$m S$^4$G images. 
An example of a galaxy  with an active inner ring (NGC~5850) is shown in Fig.~\ref{Fig_classB}. 
In Table~\ref{table_SF_class_sources} we indicate whether inner rings are active (rA) or passive (rP). 
Using FUV images the number of barred galaxies classified as rA is 253 ($90.4 \pm 1.8 \%$); 
only 23 ($8.2 \pm 1.6\%$) belong to class rP, while 4 ($1.4 \pm 0.7\%$) are uncertain (rU). 
When using H$\alpha$ these numbers are 175 ($80.6 \pm 2.7\%$), 35 ($16.1 \pm 2.5\%$), and 7 ($3.2 \pm 1.2 \%$), respectively. 
Further analysis is presented and discussed in Sect.~\ref{inner_ring_disc}.

%
%
\begin{figure}
\includegraphics[width=0.5\textwidth]{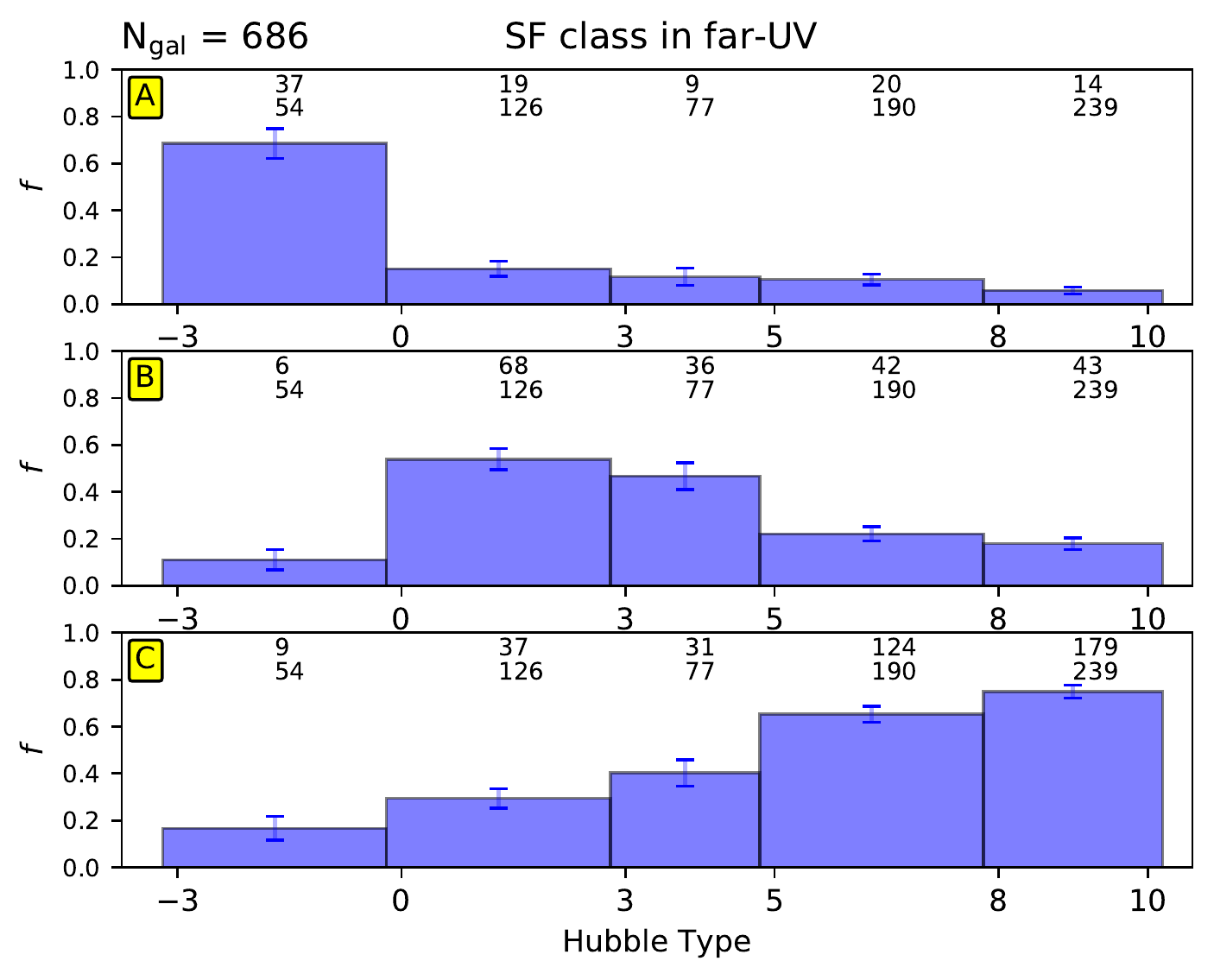}
\includegraphics[width=0.5\textwidth]{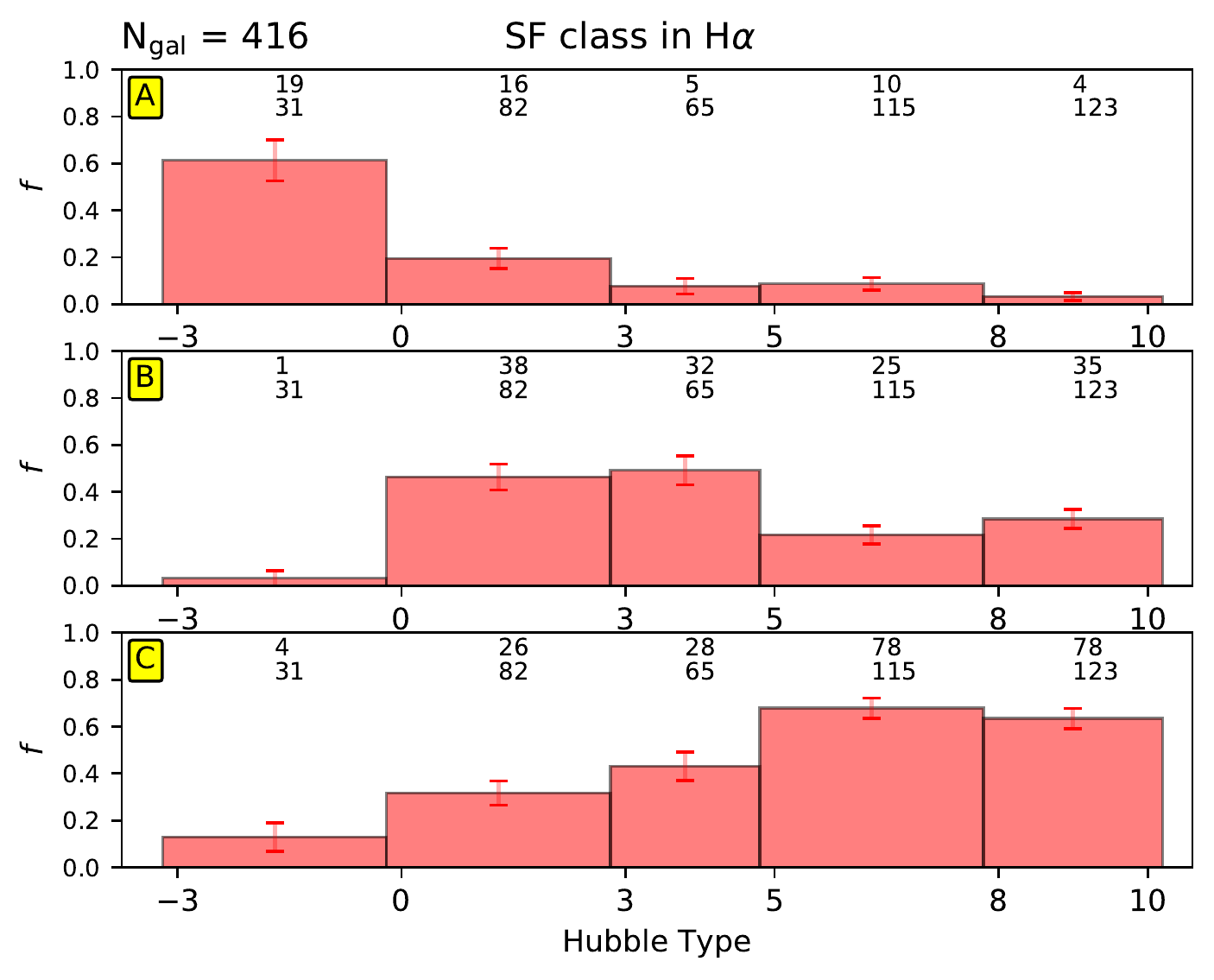}
\caption{Fraction of SF classes as a function of the revised Hubble stage, 
classified based on the FUV (upper panels, in blue) and continuum-subtracted H$\alpha$ emission (lower panels, in red). 
The subpanels correspond to SF classes A (only circumnuclear SF, \emph{top}), 
B (SF at the bar ends, but not along bar, \emph{middle}), and C (SF along bar, \emph{bottom}). 
Error bars correspond to binomial errors. 
Indicated for each category is the number of analyzed galaxies within the bin (bottom row of numbers), 
and the number of identified cases in each SF class (top row of numbers). 
In the upper left corners the total number of analyzed galaxies ($N_{\rm gal}$) are given.
}
\label{THUBBLE}
\end{figure}
%
%
\subsection{Frequency of SF categories as a function of morphological type}\label{ttype_SF}
%
%
Our goal is to determine whether different distributions of SF in bars depend on the global properties of the host galaxy, 
focusing on the galaxies with SF classes A, B, and C, and excluding from the analysis uncertain cases (U). 
In Fig.~\ref{THUBBLE} we show the fraction of SF classes as a function of the morphological type of the host galaxies from 
\citet[][]{2015ApJS..217...32B}, including binomial counting errors (error bars), 
using classifications based on  FUV and continuum-subtracted H$\alpha$ images. 

The histograms of the frequency of the three SF classes are significantly different. 
For SF class A, the distribution peaks for  S0s ($\sim 60-70 \%$) and drops among the spirals ($\le 20 \%$). 
SF class B is dominant in early- and intermediate-type spirals ($\sim 40-60 \%$), 
and is  a factor of $\sim 2$ higher than in their late-type counterparts. 
A negligible amount of lenticulars belong to SF class B. Lastly, for SF class C, the fraction increases with increasing Hubble type: 
a maximum frequency of $\sim 60-75 \%$ is found for Sc and irregular galaxies ($5 \le\emph{T}\le 10$), 
while a marginal $\sim 10-20 \%$ is found for S0s. 
In conclusion, the modes of the statistical distributions of SF classes A, B, and C are clearly segregated in the Hubble sequence, 
even though examples of all SF classes can be found for a given $T$-bin. 
These reported trends are qualitatively the same regardless of the passband used, either FUV or H$\alpha$.
%
%
\subsection{Frequency of SF categories as a function of total stellar mass}\label{mstar_SF}
\begin{figure}
\centering
\includegraphics[width=0.5\textwidth]{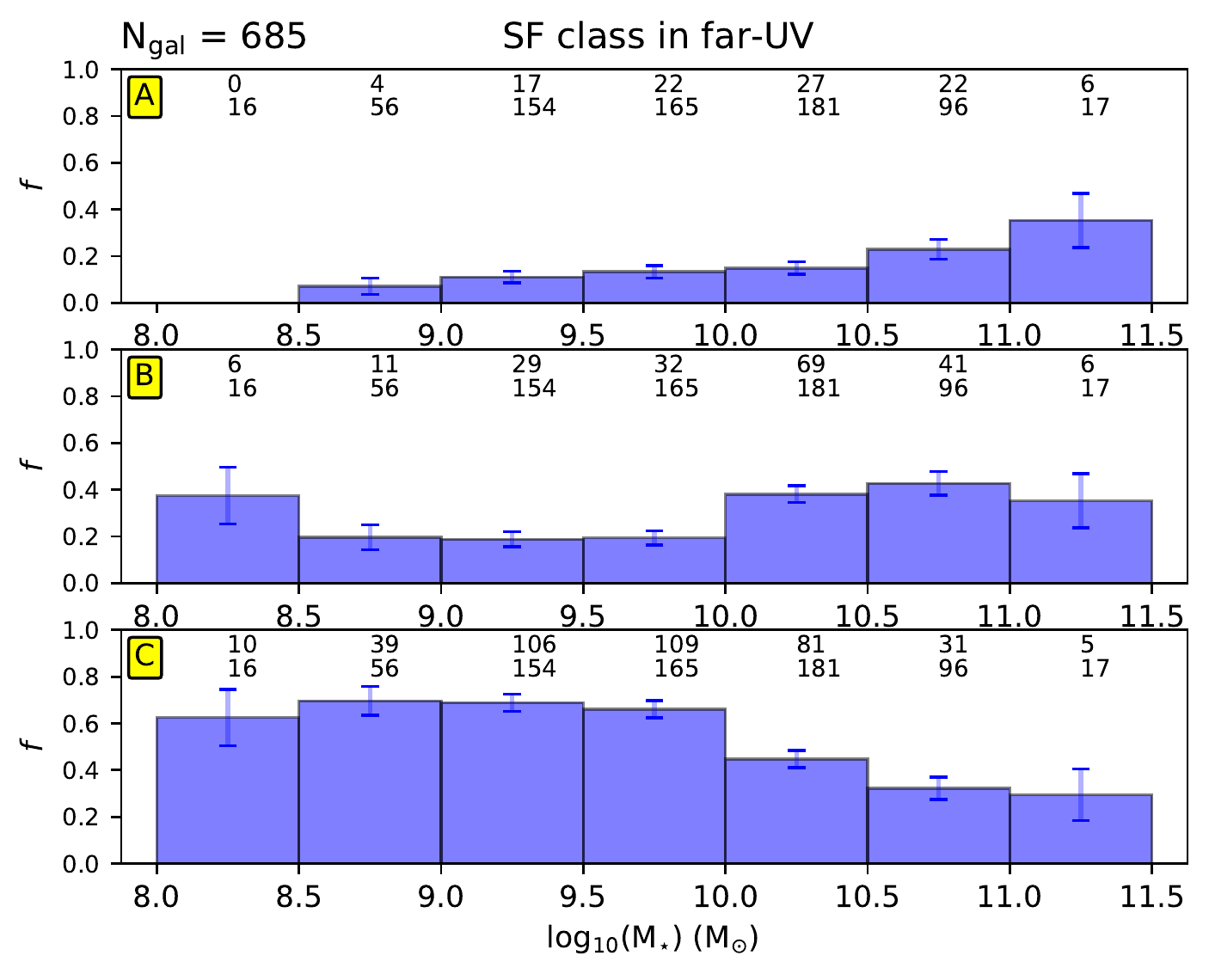}
\includegraphics[width=0.5\textwidth]{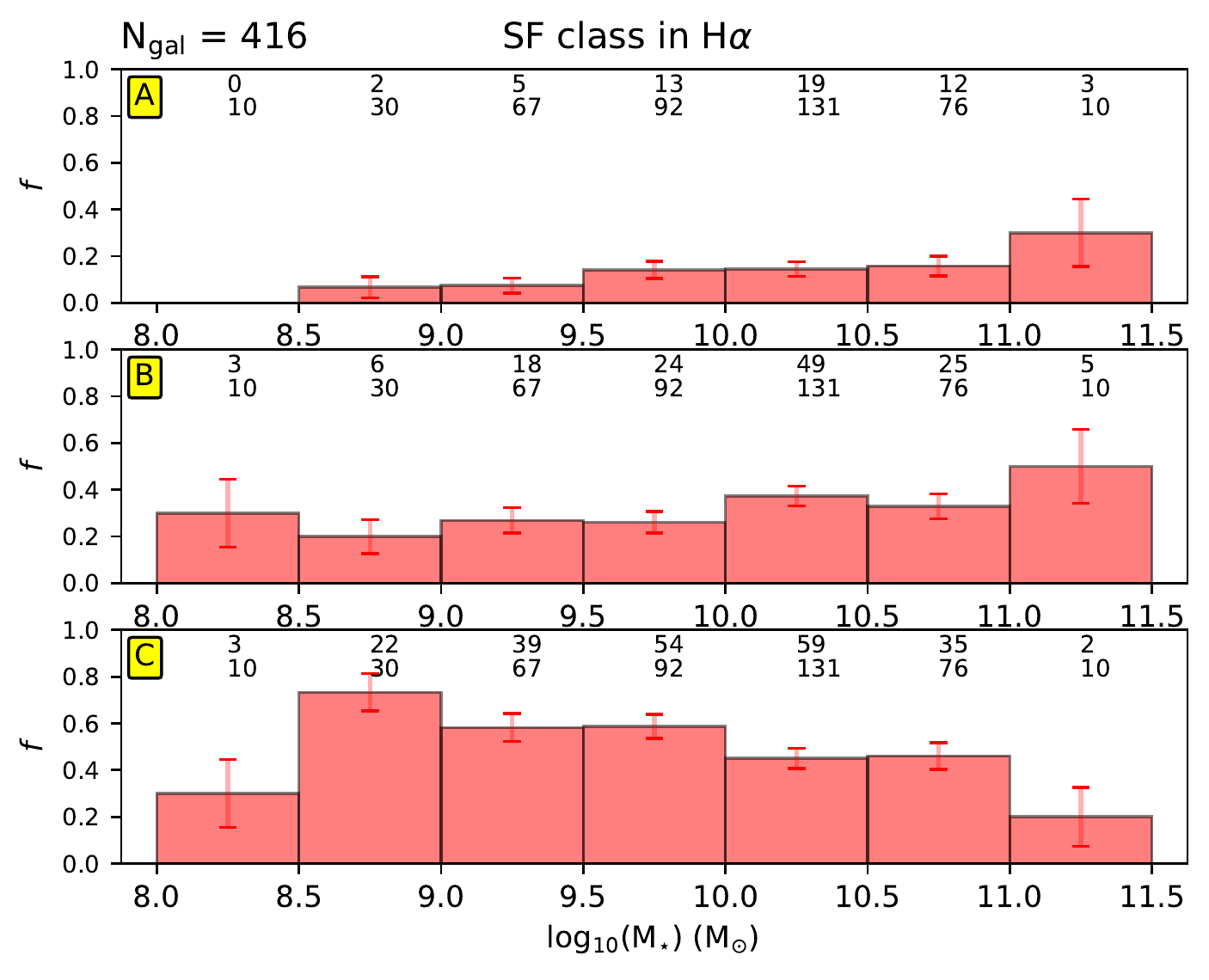}
\caption{
As in Fig.~\ref{THUBBLE}, but as a function of total stellar mass of the host galaxy, in bins of 0.5 dex and in units of solar masses.
}
\label{MSTAR}
\end{figure}

%
%
While much can be learned from the study of galaxy properties in the Hubble sequence, 
it is also convenient to use quantifiable physical parameters such as the total stellar mass. 
In Fig.~\ref{MSTAR} we show the frequency of SF classes as a function of $M_{\ast}$, 
where clear trends stand out: e.g., the fainter the galaxy, the more frequent the SF class C is. 
That is, low-mass galaxies tend to host bars that are actively forming stars along the whole extent of the bar. 

Specifically, $\gtrsim 60 \%$ of the galaxies with $M_{\ast}<10^{10}M_{\sun}$ belong to SF class C, 
and the fraction declines with increasing $M_{\ast}$, very clearly in the FUV sample. 
Even so, $\sim 50 \%$ of the galaxies with $10^{10}M_{\odot}<M_{\ast}<10^{11}M_{\odot}$ are classified as C in the H$\alpha$ sample. 
The fraction of SF class A peaks for the highest $M_{\ast}$-bin. In general, the frequency of class B is higher among massive systems 
($39.5\pm2.9\%$ and $36.4\pm3.3\%$ at FUV and H$\alpha$, respectively, when $M_{\ast}>10^{10}M_{\sun}$) than 
among their faint counterparts (fractions of $20\pm 2\%$ and $25.6\pm 3.1\%$), but the histogram is not peaked.
%
%
%
%
\subsection{Frequency of SF categories as a function of gas fraction}\label{gas_frac}
\begin{figure}
\centering
\includegraphics[width=0.5\textwidth]{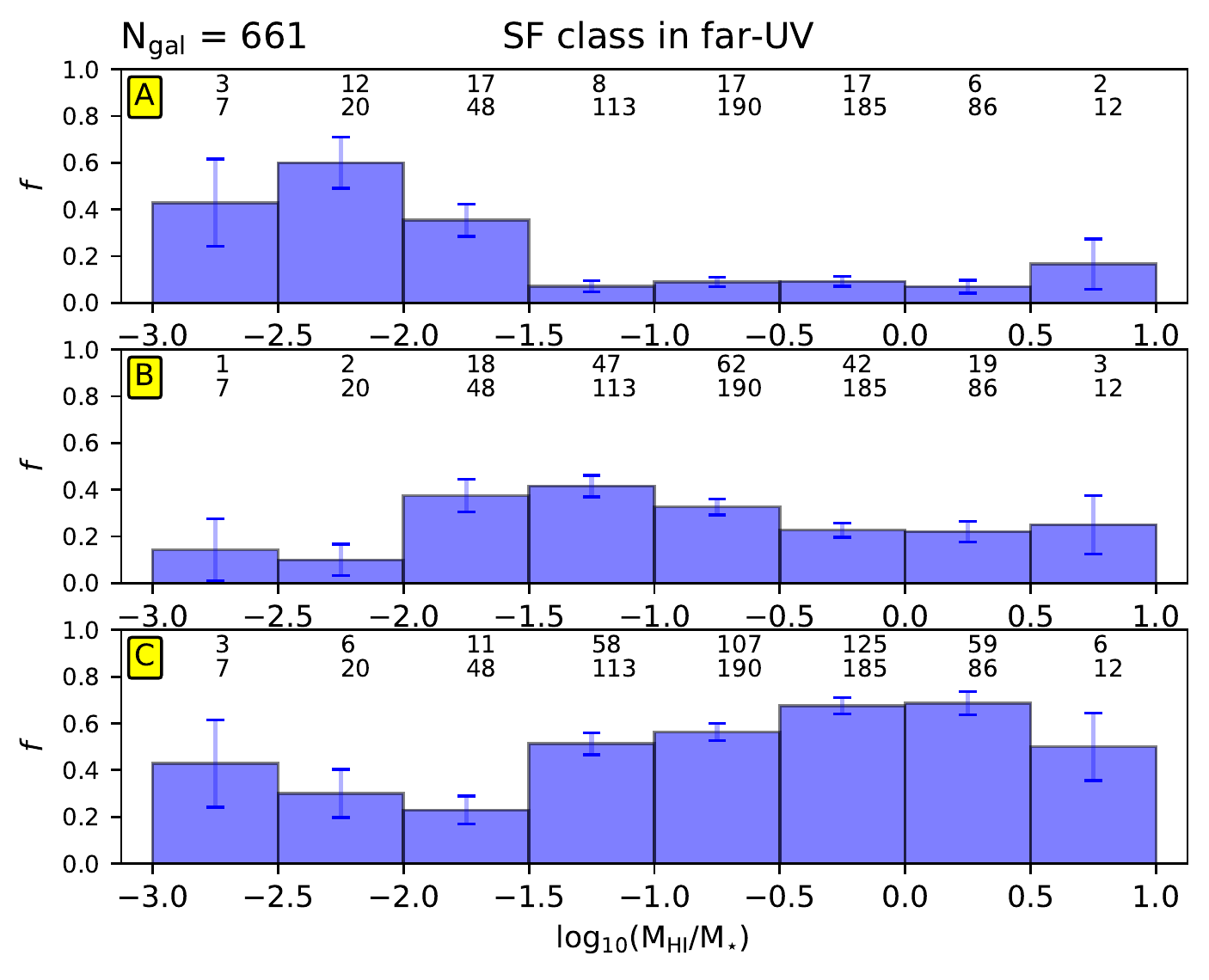}
\includegraphics[width=0.5\textwidth]{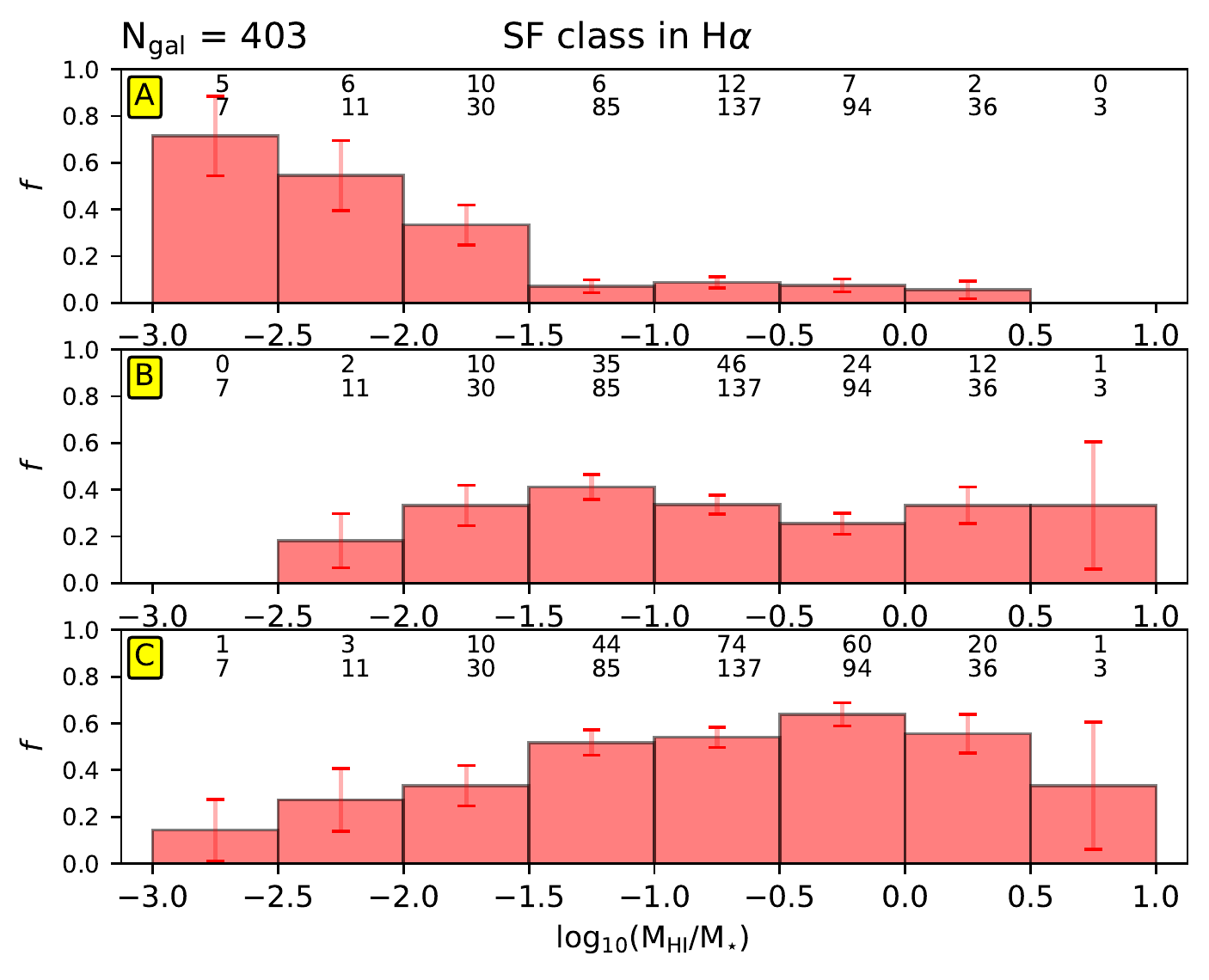}
\caption{
As in Fig.~\ref{THUBBLE}, but as a function of the H{\sc\,i} gas fraction (relative to the total stellar mass), in bins of 0.5 dex.
}
\label{MHIMSTAR}
\end{figure}

%
%
The global content of atomic hydrogen in the galactic disk indicates 
the principal fuel reservoir for SF, even though the main fuel for SF is molecular gas. 
In Fig. \ref{MHIMSTAR} we show the frequency of SF classes against the relative content of atomic gas (i.e., 
the mass of H{\sc\,i} gas normalized by the total stellar mass). 
Atomic gas masses are estimated as \citep[e.g.,][]{1988gera.book..522G,2018MNRAS.474.5372E,2019A&A...625A.146D}
\begin{equation}\label{gasfrac}
M_{\rm HI}=2.356 \cdot 10^5 \cdot D^2 \cdot 10^{0.4 \cdot (17.4-m21c)},
\end{equation}
where $m21c$ is the corrected 21 cm line flux in magnitude from HyperLEDA and 
$D$ is the distance to the galaxy (in megaparsecs) adopted by \citet[][]{2015ApJS..219....3M}. 
We confirmed the good agreement between our heterogenous $M_{\rm HI}$ estimates and those available 
from The Arecibo Legacy Fast ALFA Survey \citep[ALFALFA;][]{2011AJ....142..170H} for 
the overlap between the two samples (less than $25\%$ of our galaxies).

The distribution of the three main SF categories are somewhat different when studied versus 
$M_{\rm HI}$/$M_{\ast}$, resembling the behavior in the Hubble sequence (Sect.~\ref{ttype_SF}). 
The frequency of SF class A (C) decreases (increases) with increasing gas fraction. 
The mode of SF class B ($\sim 40 \%$) occurs for intermediate gas 
fractions with $-1.5 \le {\rm log_{10}} (M_{\rm HI}/M_{\ast}) \le -1.0$, 
and its distribution is rather flat ($\sim 20-30\%$) among the gas-rich galaxies.
%
%
%
%
\subsection{Frequency of SF categories as a function of gravitational torque}\label{Qb_sec}
%
%
We finally study the frequency of SF categories as a function of the gravitational torque measured by the 
tangential-to-radial force ratio ($F_{\rm T}/\left<F_{\rm R}\right>$) 
\citep[e.g.,][]{2001ApJ...550..243B,2002MNRAS.331..880L,2002MNRAS.337.1118L,2004ApJ...607..103L}. 
This test is especially relevant in that it sheds light on the physics driving SF across bars (see Sect.~\ref{dist_SF}): 
tangential forces trace bar-induced gas motions, while radial forces control circular velocities in the inner parts, 
and thus the degree of shear \citep[e.g.,][]{2005MNRAS.359.1065S}. 
In particular, we use the radial force profiles derived by \citet[][]{2016A&A...587A.160D} from $3.6\,\mu$m S$^4$G imaging, 
following \citet[][]{1981A&A....96..164C}: 
\begin{equation}\label{torquerad}
Q_{\rm T}(r)=\frac{{\rm max}\left( |F_{\rm T}(r,\phi)| \right)}{\langle |F_{\rm R}(r,\phi)|\rangle}.
\end{equation}
Here $r$ and $\phi$ refer to the radial distance and azimuthal angle, respectively. 
Specifically, the maximum of $Q_{\rm T}$ at the bar region is used as proxy of the bar-induced perturbation 
strength \citep[e.g.,][]{2001ApJ...550..243B,2019A&A...631A..94D}, called $Q_{\rm b}$. 
We note that at the bar region the unaccounted dark halo contribution to radial forces is likely to be only minor, 
and becomes somewhat important for later types, implying a reduction of $\sim 20-25\%$ on $Q_{\rm b}$ for $T = 7-10$ \citep[][]{2016A&A...587A.160D}.

\begin{figure}
\centering
\includegraphics[width=0.5\textwidth]{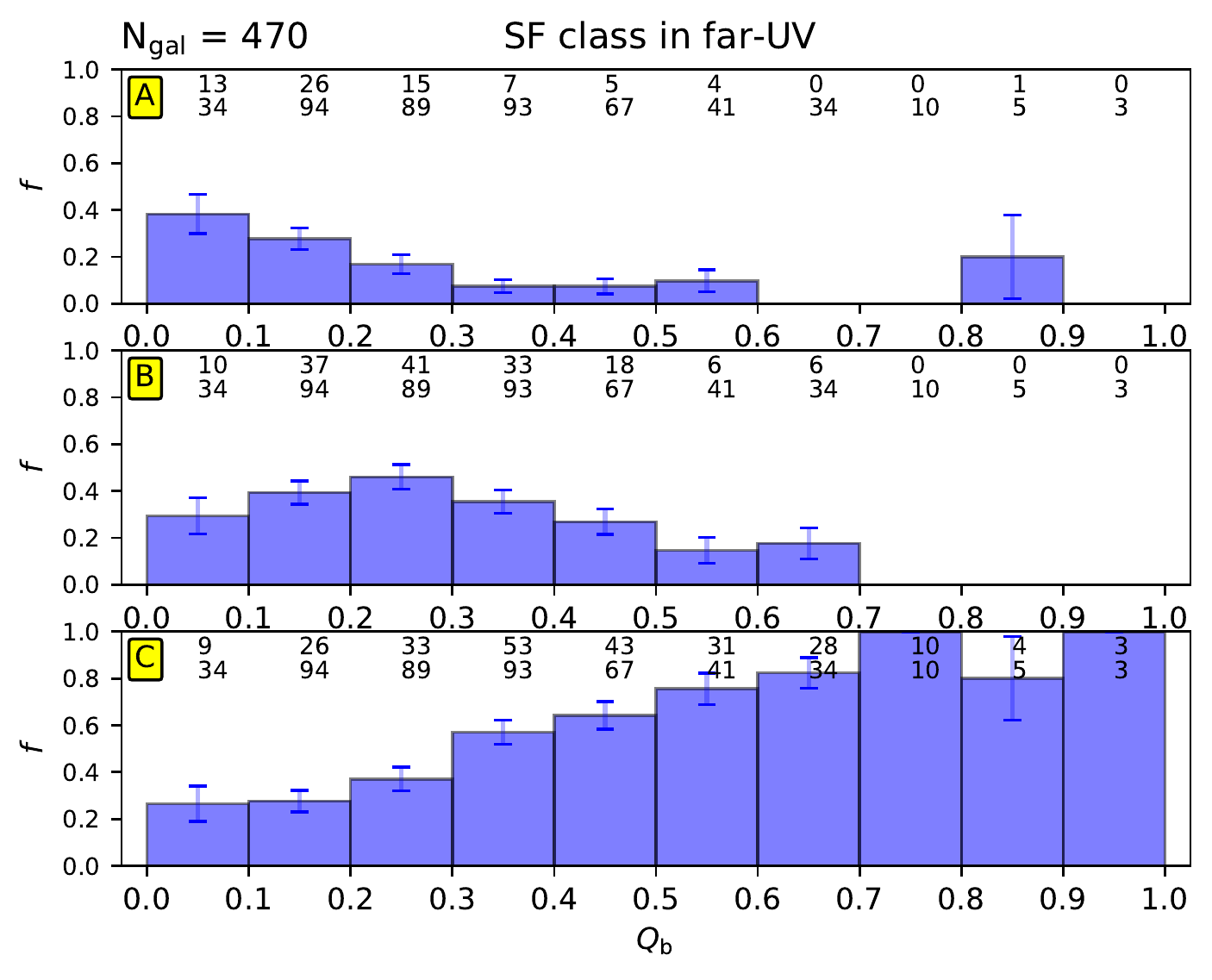}
\includegraphics[width=0.5\textwidth]{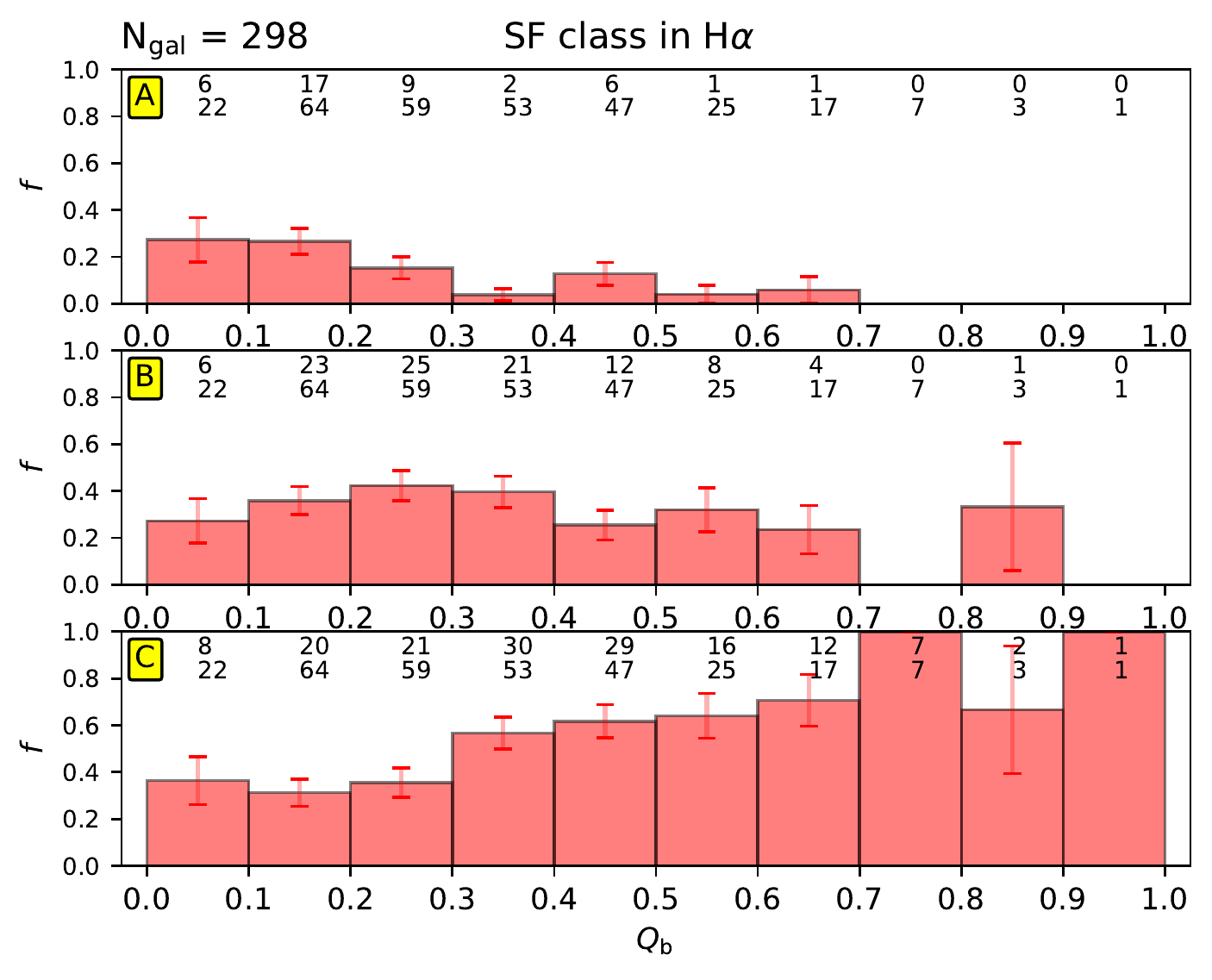}
\caption{
As in Fig. \ref{THUBBLE}, but as a function of the bar strength, measured from the maximum of the 
tangential-to-radial force ratio at the bar region, using bins of 0.1.
}
\label{QB}
\end{figure}

The fraction of SF categories versus $Q_{\rm b}$ is shown in Fig.~\ref{QB}, 
confirming differences in the distribution of SF classes identified earlier. 
The occurrence of SF class B peaks at $\sim 40 \%$ for $0.2 \le$ $Q_{\rm b} \le 0.3$, and smoothly 
decreases towards both weaker and stronger bars. 
This is an intermediate case between SF classes A and C, which hints at a physical transition of SF and local dynamical conditions in bars. 
SF class C is more typical of barred galaxies with high gravitational torques (strong bars); 
there is a sharp increase in its frequency with increasing $Q_{\rm b}$.
%
%
%
%
\section{Discussion}\label{discussion_chapter}
%
%
We report differences in the statistical distributions of star-forming and passive bars in the S$^4$G survey, 
pointing to the influence of global morphological and physical properties on the distribution of SF activity in the central regions of galaxies. 
The S$^4$G is representative of the local Universe despite not being complete in any quantitative form (e.g., volume) in its current version. 
It is currently under completion with analysis of new \emph{Spitzer} 3.6 $\mu$m for early-type galaxies 
with $T \le 0$ \citep[][]{2013sptz.prop10043S}, and new ground-based $i$-band imaging for relatively gas-poor late-type galaxies.

We use archival GALEX FUV and NUV imaging of 772 barred galaxies and a 
compilation of 433 continuum-subtracted H$\alpha$ images combining CALIFA and MUSE IFU data cubes with archival imaging of better resolution 
and employ both stacking techniques and visual classifications. 
Here, we discuss the statistical trends reported in Sect.~\ref{stacks} (stacking techniques) and Sect.~\ref{individual} 
(visual inspection of images), which consistently yield similar results in all used passbands (H$\alpha$, NUV, and FUV), 
and their importance in  shedding light on the regulation of the SF activity by bars. 
%
%
\subsection{Evidence for bar-induced secular evolution in the central regions of disk galaxies}
%
%
The torques exerted by stellar bars are expected to provoke the flow of gaseous and stellar material within the disk, 
driving the secular evolution of the inner parts of the galaxy, as established in simulation models since 
the 1990s \citep[e.g.,][]{1992MNRAS.259..328A,1992MNRAS.258...82W,1993A&A...268...65F,1993RPPh...56..173S,1995ApJ...454..623K,2004A&A...424..799P,2016MNRAS.462L..41F}. 
Dark matter bars \citep[][]{2016MNRAS.463.1952P,2019MNRAS.488.5788C}, if present, might also contribute to this effect. 
Evidence of gas streams in barred spirals was reported in the 1960s \citep[e.g.,][studying NGC$\,$4027 and NGC$\,$7741]{1963AJ.....68..278D}. 
The funneled gas might eventually be spent in central starbursts, 
and contribute to the buildup and evolution of disk-like bulges \citep[][]{2004ARA&A..42..603K}.

Observational evidence of secular evolution within the bar region 
comes from detections of inner rotating stellar and gaseous substructures in the center of barred 
galaxies \citep[e.g.,][and references therein]{2008A&A...485..695C,2009A&A...495..775P,2013seg..book....1K,2014A&A...572A..25M,2015MNRAS.451..936S}, 
including nuclear rings \citep[e.g.,][]{1995ApJ...454..623K,2005A&A...429..141K,2010MNRAS.402.2462C,2019MNRAS.488.3904L} and 
inner bars \citep[e.g.,][]{2002AJ....124...65E,2019MNRAS.484.5296D,2019MNRAS.482L.118M}; from enhanced central SF 
and chemical abundances \citep[e.g.,][]{2011MNRAS.416.2182E,2012ApJS..198....4O,2015A&A...584A..88F,2016A&A...595A..63V,2017ApJ...848...87C,2020MNRAS.tmp.2737L}; 
and in general from stellar populations analyses \citep[e.g.,][]{2007A&A...465L...9P,2009A&A...495..775P,2011ApJ...743L..13C,2011MNRAS.415..709S,2011A&A...529A..64P,
2012MNRAS.420.1092D,2013MNRAS.431.2397D,2014A&A...570A...6S,2015A&A...584A..90G,2016MNRAS.460.3784S,2017MNRAS.470L.122P,2019MNRAS.488L...6F,
2019MNRAS.482..506G,2020A&A...637A..56N}.

On the other hand, the gas swept by bars can eventually fuel active galactic nuclei (AGN). However, in spite of 
the proposed mechanisms based on numerical simulations \citep[e.g.,][]{1989Natur.338...45S,2015MNRAS.446.2468E}, 
it remains unclear how galaxies drive this gas to the central $\sim 100$~pc to feed the supermassive black holes 
\citep[e.g.,][and references therein]{2000ApJ...529...93K,2004cbhg.symp..186W,2009ASPC..419..402H,2012ApJ...750..141L,2013ApJ...776...50C}.
%
%
\subsection{Insights from UV stacking: strong bars redistribute gas and nourish central star formation}
%
%
Based on the analysis of mean stellar density profiles derived from 3.6 $\mu$m images with a large unbiased sample, 
\citet[][]{2016A&A...596A..84D} and \citet[][]{2017ApJ...835..252S} provide evidence for bar-induced secular evolution of disk galaxies in 
terms of enhanced central mass concentration. In Sect.~\ref{bar_uv_stack} we describe how we applied these same averaging techniques 
to obtain mean bars and disks at UV wavelengths, tracing SF up to $10^{8}$~yr. 
We used GALEX images of S$^4$G galaxies from \citet[][]{2018ApJS..234...18B} 
and the parameterization of bars (length, position angle, shape, and strength) 
at 3.6~$\mu$m from \citet[][]{2015A&A...582A..86H} and \citet[][]{2016A&A...587A.160D}. 

Inferences on the distribution of SF based on UV (and to a lesser extent H$\alpha$) emission are affected by extinction, 
and thus any estimated SFR is a lower boundary to the true value. The dust tends to re-radiate the absorbed UV in mid-IR wavelengths. 
This is beyond the scope of this paper, but will be assessed in future work 
(Díaz-García et al. in prep.) for a subsample of S$^4$G galaxies. 
Highly inclined galaxies (where dust absorption is greatest) are not included in our analysis. 
We have not deconvolved the averaged luminosity profiles with the GALEX point spread function ($\sim 6$ arcsec), 
whose wings should produce the largest uncertainties in the outermost regions of the galaxies, which are not probed here. 

Stacking techniques allow us to average hundreds of UV images per sample bin and detect low levels of SF (Sect.~\ref{stacks}). 
This is an important step forward in the study of SF in bars, 
as we use a significantly larger sample compared to previous works (see Sect.~\ref{introduction}), resulting in a more in-depth analysis. 
Sample bins were defined based on detailed visual estimates of bar strength from \citet[][]{2015ApJS..217...32B}, 
thus testing with high statistical significance the role of bars in triggering or preventing SF. 

We showed that, among early-type disk galaxies, the average central UV emission is $\sim$ 0.5 mag brighter 
(i.e., $\gtrsim 50 \%$ larger $\Sigma_{\rm SFR}$) 
when only strongly barred galaxies are considered, relative to their weakly barred counterparts 
(the latter, in turn, present slightly higher levels of  UV emission in the middle and end parts of the bar)  
(Figs.~\ref{Fig_family_bars_UV_1D}~and~\ref{Fig_family_bars_NUV_1D}). 
This is most likely related to the efficiency of strong bars sweeping the disk gas and nourishing central starbursts. 
The latter is predicted in numerical models whose output resemble early-type disk galaxies \citep[see][and references therein]{1993RPPh...56..173S}. 

On the other hand, the UV central surface brightness of barred galaxies is not significantly brighter than that of non-barred galaxies 
(Figs.~\ref{Fig_mass_bars_NUV_FUV_1Dstack_bars_separated}~and~\ref{Fig_mass_bars_NUV_1Dstack_bars_separated}), which may cast doubt on 
the role of bars in central SF enhancement. Among the massive galaxies ($M_{\ast}\in [10^{10}-10^{11}]\cdot M_{\odot}$) the central UV emission 
in barred galaxies is however higher relative to the underlying exponential disk. Following  this line, 
\citet[][]{2016A&A...596A..84D} showed that the central deviation from an exponential slope of the mean stellar density profile 
is also larger in barred galaxies (see their Fig.~8). 
Moreover, since UV traces timescales that are about the same as dynamical ones (and even longer in the central parts of galaxies) 
the bar potential might have changed substantially after enhancing the central UV emission. 
In other words, some bars might have weakened or even dissolved after feeding gas to 
the circumnuclear regions \citep[e.g.,][]{2004ApJ...604..614S}, while 
SF still takes place at $z \approx 0$ out of gas reservoirs that can last for hundreds of Myr. 
However, bar dissolution is  implausible according to most modern simulations \citep[e.g.,][and references therein]{2013seg..book..305A}.

Barred galaxies present higher UV emission relative to their non-barred counterparts when $M_{\ast}<10^{10}M_{\odot}$. 
As for the largest $M_{\ast}$-bins, barred galaxies have somewhat brighter UV profiles beyond the 
radii where bars typically occur (Sect.~\ref{1-Dstacks}). 
Likewise, \citet[][]{2016A&A...596A..84D} showed that, on average, barred galaxies have disks with longer scale lengths and 
fainter extrapolated central surface brightnesses than non-barred galaxies \citep[see also][]{2013MNRAS.432L..56S,2019MNRAS.489.3553E}. 
This is probably related to bars causing a mixing of gas and stars and the spread of the disk 
\citep[e.g.,][]{2002MNRAS.330...35A,2011A&A...527A.147M,2012MNRAS.426L..46A,2015MNRAS.451..936S}\footnote{
For a recent analysis of the dependence on $M_{\ast}$ of bar-induced radial 
distribution of metals in the gas phase of spirals, see \citet[][and references therein]{2020MNRAS.tmp.2303Z,2020arXiv200712289Z}.}, 
perhaps as a result of the coupling between bar and 
spiral amplitudes \citep[][]{2010ApJ...715L..56S,2012A&A...548A.126M,2019A&A...631A..94D,2020MNRAS.497..933H}. 
In other words,  spiral arms are loci of active SF, and their amplitudes are larger in barred galaxies than in 
non-barred ones \citep[see Fig.~9 in][]{2016A&A...596A..84D}. This translates into a higher UV emission beyond the bar radius. 
We conclude that bars are important agents in the regulation of the SF in disk galaxies.
%
%
\subsection{Is SF quenching in galaxies bar-driven?}
%
%
Bar-driven central starbursts have been proposed as the mechanism that eventually depletes the gas in barred galaxies, 
unless it is replenished from the outside. 
However, whether the presence of a bar is connected to the total SFR in a galaxy 
remains a matter of debate \citep[e.g.,][]{1986MNRAS.221P..41H,1988ApJ...329L..69D,1988MNRAS.231..465P,
1999A&A...351...43A,2002AJ....124.2581S,2007A&A...474...43V,2020ApJ...893...19W}. 
For instance, using H{$\alpha$} imaging of galaxies in the Coma \citep[][]{2015A&A...576A..16G} and 
Local superclusters \citep[][]{2012A&A...545A..16G}, \citet[][]{2015A&A...580A.116G} 
proposed that strong bars play an important role in the quenching of the SF of massive galaxies since $z=3$. 
This is supported by their observations at different $z$ of a declining bar fraction for non-quenched galaxies, 
and is also consistent with the study by \citet[][]{2013ApJ...779..162C}, 
who found a larger bar fraction among galaxies with a low total specific SFR (i.e., SFR divided by $M_{\ast}$). 
Similar trends have been found in the local Universe: 
a drop in the bar fraction among gas-rich galaxies was reported by \citet[][]{2012MNRAS.424.2180M} 
\citep[see also][]{2012MNRAS.423.3486W,2018MNRAS.473.4731K} based on the Sloan Digital Sky Survey \citep[SDSS;][]{2006AJ....131.2332G}. 
Further supporting this picture, \citet[][]{2020MNRAS.495.4158F} find a segregation in the 
SFR-$M_{\ast}$ relation as a function of scaled bar length, 
where SF classes (very similar to those  used in this paper) also separate clearly. 
If true, the interpretation of these statistical trends is affected by a chicken or egg causality dilemma: 
are strong bars responsible for galaxy quenching \citep[e.g.,][]{2018A&A...609A..60K} 
or do they preferentially form in red gas-poor galaxies \citep[see, e.g.,][]{2013MNRAS.429.1949A,2010ApJ...719.1470V}? 

The causality between bars and quenching might also be linked to the observation that in lenticulars ($T\le0$), 
and in gas-poor massive galaxies in general, the UV emission is scant across the disk and 
only circumnuclear (Sect.~\ref{bar_uv_stack}) and does not follow the bar (see also Sects.~\ref{ttype_SF}, \ref{mstar_SF}, and \ref{gas_frac}). 
However, UV might not be the most reliable tracer of SF or young populations among the reddest galaxies, 
as discussed in \citet[][]{2018ApJS..234...18B}, i.e., the emission can also be coming from evolved stars (UV-upturn), 
for instance main-sequence turnoff or extreme horizontal branch stars \citep[e.g.,][]{2005ApJ...619L.111Y,2011ApJS..195...22Y}. 
However, elliptical galaxies are not included in our analysis, and the discussed effect is not expected to be so severe among S0s. 
In addition, the statistical trends when only using H$\alpha$ are basically the same. 
In Appendix~\ref{app_AIS_Agn} we check and confirm that the statistical trends for the frequency of SF classes are not determined by 
the presence of AGN \citep[as reported by][]{2010A&A...518A..10V}, 
which is a source of photoionization. Likewise, \citet[][]{2020MNRAS.495.4158F} 
used Baldwin, Phillips and Terlevich \citep[BPT;][]{1981PASP...93....5B} diagrams 
reproduced from IFU data across the bar spaxels from 
MaNGA to conclude that the bulk of the H$\alpha$ emission in barred galaxies is associated with SF and not with AGN emission.

The connection between bars and quenching reviewed above is challenged by the fact that 
bars among late-type galaxies in the S$^4$G (typically gas rich) are unexpectedly frequent \citep[][]{2015ApJS..217...32B} 
and long \citep[][]{2016A&A...587A.160D} relative to the sizes of their host disks \citep[see also][]{2019MNRAS.489.3553E}, 
yet their age and exchange of angular momentum might be much different from earlier types. 
\citet[][]{2016A&A...587A.160D} speculated that many of the late-type bars identified in the S$^4$G 
would possibly be overlooked if they were observed at higher redshift, given their faint disks (see their Sect. 5.1). 
On the other hand, as discussed by \citet[][]{2015ApJS..217...32B}, 
the types of bars seen in nearby late-type galaxies may not necessarily be the ones we see at high redshift. 

\citet[][]{2018MNRAS.474.5372E} showed that SDSS-based studies tend to 
underestimate the bar fraction (mainly among low-mass, blue, gas-rich galaxies) 
due to poor spatial resolution and the correlation between bar size and stellar mass. 
He also found that the bar fraction is roughly constant with $g-r$ color and atomic gas fraction. 
In addition, \citet[][]{2019A&A...625A.146D} do not find differences on SFRs, gas fraction, or [FUV]-[3.6] 
color between barred and  non-barred S$^4$G galaxies based on the use of clustering algorithms (self-organizing maps). 
On the other hand, it is known that S$^4$G missed galaxies due to sample selection based on H{\sc\,i} recessional velocities. 
However, this alone is not sufficient to explain the discrepancies between the SDSS and S$^4$G surveys, 
such as the overall lower bar fraction in the former or its sharp decrease towards low-mass gas-poor galaxies.

We argue that a definite connection between bar fraction and SF quenching is still lacking in the literature. 
A new picture may arise from forthcoming surveys in the next decade with the next generation of telescopes 
(e.g., LSST, JWST, WFIRST, EUCLID). 
This will allow us to study the cosmic bar fraction \citep[e.g.,][]{2008ApJ...675.1141S,2010ApJ...714L.260N} 
with unprecedented depth and resolution, and with the aid of 
automated bar detections that are based on neural networks \citep[e.g.,][]{2018MNRAS.476.3661D}.
%
%
\subsection{Spatial distribution of SF in galactic bars}\label{dist_SF}
%
%
The distribution of ionized gas in the bar region, traced from the H$\alpha$ emission, 
can be distributed along the bar; concentrated in the nuclear or circumnuclear regions, with little or no emission from the bar;
and in both the bar and the nuclear region \citep[][]{1997A&A...326..449M,2002AJ....124.2581S,2007A&A...474...43V,2008A&A...485....5Z}. 
Interest has emerged on this topic with the advent of large surveys \citep[e.g.,][]{2004A&A...414...23J} 
and the use of homogenous IFU data \citep[e.g.,][]{2019A&A...627A..26N,2020ApJ...898..116K}.

Most of the work attempting to classify the SF in bars has been carried out with small samples. 
In order to provide the most complete study with a large unbiased sample of objects that are  not highly inclined, 
in Sect.~\ref{class_met} we presented a simple visual classification system (outlined in Table~\ref{sf_system}) 
for the galaxies in the S$^4$G survey. \citet[][]{2020MNRAS.495.4158F} recently presented a study 
of the SFR and distribution in 684 barred galaxies surveyed in MaNGA with an approach similar to ours. 
While they study the frequency of SF categories as a function of the total stellar mass and global SFR, 
here we test how SF relates to $M_{\ast}$ and to other parameters such as $T$-type, gas fraction, and tangential-to-radial forcing. 
Their use of IFU data has advantages (e.g., analysis with a homogeneous dataset of BPT diagrams) and 
disadvantages (e.g., poorer angular resolution) compared to our data, and represents an important complement to our paper.

To date, there is no clear understanding about the influence of local dynamical conditions on the SF activity in bars. 
The formation of new stars out of molecular clouds along the bars is expected to be regulated by the effect of shear, 
which can be controlled by orbits making up the bar \citep[e.g.,][]{1992MNRAS.259..345A}. 
\citet[][]{2002ApJ...570L..55J} discussed that SF can be induced in weak bars, owing to the weaker shocks and shear, 
and mention the case of galaxies like M$\,$100 \citep[][]{1989ApJ...343..602E}, NGC$\,$4254, and NGC$\,$4303 \citep[][]{1997PhDT........11K}. 
Numerical simulations by \citet[][]{1998ApJ...508..291V} show that   weak shocks with speeds of order 20-30 km~s$^{-1}$ 
can indeed favor the collapse of gas and the formation of stars. In some cases, the distribution of molecular gas 
indicates that the SF along the bar appears to be inhibited in some locations of the dust lanes due to the high strength 
of shocks and shear stress \citep[e.g.,][]{1998A&A...337..671R}. 
This is confirmed in the fluid dynamics simulations by \citet[][]{1992MNRAS.259..345A}. 
Nevertheless, H{\sc\,ii} regions have been found under these 
conditions in other galaxies \citep[e.g.,][]{1997A&A...326..449M,2002AJ....124.2581S,2008A&A...485....5Z}. 
Furthermore, observations of H$\alpha$ velocity gradients showed that shear makes SF drop, 
whereas shocks enhance it in general \citep[][]{2004A&A...413...73Z}.

In Sect.~\ref{bar_uv_stack} we showed that in bar stacks of spiral galaxies ($0<T<8$) the UV emission traces 
the stellar bars and dominates on their leading side, 
a behavior expected from simulations that model bar-triggered gas inflow \citep[e.g.,][]{1992MNRAS.259..345A}. 
H{\sc\,ii} regions on the leading side of the bars have been detected 
\citep[e.g.,][]{2002AJ....124.2581S,2010A&A...521A...8P,2020MNRAS.495.4158F}, 
and are expected to be due to the combined effect of shear and turbulence forces inhibiting SF in most places, 
but not on the leading side of the bar \citep[][]{2015MNRAS.446.2468E,2015MNRAS.454.3299R} {\citep[for pioneering theoretical 
input on the interplay between shear, shocks, and SF, see][]{1992MNRAS.259..328A}}.
%
%
\subsection{Differences in the distributions of SF classes A-B-C}
%
%
We find distinct distributions in the Hubble sequence of the loci of SF within bars, using 
both stacking techniques (Sect.~\ref{UV_emission_spatial}) and visual classifications (Sect.~\ref{ttype_SF}). 
Differences in the statistical distributions of the star-forming and passive bars are also reported as a function 
of physical properties, such as $M_{\rm HI}/M_{\ast}$ (Sect.~\ref{gas_frac}) and $F_{\rm T}/\left<F_{\rm R}\right>$ (Sect.~\ref{Qb_sec}). 
However, the segregation of SF classes (especially between B and C) is less clear as a function of $M_{\ast}$ (Sect.~\ref{mstar_SF}), 
which is not surprising as the Hubble sequence is not a mass sequence \citep[e.g., Fig.~1 in][]{2016A&A...596A..25L}. 
We also note that SF classes are not clustered (e.g., examples of any SF class can be found for any given $T$-bin). 

We find that bar stacks comprising late-type galaxies ($T \ge 5$) have  
SF that is more evenly distributed along the bar major-axis, and that the UV emission is higher for strong bars at all bar radii (Sect.~\ref{bar_uv_stack}). 
Likewise, by studying individual objects we show that the fraction of star-forming bars (category C) is larger for later types (Sect.~\ref{ttype_SF}). 
In Appendix~\ref{app_AIS_Agn} we checked that the depth of FUV imaging does not affect the statistical trends presented in this work: 
limiting the analysis to the deepest GALEX images yields the same results in the Hubble sequence. 

The correlation between Hubble type (in a narrower $T$ range) and the presence of SF along the bar has previously been reported 
from smaller samples that did not probe the plentiful galaxies at the end of the Hubble sequence \citep[e.g.,][]{1996ASPC...91...44P}. 
\citet[][]{1996RMxAA..32...89G} showed that SBb galaxies tend to host less SF along bars than SBc. 
Among the spirals, the dissimilarity in the distributions of SF classes B and C in the Hubble sequence is likely related to 
general differences in the mass distribution and photometric/kinematic properties of disks in galaxies with $T$-type higher or 
lower than $\sim 5$ \citep[][]{2016A&A...596A..84D}. The former have larger central mass concentrations than the latter, 
among which many galaxies are bulge-less \citep[][]{2015ApJS..219....4S,2016A&A...596A..84D}, 
and the shape of their rotation curves and mass distribution is remarkably different. 

Late-type gas-rich galaxies are characterised by low amplitude, slowly rising rotation curves \citep[e.g.,][]{1991ApJ...368...60P}. 
Thus, the shear ($\Gamma$) in these galaxies might be lower at the bar region 
\citep[favoring SF;][]{2005MNRAS.361L..20S} compared to their early-type counterparts, 
which can be estimated from the slope of the rotation curves ($V$) in the central regions: 
$\Gamma=-d {\rm ln} \Omega/d {\rm ln} r$, where $\Omega(r)=V/r$ is the angular velocity \citep[e.g.,][]{2006ApJ...645.1012S,2018MNRAS.477.1451F} 
at a given radius $r$, and $\Gamma=1$ in the flat regime. 
This interpretation is also favored by our observations that star-forming bars 
are typically hosted by disk galaxies with high tangential-to-radial force ratios (Sect.~\ref{Qb_sec}). 
$F_{\rm T}$ traces the bar-induced gravitational torques and the efficiency of the 
bar potential controlling the orbits of the gas \citep[][]{2015MNRAS.451..936S}. 
$\left<F_{\rm R}\right>$ determines the stellar contribution to the circular velocity \citep[e.g.,][]{2016A&A...587A.160D,2019A&A...625A.146D}, 
which in turn gives a lower bound for the rotation curve of the galaxy in the inner parts 
\citep[the nuclear regions tend to be baryon dominated according to, e.g.,][]{2016MNRAS.458.1199E}. 
For a given galactocentric radius and galaxy size, the higher the $\left<F_{\rm R}\right>$ values the larger the shear, 
and thus the torque parameter ($Q_{\rm b}$) is related to the SF activity.

In conclusion, a lower shear is likely in Sc-irregular and in low-mass galaxies in general, 
where the inner slope and amplitude of the rotation curve are lowest \citep[e.g.,][]{2016A&A...587A.160D,2020A&A...635A.197D} 
and $Q_{\rm b}$ is largest \citep[e.g.,][]{2016A&A...587A.160D}. 
The latter is mainly due to the dilution of bar gravitational torques by the bulge contribution to the overall radial force field 
\citep[][]{2001A&A...375..761B,2002MNRAS.337.1118L}, which dominates over the dark matter halo dilution \citep[][]{2016A&A...587A.160D}. 
This is in spite of the fact that bar-induced tangential forces are probably  stronger among the largest galaxies 
with the most massive bars \citep[massive disks host bars with large $m=2$ Fourier density amplitudes;][]{2016A&A...587A.160D}. 
On the other hand, for a given $M_{\ast}$, a higher $F_{\rm T}/\left<F_{\rm R}\right>$ can cause a twist 
on stellar orbits, enhancing the local shear. 
In addition, in Sect.~\ref{1-Dstacks} we showed that, on average, non-barred galaxies 
are characterized by exponentially decaying UV luminosity profiles without any light deficit in the central regions (unlike in barred ones), 
in agreement with reports by \citet[][]{2009A&A...501..207J} that were based on (1D) H$\alpha$ averaging. 
Altogether, this implies that the dynamical conditions determined by the axisymmetric stellar components alone cannot 
explain  the inhibition of SF, and hence bars play a major role.

We confirm the drop in the frequency of star-forming bars in galaxies with $M_{\ast}>10^{10}M_{\sun}$ (Sect.~\ref{mstar_SF}) 
reported by \citet[][]{2020MNRAS.495.4158F}. This is seen both in FUV and in H$\alpha$ ($\sim 40\%$ smaller sample); 
however, $\sim 1/2$ of the analyzed galaxies in the H$\alpha$ sample 
with $10^{10}M_{\odot}<M_{\ast}<10^{11}M_{\odot}$ belong to SF class C. 
Among the most massive galaxies, physical processes other than SF, such as gas shocks, can also account for the H$\alpha$ emission. 
\citet{2019A&A...627A..26N} used a sample of 16 galaxies (with $M_{\ast} \gtrsim 10^{10}M_{\odot}$) from 
the Close AGN Reference Survey \citep[CARS;][]{2017Msngr.169...42H} 
to study the properties of star-forming bars and non-star-forming bars using IFU MUSE data, and report that the SF along the bar is linked to 
the flatness of the surface brightness profile: the flattest bars are star-forming. 
The latter is not easy to reconcile with our report of a low fraction of SF bars in early-type galaxies 
\citep[see also][]{2020MNRAS.495.4158F}, 
which typically host flat bars \citep[e.g.][]{1985ApJ...288..438E,1996AJ....111.2233E,2015ApJ...799...99K,2016A&A...596A..84D}. 
This may be due to our use of a different and larger sample, and thus further analysis is needed.

Galaxies of $T$-types $0 \le T < 5$, which are characterized by intermediate gas 
fractions and gravitational torques, predominantly have SF regions at the bar ends, 
but not along the bar. As discussed by \citet[][]{2020MNRAS.495.4158F}, the occurrence of intense H$\alpha$ at the bar ends 
has been postulated to be a consequence of the gas flows and shear at the kpc level, and 
cloud-cloud collisions and turbulence on a parsec scale \citep[][]{2015MNRAS.454.3299R}. 
These favorable physical conditions are likely to be present in early- and intermediate-type spirals. 
This trait is in principle not related to ansae structure \citep[stellar blobs at the end of the bars;][]{1965AJ.....70..501D} 
since most ansae are detected in early-type galaxies  \citep[e.g.,][]{2007MNRAS.381..401L}. 
However, recent work by \citet[][]{2019MNRAS.488..590B} reports the detection of 
blue bar ansae in late-type galaxies, which could indeed be related to some of the UV enhancements (SF class B) 
characterized in this work or to highly oval star-forming inner rings. 
\citet[][]{2007AJ....134.1863M} argue that the nature of this ansae structure is in principle stellar dynamical in origin, 
yet one example of an ansae harboring SF is reported (NGC$\,$4151).
 
The likelihood of a bar to host SF is correlated with the total relative content of H{\sc\,i} gas 
(Sect.~\ref{gas_frac}), which is explained by the behavior of SF classes in the Hubble sequence. 
However, the availability of gas is not sufficient to explain the statistical trends: quite a few early- and 
intermediate-type spirals have plenty of cold gas and host H{\sc\,ii} regions everywhere except the bar where dynamical conditions must be different. 
Galaxies with $T \ge 5$ are known to be dark matter dominated within the optical disk \citep[see Fig.~6 in][]{2016A&A...587A.160D}; 
thus, the distinct disk stability properties and the interplay between dark matter, disk temperature, 
and SF must play a role to explain our observations.

Last but not least, one important ingredient that is missing from our analysis is the content of molecular gas. 
Ideally, we should use observations of the CO(1-0) line (115 GHz) and infer H$_2$ gas masses from the velocity-integrated line 
intensities across the bars \citep[e.g.,][]{1999ApJ...526...97R,2015ApJ...815...59P,2019A&A...621L...4G,2020MNRAS.tmp.1411M}, 
but such data are scarce for our large galaxy sample. 
In future work (Díaz-García et al. in prep.) we will study the relationship between the molecular gas column density and 
the surface density of the SFR  (derived 
from GALEX UV and H$\alpha$ imaging compiled in this work, correcting for extinction using 22~$\mu$m WISE photometry) s
for a subsample of S$^4$G galaxies, using the CO emission along the bars observed with the IRAM-30 m single dish.
%
%
\subsection{Bars in late-type galaxies: resolution and non-stellar contaminants}
%
%
Gas-rich galaxies often host clumpy bars (e.g., NGC$\,$3023 in Fig.~\ref{Fig_classC}) and low-quality imaging and the consequent blurring of SF regions can lead to a missclassification of 
bars among the latest types \citep[e.g.,][]{2014AAS...22320502S}, 
even at near-IR wavelengths, but this is not expected to be severe in the S$^4$G 
given the good quality of the data \citep[see, e.g., discussion in][]{2015ApJS..217...32B}. 
It is  worth noting, however,  that non-stellar emission (hot dust, polycyclic aromatic hydrocarbons, 
or asymptotic giant and red super-giant stars) can contaminate the 3.6~$\mu$m flux \citep[][]{2014ApJ...788..144M,2015ApJS..219....5Q}. 
This can cast doubt on whether there is actually an underlying bar pattern in the old stellar populations of some late-type galaxies, 
i.e., whether self-gravity can alone make SF clumps aligning without presence of $x_1$ orbits characterizing bars. 
While this might be the case for some bars, we argue that this does not explain the general picture for bars in late-type galaxies 
\citep[see discussion and observational characterization of bars in late-type galaxies in, e.g.,][]{2016A&A...596A..84D}. 

Non-stellar contaminants could also contribute to the greater FUV and NUV emission in strong bars (Sect.~\ref{bar_uv_stack}), 
as seen in bar stacks comprising late-type galaxies. 
In other words, the visual identification of strong bars in 3.6~$\mu$m images of clumpy gas-rich galaxies 
can be biased due to the contribution of non-stellar emission 
at the bar region \citep[for the analysis of the impact of non-stellar contaminant 
on bar forcing, see Appendix C in][]{2016A&A...587A.160D}. 
We also checked that our assignment of SF classes is not affected by the resolution of the employed imaging: 
the distribution of SF classes is uncorrelated with sizes of bars in pixels.
%
%
\subsection{Gas inflow slowed down at the 1/4 ultraharmonic resonance} \label{inner_ring_disc}
%
%
Many rings present recent SF and host young stars \citep[][]{1993AJ....105.1344B,1995ApJ...454..623K,2009A&A...501..207J}. 
However, rings lacking SF activity have also been found \citep[e.g.,][]{1991AJ....102.1715B,2013A&A...555L...4C}. 
\citet[][]{2010AJ....139.2465G} used a sample of 44 galaxies (26 non-barred or weakly barred, and 18 strongly barred) 
to show that the SFR within rings does not depend on the amplitude of the non-axisymmetric perturbation strength. 
More recently, \citet[][]{2019A&A...627A..26N} report a correlation between the lack of SF in a bar and the presence of an inner ring. 
This suggests that gas is caught in inner resonance rings 
\citep[that tend to live at the 1/4 ultraharmonic resonance; e.g.,][]{1993RPPh...56..173S,2000A&A...362..465R,2019A&A...625A.146D} 
and is prevented from funneling towards the center; this hypothesis is tested and discussed here.

\citet[][]{2015ApJS..217...32B} identified 73 galaxies with inner rings and 268 with inner pseudo-rings (i.e., made of tightly wrapped spiral arms) 
in our sample of barred galaxies. Of the 73 galaxies with closed inner rings and available imaging (Sect.~\ref{class_met}), 
only $20.5 \pm 6.1\%$ and $18.5 \pm 4.8\%$ belong to class C (SF-bar) in the H$\alpha$ and FUV samples, respectively, 
which is much lower than the overall frequency of SF class C ($\sim 50 \%$) of the parent sample. 

The lower fraction of star-forming bars in galaxies with inner rings is 
qualitatively in agreement with the reports by \citet[][]{2019A&A...627A..26N}, 
whose subsample of four inner-ringed galaxies host non-star-forming bars. 
This can however be a consequence of inner rings living in 
massive galaxies \citep[e.g.,][]{2019A&A...625A.146D} where the suppression of SF in bars is greatest (Sect.~\ref{mstar_SF}). 
\citet[][]{2020MNRAS.495.4158F} also report that inner rings in H$\alpha$ maps are mainly detected in galaxies 
with total stellar masses higher than $10^{10}M_{\odot}$, which is not surprising as this is 
the $M_{\ast}$-threshold where the fraction of rings (as detected at 
near-IR wavelengths) starts to rise \citep[e.g.,][]{2015A&A...582A..86H,2019A&A...625A.146D}. 
On the other hand, we checked and confirmed that the inclusion of 
pseudorings makes the connection between SF in bars and presence of inner rings less clear: 
the fraction of inner-ringed galaxies with SF class C becomes $43 \pm 3.4 \%$ (H$\alpha$) and $42.6 \pm 2.8 \%$ (FUV); 
this may be explained by the fact that many pseudorings do not have a resonance origin (and hence trap less gas) 
or are hosted by late-type galaxies (with SF bars).

In addition, part of the picture above is also consistent with our findings in Sect.~\ref{1-Dstacks}, 
in which we studied the mean UV radial profiles for galaxies hosting inner rings, 
normalized to the ring SMA. The radial distribution of SF relative to the inner ring loci is similar for barred and non-barred galaxies. 
We confirmed the inner dip in the UV emission for the subsample of barred galaxies hosting inner rings. 
Interestingly, this lack of UV emission is also detected in non-barred ringed galaxies, which are not expected to have their SF strongly suppressed 
in their central regions (as shown in Figs.~\ref{Fig_mass_bars_NUV_FUV_1Dstack_bars_separated}~and~\ref{Fig_mass_bars_NUV_1Dstack_bars_separated}). 
That is, SF is on average suppressed at radii smaller than the inner ring SMA, irrespective of the presence of a bar. 

Passive rings (i.e., lacking SF) are found only in early-type disk galaxies ($-3 \le T \le 2$), 
with a large fraction corresponding to ringlenses ($30-40 \%$) \citep[][]{2013A&A...555L...4C}. 
We updated the classifications by \citet[][]{2013A&A...555L...4C} (Sect.~\ref{class_met}) 
and confirm his results (Fig.~\ref{appendixrings} in Appendix~\ref{inner_rings_appendix}) 
by including the new H$\alpha$ images of barred galaxies from Sect.~\ref{compilations_halpha}, 
showing that passive rings are mainly hosted by lenticular galaxies, in which the fraction of active rings is $\lesssim 50 \%$. 
Naturally, this is a consequence of passive rings being hosted by 
galaxies with low relative contents of H{\sc\,i} gas, as shown in Fig.~\ref{appendixrings2}.

Curiously enough, there are a number of late-type galaxies ($T\ge5$) 
hosting passive rings as well,  two in FUV (NGC$\,$3389 and NGC$\,$3906) and six in H$\alpha$ (NGC$\,$3906, NGC$\,$4504, NGC$\,$7437, UGC$\,$04867, UGC$\,$09245, and UGC$\,$10791). 
It is also interesting that 16 barred galaxies in our sample host inner rings that are passive in H$\alpha$ but not in FUV. 
As discussed in \citet[][]{2013A&A...555L...4C}, it is possible to  infer quenching timescales on the order of 20-100 Myr from rings 
presenting FUV emission (tracing SF up to 100 Myr), but not H$\alpha$ (tracing SF up to 20 Myr) \citep[][]{1998ARA&A..36..189K}. 
In particular, \citet[][]{2013A&A...555L...4C} estimated 200 Myr to be a lower bound for the dissolution timescale of inner rings 
(on the order of one orbital period at the ring SMA).

We conclude that the gas funneled by non-axisymmetries, such as spiral arms, gets partially trapped at the inner rings. 
The gas no longer migrates to the nuclear regions, explaining the diminished UV and H$\alpha$ 
emission within the rings' SMA and along the bar. 
Nevertheless, the fact that a peak of UV and H$\alpha$ emission is still detected in the circumnuclear 
regions implies that the presence of inner rings does not control circumnuclear SF, 
nourished by gas reservoirs accumulated for several (hundreds of) Myr.
%
%
\section{Summary and conclusions}\label{summarysection}
%
%
The main goals of this study are to shed light on the role of galactic bars regulating the SF 
activity across disks, and to link the distribution of SF in bars to the global properties of the host galaxies. 
With unprecedented statistical significance, we studied the spatial distribution of SF regions in the inner parts of more than 800 nearby 
disk galaxies (within $\sim$40 Mpc) with inclinations lower than 65$^{\circ}$, drawn from the S$^4$G survey \citep[][]{2010PASP..122.1397S}. 
Two complementary methods were used:
\begin{enumerate}
\item We applied the stacking techniques developed in \citet[][]{2016A&A...596A..84D} to GALEX NUV and FUV imaging from the 
GALEX/S$^{4}$G Surface Brightness and Color Profiles Catalog \citep[][]{2018ApJS..234...18B}. Prior to averaging, 
subsamples were defined based on global physical properties such as total stellar mass ($M_{\ast}$), Hubble stage ($T$), and morphological family.
\begin{enumerate}
\item Bar stacks (2D) were built from co-added UV images (Figs.~\ref{Fig_ttype_bars_NUV}~and~\ref{Fig_mass_bars_NUV}) 
that were uniformly scaled and re-oriented with respect to the stellar bars, using bar parameters at 3.6 $\mu$m 
from \citet[][]{2015A&A...582A..86H} and \citet[][]{2016A&A...587A.160D}. 
The winding direction of the spiral arms was also systematically corrected to differentiate the leading and trailing sides of the bar.
\item UV luminosity profiles were scaled to a common framework defined by the extent of the disks in physical units 
(and that of the sizes of inner rings) followed by the calculation of the radial 1D average and dispersion 
(Figs.~\ref{Fig_mass_bars_NUV_FUV_1Dstack}, \ref{Fig_mass_bars_FUV_1Dstack_disp}, and \ref{Fig_inner_rings_SMA}), 
so that we could study differences in SF between barred and non-barred galaxies 
(Fig.~\ref{Fig_mass_bars_NUV_FUV_1Dstack_bars_separated}).
\end{enumerate}
\item We classified the spatial distribution of SF regions by visually inspecting H$\alpha$ and GALEX FUV images. 
Our classification system devises three main categories (Table~\ref{sf_system}), namely:
\begin{itemize}
\item SF class A): only circumnuclear SF (accounting for $\sim 1/8$ of the galaxies in our sample) (Fig.~\ref{Fig_classA}), 
\item SF class B): SF at the bars ends, but not along the bar ($\sim 1/4$ of the sample) (Fig.~\ref{Fig_classB}),
\item SF class C): SF along the bar ($\sim 1/2$ of the sample) (Fig.~\ref{Fig_classC}). 
\end{itemize}
For this purpose, we   assembled the largest compilation of continuum-subtracted H$\alpha$ images in the S$^4$G, 
comprising 433 galaxies (see Table~\ref{table_SF_class_sources}), and made them publicly available (via CDS). For 70 galaxies, we processed the 
continuum-subtraction ourselves from archival imaging and integral field unit datacubes (e.g., from the CALIFA and ESO archives).
\end{enumerate}
%
%
The main results of this paper are the following:
%
%
\begin{itemize}
\item Among massive galaxies with $M_{\ast}>10^{10}M_{\odot}$ (typically S0/a-Sbc), barred galaxies 
are characterized by a dip in the radial distribution of SF that is not seen in  non-barred systems \citep[see also][]{2009A&A...501..207J} 
(Figs.~\ref{Fig_mass_bars_NUV_FUV_1Dstack},~\ref{Fig_mass_bars_NUV_FUV_1Dstack_bars_separated},~and~\ref{Fig_mass_bars_NUV_1Dstack_bars_separated}). 
This shows that bars are loci of SF suppression, quite plausibly because of 
the combined effect of gas flows and shear \citep[e.g.,][]{2015MNRAS.454.3299R}.
\item The UV emission traces the stellar bars and mainly appears on their leading side of the bar stacks in spiral galaxies 
(S0a-Sdm) (Fig.~\ref{Fig_ttype_bars_NUV}). This is in agreement with the expectation from numerical models \citep[e.g.,][]{1992MNRAS.259..345A}.
\item By studying individual galaxies we show that the distributions of SF classes A, B, and C are 
significantly different in the Hubble sequence. Whether a bar is star-forming or passive is likewise linked to 
global physical properties of the host galaxies (Figs.~\ref{THUBBLE}, \ref{MSTAR}, \ref{MHIMSTAR}, and \ref{QB}).
\item In particular, massive, gas-poor, S0 galaxies tend to host SF exclusively in the circumnuclear regions (category A), 
which is probably linked to the role of bars in galaxy quenching 
postulated from studies at high-$z$ \citep[e.g.,][]{2015A&A...580A.116G} and simulations \citep[e.g.,][]{2018A&A...609A..60K}.
\item The SF in late-type galaxies (Sc-Im) is evenly distributed along the bar major-axis. 
The UV emission is on average higher at all bar radii among strong bars, relative to their weakly barred counterparts 
(Figs.~\ref{Fig_family_bars_UV_1D}~and~\ref{Fig_family_bars_NUV_1D}). The fraction of star-forming bars (class C) 
is larger for later morphological types, larger H{\sc\,i} gas fractions, 
and higher tangential-to-radial force ratios, at both H$\alpha$ and UV wavelengths. 
We argue that shear has the smallest effect in these late-type galaxies, favoring SF \citep[e.g.,][]{2005MNRAS.361L..20S}.
\item The SF activity dominates at the bar ends and the circumnuclear regions in bar stacks comprising galaxies of morphological types 
ranging between S0/a and Sbc (Fig.~\ref{Fig_ttype_bars_NUV_1D}). The UV emission gets weaker, 
relative to the outer exponential disk, in the intermediate parts of the bar. 
We confirm that SF class B) is typical of early- and intermediate-type spirals (Fig.~\ref{THUBBLE}), 
with distributions of gas fraction (Fig.~\ref{MHIMSTAR}) and torque parameter (Fig.~\ref{QB}) that peak between those of classes A and C, 
likely due to a higher shear in galaxies with larger central mass concentrations and bar amplitudes.
\item Strongly barred early-type spiral galaxies are characterized by a $\sim$0.5 mag 
brighter central UV emission (Figs.~\ref{Fig_family_bars_UV_1D}~and~\ref{Fig_family_bars_NUV_1D}) 
(i.e., $\gtrsim 50 \%$ larger $\Sigma_{\rm SFR}$), 
 compared to their weakly barred counterparts (that show a somewhat higher UV emission in the middle and end parts of the bar). 
These observations can be explained by the effect of the bar-induced gravitational torques 
sweeping the gas in the disk that eventually fuels starbursts in the central regions \citep[e.g.,][and references therein]{1993RPPh...56..173S}. 
\item In galaxies hosting inner rings, the mean radial UV luminosity profiles 
are similar for barred and non-barred galaxies. They show a local maximum close to the ring SMA, 
diminished UV emission in the region 0.3-0.7 SMA, and a nuclear peak (Fig.~\ref{Fig_inner_rings_SMA}). 
This can be explained by gas being partially trapped at the 1/4 ultraharmonic resonance 
\citep[][]{1984MNRAS.209...93S,1996FCPh...17...95B}, causing a halt in its migration to the nuclear regions, 
irrespective of the presence of a bar.
\item The latter is further supported by the low fraction ($<1/3$) of galaxies with closed inner rings belonging to class C. 
We also confirm that, while most inner rings detected at 3.6 $\mu$m are active in FUV and H$\alpha$ passbands, 
the frequency of passive rings is highest among S0s, accounting for $> 50 \%$ \citep[][]{2013A&A...555L...4C}.
\end{itemize}
%
%
This work highlights the connection between bars and SF activity in the central parts of local disk galaxies, 
using an unprecedentedly large unbiased sample based on the analysis of GALEX UV and continuum-subtracted H$\alpha$ imaging. 
Differences in the typical spatial distribution of SF in galactic bars 
are dependent on physical and morphological global properties of the host galaxies. 
We encourage these trends to be further studied elsewhere with numerical models. 
%
%
\begin{acknowledgements}
We thank the anonymous referee for comments that improved this paper. 
We thank Stéphane Courteau, Jesús Falcón-Barroso, Estrella Florido, Ryan Leaman, Ute Lisenfeld, Isabel Pérez, 
Glenn van de Ven, Simon Verley, and Almudena Zurita for useful discussions. 
We thank Serafim Kaisin for providing continuum-subtracted H$\alpha$ images for the galaxies NGC$\,$3384, UGC$\,$07257, and UGC$\,$07534.
This project has received funding from the European Union’s Horizon 2020 research and innovation programme 
under the Marie Sk$\l$odowska-Curie grant agreement No 893673. SDG acknowledges support from the Spanish Public Employment Service (SEPE). 
We acknowledge financial support from the European Union's Horizon 2020 research and innovation programme under 
Marie Sk$\l$odowska-Curie grant agreement No 721463 to the SUNDIAL ITN network, 
from the State Research Agency (AEI-MCINN) of the Spanish Ministry of Science and Innovation under the grant 
"The structure and evolution of galaxies and their central regions" with reference PID2019-105602GB-I00/10.13039/501100011033, 
and from the IAC project P/300724 which is financed by the Ministry of Science and Innovation, through the State Budget and by the 
Canary Islands Department of Economy, Knowledge and Employment, through the Regional Budget of the Autonomous Community. 
AYKB acknowledges financial support from the Spanish Ministry of Economy and Competitiveness (MINECO), 
project Estallidos AYA2016-79724-C4-2-P. This research makes use of 
IDL (\href{https://www.harrisgeospatial.com/docs/using_idl_home.html}{https://www.harrisgeospatial.com/docs/using$\_$idl$\_$home.html}), 
python (\href{http://www.python.org}{http://www.python.org}), 
Matplotlib \citep[][]{Hunter2007}, and Astropy \citep[][]{2013A&A...558A..33A,2018AJ....156..123A}.
%
%
{\it Facilities}: GALEX, \emph{Spitzer} (IRAC).
%
%
\end{acknowledgements}
%
%
\bibliographystyle{aa}
\bibliography{bibliography}

\begin{thebibliography}{242}
\expandafter\ifx\csname natexlab\endcsname\relax\def\natexlab#1{#1}\fi

\bibitem[{{Aguerri}(1999)}]{1999A&A...351...43A}
{Aguerri}, J.~A.~L. 1999, \aap, 351, 43

\bibitem[{{Aguerri} {et~al.}(2009){Aguerri}, {M{\'e}ndez-Abreu}, \&
  {Corsini}}]{2009A&A...495..491A}
{Aguerri}, J.~A.~L., {M{\'e}ndez-Abreu}, J., \& {Corsini}, E.~M. 2009, \aap,
  495, 491

\bibitem[{{Astropy Collaboration} {et~al.}(2018){Astropy Collaboration},
  {Price-Whelan}, {Sip{\H{o}}cz}, {G{\"u}nther}, {Lim}, {Crawford}, {Conseil},
  {Shupe}, {Craig}, {Dencheva}, {Ginsburg}, {Vand erPlas}, {Bradley},
  {P{\'e}rez-Su{\'a}rez}, {de Val-Borro}, {Aldcroft}, {Cruz}, {Robitaille},
  {Tollerud}, {Ardelean}, {Babej}, {Bach}, {Bachetti}, {Bakanov}, {Bamford},
  {Barentsen}, {Barmby}, {Baumbach}, {Berry}, {Biscani}, {Boquien}, {Bostroem},
  {Bouma}, {Brammer}, {Bray}, {Breytenbach}, {Buddelmeijer}, {Burke},
  {Calderone}, {Cano Rodr{\'\i}guez}, {Cara}, {Cardoso}, {Cheedella}, {Copin},
  {Corrales}, {Crichton}, {D'Avella}, {Deil}, {Depagne}, {Dietrich}, {Donath},
  {Droettboom}, {Earl}, {Erben}, {Fabbro}, {Ferreira}, {Finethy}, {Fox},
  {Garrison}, {Gibbons}, {Goldstein}, {Gommers}, {Greco}, {Greenfield},
  {Groener}, {Grollier}, {Hagen}, {Hirst}, {Homeier}, {Horton}, {Hosseinzadeh},
  {Hu}, {Hunkeler}, {Ivezi{\'c}}, {Jain}, {Jenness}, {Kanarek}, {Kendrew},
  {Kern}, {Kerzendorf}, {Khvalko}, {King}, {Kirkby}, {Kulkarni}, {Kumar},
  {Lee}, {Lenz}, {Littlefair}, {Ma}, {Macleod}, {Mastropietro}, {McCully},
  {Montagnac}, {Morris}, {Mueller}, {Mumford}, {Muna}, {Murphy}, {Nelson},
  {Nguyen}, {Ninan}, {N{\"o}the}, {Ogaz}, {Oh}, {Parejko}, {Parley}, {Pascual},
  {Patil}, {Patil}, {Plunkett}, {Prochaska}, {Rastogi}, {Reddy Janga},
  {Sabater}, {Sakurikar}, {Seifert}, {Sherbert}, {Sherwood-Taylor}, {Shih},
  {Sick}, {Silbiger}, {Singanamalla}, {Singer}, {Sladen}, {Sooley},
  {Sornarajah}, {Streicher}, {Teuben}, {Thomas}, {Tremblay}, {Turner},
  {Terr{\'o}n}, {van Kerkwijk}, {de la Vega}, {Watkins}, {Weaver}, {Whitmore},
  {Woillez}, {Zabalza}, \& {Astropy Contributors}}]{2018AJ....156..123A}
{Astropy Collaboration}, {Price-Whelan}, A.~M., {Sip{\H{o}}cz}, B.~M., {et~al.}
  2018, \aj, 156, 123

\bibitem[{{Astropy Collaboration} {et~al.}(2013){Astropy Collaboration},
  {Robitaille}, {Tollerud}, {Greenfield}, {Droettboom}, {Bray}, {Aldcroft},
  {Davis}, {Ginsburg}, {Price-Whelan}, {Kerzendorf}, {Conley}, {Crighton},
  {Barbary}, {Muna}, {Ferguson}, {Grollier}, {Parikh}, {Nair}, {Unther},
  {Deil}, {Woillez}, {Conseil}, {Kramer}, {Turner}, {Singer}, {Fox}, {Weaver},
  {Zabalza}, {Edwards}, {Azalee Bostroem}, {Burke}, {Casey}, {Crawford},
  {Dencheva}, {Ely}, {Jenness}, {Labrie}, {Lim}, {Pierfederici}, {Pontzen},
  {Ptak}, {Refsdal}, {Servillat}, \& {Streicher}}]{2013A&A...558A..33A}
{Astropy Collaboration}, {Robitaille}, T.~P., {Tollerud}, E.~J., {et~al.} 2013,
  \aap, 558, A33

\bibitem[{{Athanassoula}(1992{\natexlab{a}})}]{1992MNRAS.259..328A}
{Athanassoula}, E. 1992{\natexlab{a}}, \mnras, 259, 328

\bibitem[{{Athanassoula}(1992{\natexlab{b}})}]{1992MNRAS.259..345A}
{Athanassoula}, E. 1992{\natexlab{b}}, \mnras, 259, 345

\bibitem[{{Athanassoula}(2012)}]{2012MNRAS.426L..46A}
{Athanassoula}, E. 2012, \mnras, 426, L46

\bibitem[{{Athanassoula}(2013)}]{2013seg..book..305A}
{Athanassoula}, E. 2013, {Bars and secular evolution in disk galaxies:
  Theoretical input}, ed. J.~{Falc{\'o}n-Barroso} \& J.~H. {Knapen}, 305

\bibitem[{{Athanassoula} {et~al.}(2013){Athanassoula}, {Machado}, \&
  {Rodionov}}]{2013MNRAS.429.1949A}
{Athanassoula}, E., {Machado}, R.~E.~G., \& {Rodionov}, S.~A. 2013, \mnras,
  429, 1949

\bibitem[{{Athanassoula} \& {Misiriotis}(2002)}]{2002MNRAS.330...35A}
{Athanassoula}, E. \& {Misiriotis}, A. 2002, \mnras, 330, 35

\bibitem[{{Bacon} {et~al.}(2010){Bacon}, {Accardo}, {Adjali}, {Anwand},
  {Bauer}, {Biswas}, {Blaizot}, {Boudon}, {Brau-Nogue}, {Brinchmann},
  {Caillier}, {Capoani}, {Carollo}, {Contini}, {Couderc}, {Daguis{\'e}},
  {Deiries}, {Delabre}, {Dreizler}, {Dubois}, {Dupieux}, {Dupuy}, {Emsellem},
  {Fechner}, {Fleischmann}, {Fran{\c{c}}ois}, {Gallou}, {Gharsa}, {Glindemann},
  {Gojak}, {Guiderdoni}, {Hansali}, {Hahn}, {Jarno}, {Kelz}, {Koehler},
  {Kosmalski}, {Laurent}, {Le Floch}, {Lilly}, {Lizon}, {Loupias}, {Manescau},
  {Monstein}, {Nicklas}, {Olaya}, {Pares}, {Pasquini}, {P{\'e}contal-Rousset},
  {Pell{\'o}}, {Petit}, {Popow}, {Reiss}, {Remillieux}, {Renault}, {Roth},
  {Rupprecht}, {Serre}, {Schaye}, {Soucail}, {Steinmetz}, {Streicher}, {Stuik},
  {Valentin}, {Vernet}, {Weilbacher}, {Wisotzki}, \&
  {Yerle}}]{2010SPIE.7735E..08B}
{Bacon}, R., {Accardo}, M., {Adjali}, L., {et~al.} 2010, in Society of
  Photo-Optical Instrumentation Engineers (SPIE) Conference Series, Vol. 7735,
  \procspie, 773508

\bibitem[{{Baldwin} {et~al.}(1981){Baldwin}, {Phillips}, \&
  {Terlevich}}]{1981PASP...93....5B}
{Baldwin}, J.~A., {Phillips}, M.~M., \& {Terlevich}, R. 1981, \pasp, 93, 5

\bibitem[{{Bianchi} {et~al.}(2017){Bianchi}, {Shiao}, \&
  {Thilker}}]{2017ApJS..230...24B}
{Bianchi}, L., {Shiao}, B., \& {Thilker}, D. 2017, \apjs, 230, 24

\bibitem[{{Block} {et~al.}(2001){Block}, {Puerari}, {Knapen}, {Elmegreen},
  {Buta}, {Stedman}, \& {Elmegreen}}]{2001A&A...375..761B}
{Block}, D.~L., {Puerari}, I., {Knapen}, J.~H., {et~al.} 2001, \aap, 375, 761

\bibitem[{{B{\"o}ker} {et~al.}(1999){B{\"o}ker}, {Calzetti}, {Sparks}, {Axon},
  {Bergeron}, {Bushouse}, {Colina}, {Daou}, {Gilmore}, {Holfeltz}, {MacKenty},
  {Mazzuca}, {Monroe}, {Najita}, {Noll}, {Nota}, {Ritchie}, {Schultz}, {Sosey},
  {Storrs}, {Suchkov}, \& {StSci NICMOS Group}}]{1999ApJS..124...95B}
{B{\"o}ker}, T., {Calzetti}, D., {Sparks}, W., {et~al.} 1999, \apjs, 124, 95

\bibitem[{{Boselli} {et~al.}(2019){Boselli}, {Epinat}, {Contini},
  {Abril-Melgarejo}, {Boogaard}, {Pointecouteau}, {Ventou}, {Brinchmann},
  {Carton}, {Finley}, {Michel-Dansac}, {Soucail}, \&
  {Weilbacher}}]{2019A&A...631A.114B}
{Boselli}, A., {Epinat}, B., {Contini}, T., {et~al.} 2019, \aap, 631, A114

\bibitem[{{Boselli} {et~al.}(2015){Boselli}, {Fossati}, {Gavazzi}, {Ciesla},
  {Buat}, {Boissier}, \& {Hughes}}]{2015A&A...579A.102B}
{Boselli}, A., {Fossati}, M., {Gavazzi}, G., {et~al.} 2015, \aap, 579, A102

\bibitem[{{Bouquin} {et~al.}(2015){Bouquin}, {Gil de Paz}, {Boissier},
  {Mu{\~n}oz-Mateos}, {Sheth}, {Zaritsky}, {Laine}, {Gallego}, {Peletier},
  {R{\"o}ck}, \& {Knapen}}]{2015ApJ...800L..19B}
{Bouquin}, A.~Y.~K., {Gil de Paz}, A., {Boissier}, S., {et~al.} 2015, \apjl,
  800, L19

\bibitem[{{Bouquin} {et~al.}(2018){Bouquin}, {Gil de Paz}, {Mu{\~n}oz-Mateos},
  {Boissier}, {Sheth}, {Zaritsky}, {Peletier}, {Knapen}, \&
  {Gallego}}]{2018ApJS..234...18B}
{Bouquin}, A.~Y.~K., {Gil de Paz}, A., {Mu{\~n}oz-Mateos}, J.~C., {et~al.}
  2018, \apjs, 234, 18

\bibitem[{{Bundy} {et~al.}(2015){Bundy}, {Bershady}, {Law}, {Yan}, {Drory},
  {MacDonald}, {Wake}, {Cherinka}, {S{\'a}nchez-Gallego}, {Weijmans}, {Thomas},
  {Tremonti}, {Masters}, {Coccato}, {Diamond-Stanic}, {Arag{\'o}n-Salamanca},
  {Avila-Reese}, {Badenes}, {Falc{\'o}n-Barroso}, {Belfiore}, {Bizyaev},
  {Blanc}, {Bland-Hawthorn}, {Blanton}, {Brownstein}, {Byler}, {Cappellari},
  {Conroy}, {Dutton}, {Emsellem}, {Etherington}, {Frinchaboy}, {Fu}, {Gunn},
  {Harding}, {Johnston}, {Kauffmann}, {Kinemuchi}, {Klaene}, {Knapen},
  {Leauthaud}, {Li}, {Lin}, {Maiolino}, {Malanushenko}, {Malanushenko}, {Mao},
  {Maraston}, {McDermid}, {Merrifield}, {Nichol}, {Oravetz}, {Pan}, {Parejko},
  {Sanchez}, {Schlegel}, {Simmons}, {Steele}, {Steinmetz}, {Thanjavur},
  {Thompson}, {Tinker}, {van den Bosch}, {Westfall}, {Wilkinson}, {Wright},
  {Xiao}, \& {Zhang}}]{2015ApJ...798....7B}
{Bundy}, K., {Bershady}, M.~A., {Law}, D.~R., {et~al.} 2015, \apj, 798, 7

\bibitem[{{Buta} \& {Block}(2001)}]{2001ApJ...550..243B}
{Buta}, R. \& {Block}, D.~L. 2001, \apj, 550, 243

\bibitem[{{Buta} \& {Combes}(1996)}]{1996FCPh...17...95B}
{Buta}, R. \& {Combes}, F. 1996, Fundamentals of Cosmic Physics, 17, 95

\bibitem[{{Buta} \& {Crocker}(1991)}]{1991AJ....102.1715B}
{Buta}, R. \& {Crocker}, D.~A. 1991, \aj, 102, 1715

\bibitem[{{Buta} \& {Crocker}(1993)}]{1993AJ....105.1344B}
{Buta}, R. \& {Crocker}, D.~A. 1993, \aj, 105, 1344

\bibitem[{{Buta}(2019)}]{2019MNRAS.488..590B}
{Buta}, R.~J. 2019, \mnras, 488, 590

\bibitem[{{Buta} {et~al.}(2015){Buta}, {Sheth}, {Athanassoula}, {Bosma},
  {Knapen}, {Laurikainen}, {Salo}, {Elmegreen}, {Ho}, {Zaritsky}, {Courtois},
  {Hinz}, {Mu{\~n}oz-Mateos}, {Kim}, {Regan}, {Gadotti}, {Gil de Paz}, {Laine},
  {Men{\'e}ndez-Delmestre}, {Comer{\'o}n}, {Erroz Ferrer}, {Seibert},
  {Mizusawa}, {Holwerda}, \& {Madore}}]{2015ApJS..217...32B}
{Buta}, R.~J., {Sheth}, K., {Athanassoula}, E., {et~al.} 2015, \apjs, 217, 32

\bibitem[{{Catal{\'a}n-Torrecilla} {et~al.}(2017){Catal{\'a}n-Torrecilla}, {Gil
  de Paz}, {Castillo-Morales}, {M{\'e}ndez-Abreu}, {Falc{\'o}n-Barroso},
  {Bekeraite}, {Costantin}, {de Lorenzo-C{\'a}ceres}, {Florido},
  {Garc{\'\i}a-Benito}, {Husemann}, {Iglesias-P{\'a}ramo}, {Kennicutt}, {Mast},
  {Pascual}, {Ruiz-Lara}, {S{\'a}nchez-Menguiano}, {S{\'a}nchez}, {Walcher},
  {Bland-Hawthorn}, {Duarte Puertas}, {Marino}, {Masegosa},
  {S{\'a}nchez-Bl{\'a}zquez}, \& {CALIFA Collaboration}}]{2017ApJ...848...87C}
{Catal{\'a}n-Torrecilla}, C., {Gil de Paz}, A., {Castillo-Morales}, A.,
  {et~al.} 2017, \apj, 848, 87

\bibitem[{{Cenarro} {et~al.}(2019){Cenarro}, {Moles},
  {Crist{\'o}bal-Hornillos}, {Mar{\'\i}n-Franch}, {Ederoclite}, {Varela},
  {L{\'o}pez-Sanjuan}, {Hern{\'a}ndez-Monteagudo}, {Angulo}, {V{\'a}zquez
  Rami{\'o}}, {Viironen}, {Bonoli}, {Orsi}, {Hurier}, {San Roman}, {Greisel},
  {Vilella-Rojo}, {D{\'\i}az-Garc{\'\i}a}, {Logro{\~n}o-Garc{\'\i}a},
  {Gurung-L{\'o}pez}, {Spinoso}, {Izquierdo-Villalba}, {Aguerri}, {Allende
  Prieto}, {Bonatto}, {Carvano}, {Chies-Santos}, {Daflon}, {Dupke},
  {Falc{\'o}n-Barroso}, {Gon{\c{c}}alves}, {Jim{\'e}nez-Teja}, {Molino},
  {Placco}, {Solano}, {Whitten}, {Abril}, {Ant{\'o}n}, {Bello}, {Bielsa de
  Toledo}, {Castillo-Ram{\'\i}rez}, {Chueca}, {Civera},
  {D{\'\i}az-Mart{\'\i}n}, {Dom{\'\i}nguez-Mart{\'\i}nez},
  {Garzar{\'a}n-Calderaro}, {Hern{\'a}ndez-Fuertes}, {Iglesias-Marzoa},
  {I{\~n}iguez}, {Jim{\'e}nez Ruiz}, {Kruuse}, {Lamadrid}, {Lasso-Cabrera},
  {L{\'o}pez-Alegre}, {L{\'o}pez-Sainz}, {Ma{\'\i}cas}, {Moreno-Signes},
  {Muniesa}, {Rodr{\'\i}guez-Llano}, {Rueda-Teruel}, {Rueda-Teruel},
  {Soriano-Lagu{\'\i}a}, {Tilve}, {Valdivielso}, {Yanes-D{\'\i}az}, {Alcaniz},
  {Mendes de Oliveira}, {Sodr{\'e}}, {Coelho}, {Lopes de Oliveira}, {Tamm},
  {Xavier}, {Abramo}, {Akras}, {Alfaro}, {Alvarez-Cand al}, {Ascaso},
  {Beasley}, {Beers}, {Borges Fernandes}, {Bruzual}, {Buzzo}, {Carrasco},
  {Cepa}, {Cortesi}, {Costa-Duarte}, {De Pr{\'a}}, {Favole}, {Galarza},
  {Galbany}, {Garcia}, {Gonz{\'a}lez Delgado}, {Gonz{\'a}lez-Serrano},
  {Guti{\'e}rrez-Soto}, {Hernandez-Jimenez}, {Kanaan}, {Kuncarayakti},
  {Landim}, {Laur}, {Licandro}, {Lima Neto}, {Lyman}, {Ma{\'\i}z
  Apell{\'a}niz}, {Miralda-Escud{\'e}}, {Morate}, {Nogueira-Cavalcante},
  {Novais}, {Oncins}, {Oteo}, {Overzier}, {Pereira}, {Rebassa-Mansergas},
  {Reis}, {Roig}, {Sako}, {Salvador-Rusi{\~n}ol}, {Sampedro},
  {S{\'a}nchez-Bl{\'a}zquez}, {Santos}, {Schmidtobreick}, {Siffert}, {Telles},
  \& {Vilchez}}]{2019A&A...622A.176C}
{Cenarro}, A.~J., {Moles}, M., {Crist{\'o}bal-Hornillos}, D., {et~al.} 2019,
  \aap, 622, A176

\bibitem[{{Cheung} {et~al.}(2013){Cheung}, {Athanassoula}, {Masters}, {Nichol},
  {Bosma}, {Bell}, {Faber}, {Koo}, {Lintott}, {Melvin}, {Schawinski}, {Skibba},
  \& {Willett}}]{2013ApJ...779..162C}
{Cheung}, E., {Athanassoula}, E., {Masters}, K.~L., {et~al.} 2013, \apj, 779,
  162

\bibitem[{{Cisternas} {et~al.}(2013){Cisternas}, {Gadotti}, {Knapen}, {Kim},
  {D{\'{\i}}az-Garc{\'{\i}}a}, {Laurikainen}, {Salo},
  {Gonz{\'a}lez-Mart{\'{\i}}n}, {Ho}, {Elmegreen}, {Zaritsky}, {Sheth},
  {Athanassoula}, {Bosma}, {Comer{\'o}n}, {Erroz-Ferrer}, {Gil de Paz}, {Hinz},
  {Holwerda}, {Laine}, {Meidt}, {Men{\'e}ndez-Delmestre}, {Mizusawa},
  {Mu{\~n}oz-Mateos}, {Regan}, \& {Seibert}}]{2013ApJ...776...50C}
{Cisternas}, M., {Gadotti}, D.~A., {Knapen}, J.~H., {et~al.} 2013, \apj, 776,
  50

\bibitem[{{Coelho} \& {Gadotti}(2011)}]{2011ApJ...743L..13C}
{Coelho}, P. \& {Gadotti}, D.~A. 2011, \apjl, 743, L13

\bibitem[{{Collier} {et~al.}(2019){Collier}, {Shlosman}, \&
  {Heller}}]{2019MNRAS.488.5788C}
{Collier}, A., {Shlosman}, I., \& {Heller}, C. 2019, \mnras, 488, 5788

\bibitem[{{Combes} \& {Gerin}(1985)}]{1985A&A...150..327C}
{Combes}, F. \& {Gerin}, M. 1985, \aap, 150, 327

\bibitem[{{Combes} \& {Sanders}(1981)}]{1981A&A....96..164C}
{Combes}, F. \& {Sanders}, R.~H. 1981, \aap, 96, 164

\bibitem[{{Comer{\'o}n}(2013)}]{2013A&A...555L...4C}
{Comer{\'o}n}, S. 2013, \aap, 555, L4

\bibitem[{{Comer{\'o}n} {et~al.}(2008){Comer{\'o}n}, {Knapen}, \&
  {Beckman}}]{2008A&A...485..695C}
{Comer{\'o}n}, S., {Knapen}, J.~H., \& {Beckman}, J.~E. 2008, \aap, 485, 695

\bibitem[{{Comer{\'o}n} {et~al.}(2010){Comer{\'o}n}, {Knapen}, {Beckman},
  {Laurikainen}, {Salo}, {Mart{\'{\i}}nez-Valpuesta}, \&
  {Buta}}]{2010MNRAS.402.2462C}
{Comer{\'o}n}, S., {Knapen}, J.~H., {Beckman}, J.~E., {et~al.} 2010, \mnras,
  402, 2462

\bibitem[{{Comer{\'o}n} {et~al.}(2014){Comer{\'o}n}, {Salo}, {Laurikainen},
  {Knapen}, {Buta}, {Herrera-Endoqui}, {Laine}, {Holwerda}, {Sheth}, {Regan},
  {Hinz}, {Mu{\~n}oz-Mateos}, {Gil de Paz}, {Men{\'e}ndez-Delmestre},
  {Seibert}, {Mizusawa}, {Kim}, {Erroz-Ferrer}, {Gadotti}, {Athanassoula},
  {Bosma}, \& {Ho}}]{2014A&A...562A.121C}
{Comer{\'o}n}, S., {Salo}, H., {Laurikainen}, E., {et~al.} 2014, \aap, 562,
  A121

\bibitem[{{Dale} {et~al.}(2009){Dale}, {Cohen}, {Johnson}, {Schuster},
  {Calzetti}, {Engelbracht}, {Gil de Paz}, {Kennicutt}, {Lee}, {Begum},
  {Block}, {Dalcanton}, {Funes}, {Gordon}, {Johnson}, {Marble}, {Sakai},
  {Skillman}, {van Zee}, {Walter}, {Weisz}, {Williams}, {Wu}, \&
  {Wu}}]{2009ApJ...703..517D}
{Dale}, D.~A., {Cohen}, S.~A., {Johnson}, L.~C., {et~al.} 2009, \apj, 703, 517

\bibitem[{{Danby}(1965)}]{1965AJ.....70..501D}
{Danby}, J.~M.~A. 1965, \aj, 70, 501

\bibitem[{{de Lorenzo-C{\'a}ceres} {et~al.}(2013){de Lorenzo-C{\'a}ceres},
  {Falc{\'o}n-Barroso}, \& {Vazdekis}}]{2013MNRAS.431.2397D}
{de Lorenzo-C{\'a}ceres}, A., {Falc{\'o}n-Barroso}, J., \& {Vazdekis}, A. 2013,
  \mnras, 431, 2397

\bibitem[{{de Lorenzo-C{\'a}ceres} {et~al.}(2019){de Lorenzo-C{\'a}ceres},
  {S{\'a}nchez-Bl{\'a}zquez}, {M{\'e}ndez-Abreu}, {Gadotti},
  {Falc{\'o}n-Barroso}, {Mart{\'\i}nez-Valpuesta}, {Coelho}, {Fragkoudi},
  {Husemann}, {Leaman}, {P{\'e}rez}, {Querejeta}, {Seidel}, \& {van de
  Ven}}]{2019MNRAS.484.5296D}
{de Lorenzo-C{\'a}ceres}, A., {S{\'a}nchez-Bl{\'a}zquez}, P.,
  {M{\'e}ndez-Abreu}, J., {et~al.} 2019, \mnras, 484, 5296

\bibitem[{{de Lorenzo-C{\'a}ceres} {et~al.}(2012){de Lorenzo-C{\'a}ceres},
  {Vazdekis}, {Aguerri}, {Corsini}, \& {Debattista}}]{2012MNRAS.420.1092D}
{de Lorenzo-C{\'a}ceres}, A., {Vazdekis}, A., {Aguerri}, J.~A.~L., {Corsini},
  E.~M., \& {Debattista}, V.~P. 2012, \mnras, 420, 1092

\bibitem[{{de Vaucouleurs}(1963)}]{1963ApJS....8...31D}
{de Vaucouleurs}, G. 1963, \apjs, 8, 31

\bibitem[{{de Vaucouleurs} \& {de Vaucouleurs}(1963)}]{1963AJ.....68..278D}
{de Vaucouleurs}, G. \& {de Vaucouleurs}, A. 1963, \aj, 68, 278

\bibitem[{{Devereux}(1987)}]{1987ApJ...323...91D}
{Devereux}, N. 1987, \apj, 323, 91

\bibitem[{{D{\'\i}az-Garc{\'\i}a}
  {et~al.}(2019{\natexlab{a}}){D{\'\i}az-Garc{\'\i}a}, {D{\'\i}az-Su{\'a}rez},
  {Knapen}, \& {Salo}}]{2019A&A...625A.146D}
{D{\'\i}az-Garc{\'\i}a}, S., {D{\'\i}az-Su{\'a}rez}, S., {Knapen}, J.~H., \&
  {Salo}, H. 2019{\natexlab{a}}, \aap, 625, A146

\bibitem[{{D{\'\i}az-Garc{\'\i}a} \& {Knapen}(2020)}]{2020A&A...635A.197D}
{D{\'\i}az-Garc{\'\i}a}, S. \& {Knapen}, J.~H. 2020, \aap, 635, A197

\bibitem[{{D{\'\i}az-Garc{\'\i}a}
  {et~al.}(2019{\natexlab{b}}){D{\'\i}az-Garc{\'\i}a}, {Salo}, {Knapen}, \&
  {Herrera-Endoqui}}]{2019A&A...631A..94D}
{D{\'\i}az-Garc{\'\i}a}, S., {Salo}, H., {Knapen}, J.~H., \& {Herrera-Endoqui},
  M. 2019{\natexlab{b}}, \aap, 631, A94

\bibitem[{{D{\'{\i}}az-Garc{\'{\i}}a}
  {et~al.}(2016{\natexlab{a}}){D{\'{\i}}az-Garc{\'{\i}}a}, {Salo}, \&
  {Laurikainen}}]{2016A&A...596A..84D}
{D{\'{\i}}az-Garc{\'{\i}}a}, S., {Salo}, H., \& {Laurikainen}, E.
  2016{\natexlab{a}}, \aap, 596, A84

\bibitem[{{D{\'{\i}}az-Garc{\'{\i}}a}
  {et~al.}(2016{\natexlab{b}}){D{\'{\i}}az-Garc{\'{\i}}a}, {Salo},
  {Laurikainen}, \& {Herrera-Endoqui}}]{2016A&A...587A.160D}
{D{\'{\i}}az-Garc{\'{\i}}a}, S., {Salo}, H., {Laurikainen}, E., \&
  {Herrera-Endoqui}, M. 2016{\natexlab{b}}, \aap, 587, A160

\bibitem[{{Dom{\'\i}nguez S{\'a}nchez} {et~al.}(2018){Dom{\'\i}nguez
  S{\'a}nchez}, {Huertas-Company}, {Bernardi}, {Tuccillo}, \&
  {Fischer}}]{2018MNRAS.476.3661D}
{Dom{\'\i}nguez S{\'a}nchez}, H., {Huertas-Company}, M., {Bernardi}, M.,
  {Tuccillo}, D., \& {Fischer}, J.~L. 2018, \mnras, 476, 3661

\bibitem[{{Donohoe-Keyes} {et~al.}(2019){Donohoe-Keyes}, {Martig}, {James}, \&
  {Kraljic}}]{2019MNRAS.489.4992D}
{Donohoe-Keyes}, C.~E., {Martig}, M., {James}, P.~A., \& {Kraljic}, K. 2019,
  \mnras, 489, 4992

\bibitem[{{Dressel}(1988)}]{1988ApJ...329L..69D}
{Dressel}, L.~L. 1988, \apjl, 329, L69

\bibitem[{{Ellison} {et~al.}(2011){Ellison}, {Nair}, {Patton}, {Scudder},
  {Mendel}, \& {Simard}}]{2011MNRAS.416.2182E}
{Ellison}, S.~L., {Nair}, P., {Patton}, D.~R., {et~al.} 2011, \mnras, 416, 2182

\bibitem[{{Elmegreen} \& {Elmegreen}(1985)}]{1985ApJ...288..438E}
{Elmegreen}, B.~G. \& {Elmegreen}, D.~M. 1985, \apj, 288, 438

\bibitem[{{Elmegreen} {et~al.}(1996){Elmegreen}, {Elmegreen}, {Chromey},
  {Hasselbacher}, \& {Bissell}}]{1996AJ....111.2233E}
{Elmegreen}, B.~G., {Elmegreen}, D.~M., {Chromey}, F.~R., {Hasselbacher},
  D.~A., \& {Bissell}, B.~A. 1996, \aj, 111, 2233

\bibitem[{{Elmegreen} {et~al.}(1989){Elmegreen}, {Elmegreen}, \&
  {Seiden}}]{1989ApJ...343..602E}
{Elmegreen}, B.~G., {Elmegreen}, D.~M., \& {Seiden}, P.~E. 1989, \apj, 343, 602

\bibitem[{{Emsellem} {et~al.}(2015){Emsellem}, {Renaud}, {Bournaud},
  {Elmegreen}, {Combes}, \& {Gabor}}]{2015MNRAS.446.2468E}
{Emsellem}, E., {Renaud}, F., {Bournaud}, F., {et~al.} 2015, \mnras, 446, 2468

\bibitem[{{Epinat} {et~al.}(2008){Epinat}, {Amram}, {Marcelin}, {Balkowski},
  {Daigle}, {Hernandez}, {Chemin}, {Carignan}, {Gach}, \&
  {Balard}}]{2008MNRAS.388..500E}
{Epinat}, B., {Amram}, P., {Marcelin}, M., {et~al.} 2008, \mnras, 388, 500

\bibitem[{{Erroz-Ferrer} {et~al.}(2019){Erroz-Ferrer}, {Carollo}, {den Brok},
  {Onodera}, {Brinchmann}, {Marino}, {Monreal-Ibero}, {Schaye}, {Woo},
  {Cibinel}, {Debattista}, {Inami}, {Maseda}, {Richard}, {Tacchella}, \&
  {Wisotzki}}]{2019MNRAS.484.5009E}
{Erroz-Ferrer}, S., {Carollo}, C.~M., {den Brok}, M., {et~al.} 2019, \mnras,
  484, 5009

\bibitem[{{Erroz-Ferrer} {et~al.}(2016){Erroz-Ferrer}, {Knapen}, {Leaman},
  {D{\'{\i}}az-Garc{\'{\i}}a}, {Salo}, {Laurikainen}, {Querejeta},
  {Mu{\~n}oz-Mateos}, {Athanassoula}, {Bosma}, {Comer{\'o}n}, {Elmegreen}, \&
  {Mart{\'{\i}}nez-Valpuesta}}]{2016MNRAS.458.1199E}
{Erroz-Ferrer}, S., {Knapen}, J.~H., {Leaman}, R., {et~al.} 2016, \mnras, 458,
  1199

\bibitem[{{Erwin}(2018)}]{2018MNRAS.474.5372E}
{Erwin}, P. 2018, \mnras, 474, 5372

\bibitem[{{Erwin}(2019)}]{2019MNRAS.489.3553E}
{Erwin}, P. 2019, \mnras, 489, 3553

\bibitem[{{Erwin} \& {Sparke}(2002)}]{2002AJ....124...65E}
{Erwin}, P. \& {Sparke}, L.~S. 2002, \aj, 124, 65

\bibitem[{{Eskew} {et~al.}(2012){Eskew}, {Zaritsky}, \&
  {Meidt}}]{2012AJ....143..139E}
{Eskew}, M., {Zaritsky}, D., \& {Meidt}, S. 2012, \aj, 143, 139

\bibitem[{{Florido} {et~al.}(2015){Florido}, {Zurita}, {P{\'e}rez},
  {P{\'e}rez-Montero}, {Coelho}, \& {Gadotti}}]{2015A&A...584A..88F}
{Florido}, E., {Zurita}, A., {P{\'e}rez}, I., {et~al.} 2015, \aap, 584, A88

\bibitem[{{Fragkoudi} {et~al.}(2016){Fragkoudi}, {Athanassoula}, \&
  {Bosma}}]{2016MNRAS.462L..41F}
{Fragkoudi}, F., {Athanassoula}, E., \& {Bosma}, A. 2016, \mnras, 462, L41

\bibitem[{{Fraser-McKelvie} {et~al.}(2020){Fraser-McKelvie},
  {Arag{\'o}n-Salamanca}, {Merrifield}, {Masters}, {Nair}, {Emsellem},
  {Kraljic}, {Krishnarao}, {Andrews}, {Drory}, \&
  {Neumann}}]{2020MNRAS.495.4158F}
{Fraser-McKelvie}, A., {Arag{\'o}n-Salamanca}, A., {Merrifield}, M., {et~al.}
  2020, \mnras, 495, 4158

\bibitem[{{Fraser-McKelvie} {et~al.}(2019){Fraser-McKelvie}, {Merrifield},
  {Arag{\'o}n-Salamanca}, {Peterken}, {Masters}, {Krawczyk}, {Andrews},
  {Knapen}, {Kruk}, {Schaefer}, {Smethurst}, {Riffel}, {Brownstein}, \&
  {Drory}}]{2019MNRAS.488L...6F}
{Fraser-McKelvie}, A., {Merrifield}, M., {Arag{\'o}n-Salamanca}, A., {et~al.}
  2019, \mnras, 488, L6

\bibitem[{{Freudling} {et~al.}(2013){Freudling}, {Romaniello}, {Bramich},
  {Ballester}, {Forchi}, {Garc{\'\i}a-Dabl{\'o}}, {Moehler}, \&
  {Neeser}}]{2013A&A...559A..96F}
{Freudling}, W., {Romaniello}, M., {Bramich}, D.~M., {et~al.} 2013, \aap, 559,
  A96

\bibitem[{{Friedli} \& {Benz}(1993)}]{1993A&A...268...65F}
{Friedli}, D. \& {Benz}, W. 1993, \aap, 268, 65

\bibitem[{{Fujii} {et~al.}(2018){Fujii}, {B{\'e}dorf}, {Baba}, \& {Portegies
  Zwart}}]{2018MNRAS.477.1451F}
{Fujii}, M.~S., {B{\'e}dorf}, J., {Baba}, J., \& {Portegies Zwart}, S. 2018,
  \mnras, 477, 1451

\bibitem[{{Gadotti} {et~al.}(2019){Gadotti}, {S{\'a}nchez-Bl{\'a}zquez},
  {Falc{\'o}n-Barroso}, {Husemann}, {Seidel}, {P{\'e}rez}, {de
  Lorenzo-C{\'a}ceres}, {Martinez-Valpuesta}, {Fragkoudi}, {Leung}, {van de
  Ven}, {Leaman}, {Coelho}, {Martig}, {Kim}, {Neumann}, \&
  {Querejeta}}]{2019MNRAS.482..506G}
{Gadotti}, D.~A., {S{\'a}nchez-Bl{\'a}zquez}, P., {Falc{\'o}n-Barroso}, J.,
  {et~al.} 2019, \mnras, 482, 506

\bibitem[{{Gadotti} {et~al.}(2015){Gadotti}, {Seidel},
  {S{\'a}nchez-Bl{\'a}zquez}, {Falc{\'o}n-Barroso}, {Husemann}, {Coelho}, \&
  {P{\'e}rez}}]{2015A&A...584A..90G}
{Gadotti}, D.~A., {Seidel}, M.~K., {S{\'a}nchez-Bl{\'a}zquez}, P., {et~al.}
  2015, \aap, 584, A90

\bibitem[{{Galbany} {et~al.}(2018){Galbany}, {Anderson}, {S{\'a}nchez},
  {Kuncarayakti}, {Pedraz}, {Gonz{\'a}lez-Gait{\'a}n}, {Stanishev},
  {Dom{\'\i}nguez}, {Moreno-Raya}, {Wood-Vasey}, {Mour{\~a}o}, {Ponder},
  {Badenes}, {Moll{\'a}}, {L{\'o}pez-S{\'a}nchez}, {Rosales-Ortega},
  {V{\'\i}lchez}, {Garc{\'\i}a-Benito}, \& {Marino}}]{2018ApJ...855..107G}
{Galbany}, L., {Anderson}, J.~P., {S{\'a}nchez}, S.~F., {et~al.} 2018, \apj,
  855, 107

\bibitem[{{Garc{\'\i}a-Barreto} {et~al.}(1996){Garc{\'\i}a-Barreto}, {Franco},
  {Carrillo}, {Venegas}, \& {Escalante-Ram{\'\i}rez}}]{1996RMxAA..32...89G}
{Garc{\'\i}a-Barreto}, J.~A., {Franco}, J., {Carrillo}, R., {Venegas}, S., \&
  {Escalante-Ram{\'\i}rez}, B. 1996, \rmxaa, 32, 89

\bibitem[{{Gavazzi} {et~al.}(2003){Gavazzi}, {Boselli}, {Donati}, {Franzetti},
  \& {Scodeggio}}]{2003A&A...400..451G}
{Gavazzi}, G., {Boselli}, A., {Donati}, A., {Franzetti}, P., \& {Scodeggio}, M.
  2003, \aap, 400, 451

\bibitem[{{Gavazzi} {et~al.}(2015{\natexlab{a}}){Gavazzi}, {Consolandi},
  {Dotti}, {Fanali}, {Fossati}, {Fumagalli}, {Viscardi}, {Savorgnan},
  {Boselli}, {Guti{\'e}rrez}, {Hern{\'a}ndez Toledo}, {Giovanelli}, \&
  {Haynes}}]{2015A&A...580A.116G}
{Gavazzi}, G., {Consolandi}, G., {Dotti}, M., {et~al.} 2015{\natexlab{a}},
  \aap, 580, A116

\bibitem[{{Gavazzi} {et~al.}(2018){Gavazzi}, {Consolandi}, {Pedraglio},
  {Fossati}, {Fumagalli}, \& {Boselli}}]{2018A&A...611A..28G}
{Gavazzi}, G., {Consolandi}, G., {Pedraglio}, S., {et~al.} 2018, \aap, 611, A28

\bibitem[{{Gavazzi} {et~al.}(2015{\natexlab{b}}){Gavazzi}, {Consolandi},
  {Viscardi}, {Fossati}, {Savorgnan}, {Fumagalli}, {Gutierrez}, {Hernandez
  Toledo}, {Boselli}, {Giovanelli}, \& {Haynes}}]{2015A&A...576A..16G}
{Gavazzi}, G., {Consolandi}, G., {Viscardi}, E., {et~al.} 2015{\natexlab{b}},
  \aap, 576, A16

\bibitem[{{Gavazzi} {et~al.}(2014){Gavazzi}, {Franzetti}, \&
  {Boselli}}]{2014arXiv1401.8123G}
{Gavazzi}, G., {Franzetti}, P., \& {Boselli}, A. 2014, arXiv e-prints,
  arXiv:1401.8123

\bibitem[{{Gavazzi} {et~al.}(2012){Gavazzi}, {Fumagalli}, {Galardo},
  {Grossetti}, {Boselli}, {Giovanelli}, {Haynes}, \&
  {Fabello}}]{2012A&A...545A..16G}
{Gavazzi}, G., {Fumagalli}, M., {Galardo}, V., {et~al.} 2012, \aap, 545, A16

\bibitem[{{George} {et~al.}(2019){George}, {Joseph}, {Mondal}, {Subramanian},
  {Subramaniam}, \& {Paul}}]{2019A&A...621L...4G}
{George}, K., {Joseph}, P., {Mondal}, C., {et~al.} 2019, \aap, 621, L4

\bibitem[{{Gil de Paz} {et~al.}(2007){Gil de Paz}, {Boissier}, {Madore},
  {Seibert}, {Joe}, {Boselli}, {Wyder}, {Thilker}, {Bianchi}, {Rey}, {Rich},
  {Barlow}, {Conrow}, {Forster}, {Friedman}, {Martin}, {Morrissey}, {Neff},
  {Schiminovich}, {Small}, {Donas}, {Heckman}, {Lee}, {Milliard}, {Szalay}, \&
  {Yi}}]{2007ApJS..173..185G}
{Gil de Paz}, A., {Boissier}, S., {Madore}, B.~F., {et~al.} 2007, \apjs, 173,
  185

\bibitem[{{Gil de Paz} {et~al.}(2003){Gil de Paz}, {Madore}, \&
  {Pevunova}}]{2003ApJS..147...29G}
{Gil de Paz}, A., {Madore}, B.~F., \& {Pevunova}, O. 2003, \apjs, 147, 29

\bibitem[{{Giovanelli} \& {Haynes}(1988)}]{1988gera.book..522G}
{Giovanelli}, R. \& {Haynes}, M.~P. 1988, {Extragalactic neutral hydrogen.},
  ed. K.~I. {Kellermann} \& G.~L. {Verschuur}, 522--562

\bibitem[{{Grouchy} {et~al.}(2010){Grouchy}, {Buta}, {Salo}, \&
  {Laurikainen}}]{2010AJ....139.2465G}
{Grouchy}, R.~D., {Buta}, R.~J., {Salo}, H., \& {Laurikainen}, E. 2010, \aj,
  139, 2465

\bibitem[{{Gunn} {et~al.}(2006){Gunn}, {Siegmund}, {Mannery}, {Owen}, {Hull},
  {Leger}, {Carey}, {Knapp}, {York}, {Boroski}, {Kent}, {Lupton}, {Rockosi},
  {Evans}, {Waddell}, {Anderson}, {Annis}, {Barentine}, {Bartoszek}, {Bastian},
  {Bracker}, {Brewington}, {Briegel}, {Brinkmann}, {Brown}, {Carr},
  {Czarapata}, {Drennan}, {Dombeck}, {Federwitz}, {Gillespie}, {Gonzales},
  {Hansen}, {Harvanek}, {Hayes}, {Jordan}, {Kinney}, {Klaene}, {Kleinman},
  {Kron}, {Kresinski}, {Lee}, {Limmongkol}, {Lindenmeyer}, {Long}, {Loomis},
  {McGehee}, {Mantsch}, {Neilsen}, {Neswold}, {Newman}, {Nitta}, {Peoples},
  {Pier}, {Prieto}, {Prosapio}, {Rivetta}, {Schneider}, {Snedden}, \&
  {Wang}}]{2006AJ....131.2332G}
{Gunn}, J.~E., {Siegmund}, W.~A., {Mannery}, E.~J., {et~al.} 2006, \aj, 131,
  2332

\bibitem[{{Hameed} \& {Devereux}(1999)}]{1999AJ....118..730H}
{Hameed}, S. \& {Devereux}, N. 1999, \aj, 118, 730

\bibitem[{{Hao} {et~al.}(2009){Hao}, {Jogee}, {Barazza}, {Marinova}, \&
  {Shen}}]{2009ASPC..419..402H}
{Hao}, L., {Jogee}, S., {Barazza}, F.~D., {Marinova}, I., \& {Shen}, J. 2009,
  in Astronomical Society of the Pacific Conference Series, Vol. 419, Galaxy
  Evolution: Emerging Insights and Future Challenges, ed. S.~{Jogee},
  I.~{Marinova}, L.~{Hao}, \& G.~A. {Blanc}, 402

\bibitem[{{Hawarden} {et~al.}(1986){Hawarden}, {Mountain}, {Leggett}, \&
  {Puxley}}]{1986MNRAS.221P..41H}
{Hawarden}, T.~G., {Mountain}, C.~M., {Leggett}, S.~K., \& {Puxley}, P.~J.
  1986, \mnras, 221, 41P

\bibitem[{{Haynes} {et~al.}(2011){Haynes}, {Giovanelli}, {Martin}, {Hess},
  {Saintonge}, {Adams}, {Hallenbeck}, {Hoffman}, {Huang}, {Kent}, {Koopmann},
  {Papastergis}, {Stierwalt}, {Balonek}, {Craig}, {Higdon}, {Kornreich},
  {Miller}, {O'Donoghue}, {Olowin}, {Rosenberg}, {Spekkens}, {Troischt}, \&
  {Wilcots}}]{2011AJ....142..170H}
{Haynes}, M.~P., {Giovanelli}, R., {Martin}, A.~M., {et~al.} 2011, \aj, 142,
  170

\bibitem[{{Heckman}(1980)}]{1980A&A....88..365H}
{Heckman}, T.~M. 1980, \aap, 88, 365

\bibitem[{{Heller} \& {Shlosman}(1994)}]{1994ApJ...424...84H}
{Heller}, C.~H. \& {Shlosman}, I. 1994, \apj, 424, 84

\bibitem[{{Herrera-Endoqui} {et~al.}(2015){Herrera-Endoqui},
  {D{\'{\i}}az-Garc{\'{\i}}a}, {Laurikainen}, \& {Salo}}]{2015A&A...582A..86H}
{Herrera-Endoqui}, M., {D{\'{\i}}az-Garc{\'{\i}}a}, S., {Laurikainen}, E., \&
  {Salo}, H. 2015, \aap, 582, A86

\bibitem[{{Hilmi} {et~al.}(2020){Hilmi}, {Minchev}, {Buck}, {Martig},
  {Quillen}, {Monari}, {Famaey}, {de Jong}, {Laporte}, {Read}, {Sand ers},
  {Steinmetz}, \& {Wegg}}]{2020MNRAS.497..933H}
{Hilmi}, T., {Minchev}, I., {Buck}, T., {et~al.} 2020, \mnras, 497, 933

\bibitem[{{Hoopes} {et~al.}(2001){Hoopes}, {Walterbos}, \&
  {Bothun}}]{2001ApJ...559..878H}
{Hoopes}, C.~G., {Walterbos}, R. A.~M., \& {Bothun}, G.~D. 2001, \apj, 559, 878

\bibitem[{{Hummel}(1981)}]{1981A&A....93...93H}
{Hummel}, E. 1981, \aap, 93, 93

\bibitem[{{Hunter} \& {Elmegreen}(2004)}]{2004AJ....128.2170H}
{Hunter}, D.~A. \& {Elmegreen}, B.~G. 2004, \aj, 128, 2170

\bibitem[{{Hunter}(2007)}]{Hunter2007}
{Hunter}, J.~D. 2007, Institute of Electrical and Electronic Engineers, 9, 90

\bibitem[{{Husemann} {et~al.}(2017){Husemann}, {Tremblay}, {Davis}, {Busch},
  {McElroy}, {Neumann}, {Urrutia}, {Krumpe}, {Scharw{\"a}chter}, {Powell},
  {Perez-Torres}, \& {CARS Team}}]{2017Msngr.169...42H}
{Husemann}, B., {Tremblay}, G., {Davis}, T., {et~al.} 2017, The Messenger, 169,
  42

\bibitem[{{Iodice} {et~al.}(2019){Iodice}, {Sarzi}, {Bittner}, {Coccato},
  {Costantin}, {Corsini}, {van de Ven}, {de Zeeuw}, {Falc{\'o}n-Barroso},
  {Gadotti}, {Lyubenova}, {Mart{\'\i}n-Navarro}, {McDermid}, {Nedelchev},
  {Pinna}, {Pizzella}, {Spavone}, \& {Viaene}}]{2019A&A...627A.136I}
{Iodice}, E., {Sarzi}, M., {Bittner}, A., {et~al.} 2019, \aap, 627, A136

\bibitem[{{James} {et~al.}(2009){James}, {Bretherton}, \&
  {Knapen}}]{2009A&A...501..207J}
{James}, P.~A., {Bretherton}, C.~F., \& {Knapen}, J.~H. 2009, \aap, 501, 207

\bibitem[{{James} \& {Percival}(2016)}]{2016MNRAS.457..917J}
{James}, P.~A. \& {Percival}, S.~M. 2016, \mnras, 457, 917

\bibitem[{{James} \& {Percival}(2018)}]{2018MNRAS.474.3101J}
{James}, P.~A. \& {Percival}, S.~M. 2018, \mnras, 474, 3101

\bibitem[{{James} {et~al.}(2004){James}, {Shane}, {Beckman}, {Cardwell},
  {Collins}, {Etherton}, {de Jong}, {Fathi}, {Knapen}, {Peletier}, {Percival},
  {Pollacco}, {Seigar}, {Stedman}, \& {Steele}}]{2004A&A...414...23J}
{James}, P.~A., {Shane}, N.~S., {Beckman}, J.~E., {et~al.} 2004, \aap, 414, 23

\bibitem[{{Jedrzejewski}(1987)}]{1987MNRAS.226..747J}
{Jedrzejewski}, R.~I. 1987, \mnras, 226, 747

\bibitem[{{Jogee} {et~al.}(2002){Jogee}, {Knapen}, {Laine}, {Shlosman},
  {Scoville}, \& {Englmaier}}]{2002ApJ...570L..55J}
{Jogee}, S., {Knapen}, J.~H., {Laine}, S., {et~al.} 2002, \apjl, 570, L55

\bibitem[{{Jogee} {et~al.}(2005){Jogee}, {Scoville}, \&
  {Kenney}}]{2005ApJ...630..837J}
{Jogee}, S., {Scoville}, N., \& {Kenney}, J. D.~P. 2005, \apj, 630, 837

\bibitem[{{Kaisin} \& {Karachentsev}(2006)}]{2006Ap.....49..287K}
{Kaisin}, S.~S. \& {Karachentsev}, I.~D. 2006, Astrophysics, 49, 287

\bibitem[{{Karachentsev} {et~al.}(2015){Karachentsev}, {Kaisin}, \&
  {Kaisina}}]{2015Ap.....58..453K}
{Karachentsev}, I.~D., {Kaisin}, S.~S., \& {Kaisina}, E.~I. 2015, Astrophysics,
  58, 453

\bibitem[{{Kennicutt}(1989)}]{1989ApJ...344..685K}
{Kennicutt}, Robert~C., J. 1989, \apj, 344, 685

\bibitem[{{Kennicutt}(1998)}]{1998ARA&A..36..189K}
{Kennicutt}, Robert~C., J. 1998, \araa, 36, 189

\bibitem[{{Kennicutt} \& {Evans}(2012)}]{2012ARA&A..50..531K}
{Kennicutt}, R.~C. \& {Evans}, N.~J. 2012, \araa, 50, 531

\bibitem[{{Kennicutt} {et~al.}(2003){Kennicutt}, {Armus}, {Bendo}, {Calzetti},
  {Dale}, {Draine}, {Engelbracht}, {Gordon}, {Grauer}, {Helou}, {Hollenbach},
  {Jarrett}, {Kewley}, {Leitherer}, {Li}, {Malhotra}, {Regan}, {Rieke},
  {Rieke}, {Roussel}, {Smith}, {Thornley}, \& {Walter}}]{2003PASP..115..928K}
{Kennicutt}, Jr., R.~C., {Armus}, L., {Bendo}, G., {et~al.} 2003, \pasp, 115,
  928

\bibitem[{{Khoperskov} {et~al.}(2018){Khoperskov}, {Haywood}, {Di Matteo},
  {Lehnert}, \& {Combes}}]{2018A&A...609A..60K}
{Khoperskov}, S., {Haywood}, M., {Di Matteo}, P., {Lehnert}, M.~D., \&
  {Combes}, F. 2018, \aap, 609, A60

\bibitem[{{Kim} {et~al.}(2015){Kim}, {Sheth}, {Gadotti}, {Lee}, {Zaritsky},
  {Elmegreen}, {Athanassoula}, {Bosma}, {Holwerda}, {Ho}, {Comer{\'o}n},
  {Knapen}, {Hinz}, {Mu{\~n}oz-Mateos}, {Erroz-Ferrer}, {Buta}, {Kim},
  {Laurikainen}, {Salo}, {Madore}, {Laine}, {Men{\'e}ndez-Delmestre}, {Regan},
  {de Swardt}, {Gil de Paz}, {Seibert}, \& {Mizusawa}}]{2015ApJ...799...99K}
{Kim}, T., {Sheth}, K., {Gadotti}, D.~A., {et~al.} 2015, \apj, 799, 99

\bibitem[{{Knapen}(2004)}]{2004ASSL..319..189K}
{Knapen}, J.~H. 2004, Astrophysics and Space Science Library, Vol. 319,
  {Fuelling starbursts and AGN}, ed. D.~L. {Block}, I.~{Puerari}, K.~C.
  {Freeman}, R.~{Groess}, \& E.~K. {Block}, 189

\bibitem[{{Knapen}(2005)}]{2005A&A...429..141K}
{Knapen}, J.~H. 2005, \aap, 429, 141

\bibitem[{{Knapen} {et~al.}(1995){Knapen}, {Beckman}, {Heller}, {Shlosman}, \&
  {de Jong}}]{1995ApJ...454..623K}
{Knapen}, J.~H., {Beckman}, J.~E., {Heller}, C.~H., {Shlosman}, I., \& {de
  Jong}, R.~S. 1995, \apj, 454, 623

\bibitem[{{Knapen} {et~al.}(2006){Knapen}, {Mazzuca}, {B{\"o}ker}, {Shlosman},
  {Colina}, {Combes}, \& {Axon}}]{2006A&A...448..489K}
{Knapen}, J.~H., {Mazzuca}, L.~M., {B{\"o}ker}, T., {et~al.} 2006, \aap, 448,
  489

\bibitem[{{Knapen} {et~al.}(2000){Knapen}, {Shlosman}, \&
  {Peletier}}]{2000ApJ...529...93K}
{Knapen}, J.~H., {Shlosman}, I., \& {Peletier}, R.~F. 2000, \apj, 529, 93

\bibitem[{{Knapen} {et~al.}(2004){Knapen}, {Stedman}, {Bramich}, {Folkes}, \&
  {Bradley}}]{2004A&A...426.1135K}
{Knapen}, J.~H., {Stedman}, S., {Bramich}, D.~M., {Folkes}, S.~L., \&
  {Bradley}, T.~R. 2004, \aap, 426, 1135

\bibitem[{{Koopmann}(1997)}]{1997PhDT........11K}
{Koopmann}, R.~A. 1997, PhD thesis, Yale University

\bibitem[{{Koopmann} \& {Kenney}(2006)}]{2006ApJS..162...97K}
{Koopmann}, R.~A. \& {Kenney}, J. D.~P. 2006, \apjs, 162, 97

\bibitem[{{Koopmann} {et~al.}(2001){Koopmann}, {Kenney}, \&
  {Young}}]{2001ApJS..135..125K}
{Koopmann}, R.~A., {Kenney}, J. D.~P., \& {Young}, J. 2001, \apjs, 135, 125

\bibitem[{{Kormendy}(2013)}]{2013seg..book....1K}
{Kormendy}, J. 2013, {Secular Evolution in Disk Galaxies}, ed.
  J.~{Falc{\'o}n-Barroso} \& J.~H. {Knapen}, 1

\bibitem[{{Kormendy} \& {Kennicutt}(2004)}]{2004ARA&A..42..603K}
{Kormendy}, J. \& {Kennicutt}, Jr., R.~C. 2004, \araa, 42, 603

\bibitem[{{Kostiuk} \& {Sil'chenko}(2015)}]{2015BaltA..24..426K}
{Kostiuk}, I.~P. \& {Sil'chenko}, O.~K. 2015, Baltic Astronomy, 24, 426

\bibitem[{{Kreckel} {et~al.}(2019){Kreckel}, {Ho}, {Blanc}, {Groves},
  {Santoro}, {Schinnerer}, {Bigiel}, {Chevance}, {Congiu}, {Emsellem}, {Faesi},
  {Glover}, {Grasha}, {Kruijssen}, {Lang}, {Leroy}, {Meidt}, {McElroy}, {Pety},
  {Rosolowsky}, {Saito}, {Sandstrom}, {Sanchez-Blazquez}, \&
  {Schruba}}]{2019ApJ...887...80K}
{Kreckel}, K., {Ho}, I.~T., {Blanc}, G.~A., {et~al.} 2019, \apj, 887, 80

\bibitem[{{Krishnarao} {et~al.}(2020){Krishnarao}, {Tremonti},
  {Fraser-McKelvie}, {Kraljic}, {Boardman}, {Masters}, {Benjamin}, {Haffner},
  {Jones}, {Pace}, {Zasowski}, {Bershady}, {Bizyaev}, {Brinkmann},
  {Brownstein}, {Drory}, {Pan}, \& {Zhang}}]{2020ApJ...898..116K}
{Krishnarao}, D., {Tremonti}, C., {Fraser-McKelvie}, A., {et~al.} 2020, \apj,
  898, 116

\bibitem[{{Kroupa}(2001)}]{2001MNRAS.322..231K}
{Kroupa}, P. 2001, \mnras, 322, 231

\bibitem[{{Kruk} {et~al.}(2018){Kruk}, {Lintott}, {Bamford}, {Masters},
  {Simmons}, {H{\"a}u{\ss}ler}, {Cardamone}, {Hart}, {Kelvin}, {Schawinski},
  {Smethurst}, \& {Vika}}]{2018MNRAS.473.4731K}
{Kruk}, S.~J., {Lintott}, C.~J., {Bamford}, S.~P., {et~al.} 2018, \mnras, 473,
  4731

\bibitem[{{Kuncarayakti} {et~al.}(2018){Kuncarayakti}, {Anderson}, {Galbany},
  {Maeda}, {Hamuy}, {Aldering}, {Arimoto}, {Doi}, {Morokuma}, \&
  {Usuda}}]{2018A&A...613A..35K}
{Kuncarayakti}, H., {Anderson}, J.~P., {Galbany}, L., {et~al.} 2018, \aap, 613,
  A35

\bibitem[{{Laine} {et~al.}(2016){Laine}, {Laurikainen}, \&
  {Salo}}]{2016A&A...596A..25L}
{Laine}, J., {Laurikainen}, E., \& {Salo}, H. 2016, \aap, 596, A25

\bibitem[{{Larsen} \& {Richtler}(1999)}]{1999A&A...345...59L}
{Larsen}, S.~S. \& {Richtler}, T. 1999, \aap, 345, 59

\bibitem[{{Laurikainen} \& {Salo}(2002)}]{2002MNRAS.337.1118L}
{Laurikainen}, E. \& {Salo}, H. 2002, \mnras, 337, 1118

\bibitem[{{Laurikainen} {et~al.}(2004){Laurikainen}, {Salo}, \&
  {Buta}}]{2004ApJ...607..103L}
{Laurikainen}, E., {Salo}, H., \& {Buta}, R. 2004, \apj, 607, 103

\bibitem[{{Laurikainen} {et~al.}(2007){Laurikainen}, {Salo}, {Buta}, \&
  {Knapen}}]{2007MNRAS.381..401L}
{Laurikainen}, E., {Salo}, H., {Buta}, R., \& {Knapen}, J.~H. 2007, \mnras,
  381, 401

\bibitem[{{Laurikainen} {et~al.}(2002){Laurikainen}, {Salo}, \&
  {Rautiainen}}]{2002MNRAS.331..880L}
{Laurikainen}, E., {Salo}, H., \& {Rautiainen}, P. 2002, \mnras, 331, 880

\bibitem[{{Leaman} {et~al.}(2019){Leaman}, {Fragkoudi}, {Querejeta}, {Leung},
  {Gadotti}, {Husemann}, {Falc{\'o}n-Barroso}, {S{\'a}nchez-Bl{\'a}zquez}, {van
  de Ven}, {Kim}, {Coelho}, {Lyubenova}, {de Lorenzo-C{\'a}ceres}, {Martig},
  {Martinez-Valpuesta}, {Neumann}, {P{\'e}rez}, \&
  {Seidel}}]{2019MNRAS.488.3904L}
{Leaman}, R., {Fragkoudi}, F., {Querejeta}, M., {et~al.} 2019, \mnras, 488,
  3904

\bibitem[{{Lee} {et~al.}(2012){Lee}, {Woo}, {Lee}, {Hwang}, {Lee}, {Sohn}, \&
  {Lee}}]{2012ApJ...750..141L}
{Lee}, G.-H., {Woo}, J.-H., {Lee}, M.~G., {et~al.} 2012, \apj, 750, 141

\bibitem[{{Lin} {et~al.}(2020){Lin}, {Li}, {Du}, {Wang}, {Xiao}, {Bureau},
  {Fraser-McKelvie}, {Masters}, {Lin}, {Wake}, \& {Hao}}]{2020MNRAS.tmp.2737L}
{Lin}, L., {Li}, C., {Du}, C., {et~al.} 2020, \mnras
  [\eprint[arXiv]{2005.09853}]

\bibitem[{{Lin} {et~al.}(2017){Lin}, {Li}, {He}, {Xiao}, \&
  {Wang}}]{2017ApJ...838..105L}
{Lin}, L., {Li}, C., {He}, Y., {Xiao}, T., \& {Wang}, E. 2017, \apj, 838, 105

\bibitem[{{L{\'o}pez-Cob{\'a}} {et~al.}(2020){L{\'o}pez-Cob{\'a}},
  {S{\'a}nchez}, {Anderson}, {Cruz-Gonz{\'a}lez}, {Galbany}, {Ruiz-Lara},
  {Barrera-Ballesteros}, {Prieto}, \& {Kuncarayakti}}]{2020AJ....159..167L}
{L{\'o}pez-Cob{\'a}}, C., {S{\'a}nchez}, S.~F., {Anderson}, J.~P., {et~al.}
  2020, \aj, 159, 167

\bibitem[{{Lyman} {et~al.}(2020){Lyman}, {Galbany}, {S{\'a}nchez}, {Anderson},
  {Kuncarayakti}, \& {Prieto}}]{2020MNRAS.495..992L}
{Lyman}, J.~D., {Galbany}, L., {S{\'a}nchez}, S.~F., {et~al.} 2020, \mnras,
  495, 992

\bibitem[{{Lyman} {et~al.}(2018){Lyman}, {Taddia}, {Stritzinger}, {Galbany},
  {Leloudas}, {Anderson}, {Eldridge}, {James}, {Kr{\"u}hler}, {Levan},
  {Pignata}, \& {Stanway}}]{2018MNRAS.473.1359L}
{Lyman}, J.~D., {Taddia}, F., {Stritzinger}, M.~D., {et~al.} 2018, \mnras, 473,
  1359

\bibitem[{{Lynden-Bell}(1979)}]{1979MNRAS.187..101L}
{Lynden-Bell}, D. 1979, \mnras, 187, 101

\bibitem[{{Madau} \& {Dickinson}(2014)}]{2014ARA&A..52..415M}
{Madau}, P. \& {Dickinson}, M. 2014, \araa, 52, 415

\bibitem[{{Maeda} {et~al.}(2020){Maeda}, {Ohta}, {Fujimoto}, {Habe}, \&
  {Ushio}}]{2020MNRAS.tmp.1411M}
{Maeda}, F., {Ohta}, K., {Fujimoto}, Y., {Habe}, A., \& {Ushio}, K. 2020,
  \mnras [\eprint[arXiv]{2005.03019}]

\bibitem[{{Marinova} \& {Jogee}(2007)}]{2007ApJ...659.1176M}
{Marinova}, I. \& {Jogee}, S. 2007, \apj, 659, 1176

\bibitem[{{Martin} \& {Friedli}(1997)}]{1997A&A...326..449M}
{Martin}, P. \& {Friedli}, D. 1997, \aap, 326, 449

\bibitem[{{Martinez-Valpuesta} {et~al.}(2007){Martinez-Valpuesta}, {Knapen}, \&
  {Buta}}]{2007AJ....134.1863M}
{Martinez-Valpuesta}, I., {Knapen}, J.~H., \& {Buta}, R. 2007, \aj, 134, 1863

\bibitem[{{Masters} {et~al.}(2012){Masters}, {Nichol}, {Haynes}, {Keel},
  {Lintott}, {Simmons}, {Skibba}, {Bamford}, {Giovanelli}, \&
  {Schawinski}}]{2012MNRAS.424.2180M}
{Masters}, K.~L., {Nichol}, R.~C., {Haynes}, M.~P., {et~al.} 2012, \mnras, 424,
  2180

\bibitem[{{Meidt} {et~al.}(2014){Meidt}, {Schinnerer}, {van de Ven},
  {Zaritsky}, {Peletier}, {Knapen}, {Sheth}, {Regan}, {Querejeta},
  {Mu{\~n}oz-Mateos}, {Kim}, {Hinz}, {Gil de Paz}, {Athanassoula}, {Bosma},
  {Buta}, {Cisternas}, {Ho}, {Holwerda}, {Skibba}, {Laurikainen}, {Salo},
  {Gadotti}, {Laine}, {Erroz-Ferrer}, {Comer{\'o}n}, {Men{\'e}ndez-Delmestre},
  {Seibert}, \& {Mizusawa}}]{2014ApJ...788..144M}
{Meidt}, S.~E., {Schinnerer}, E., {van de Ven}, G., {et~al.} 2014, \apj, 788,
  144

\bibitem[{{Mendes de Oliveira} {et~al.}(2019){Mendes de Oliveira}, {Ribeiro},
  {Schoenell}, {Kanaan}, {Overzier}, {Molino}, {Sampedro}, {Coelho}, {Barbosa},
  {Cortesi}, {Costa-Duarte}, {Herpich}, {Hernand ez-Jimenez}, {Placco},
  {Xavier}, {Abramo}, {Saito}, {Chies-Santos}, {Ederoclite}, {Lopes de
  Oliveira}, {Gon{\c{c}}alves}, {Akras}, {Almeida}, {Almeida-Fernandes},
  {Beers}, {Bonatto}, {Bonoli}, {Cypriano}, {Vinicius-Lima}, {de Souza},
  {Fabiano de Souza}, {Ferrari}, {Gon{\c{c}}alves}, {Gonzalez},
  {Guti{\'e}rrez-Soto}, {Hartmann}, {Jaffe}, {Kerber}, {Lima-Dias}, {Lopes},
  {Menendez-Delmestre}, {Nakazono}, {Novais}, {Ortega-Minakata}, {Pereira},
  {Perottoni}, {Queiroz}, {Reis}, {Santos}, {Santos-Silva}, {Santucci},
  {Barbosa}, {Siffert}, {Sodr{\'e}}, {Torres-Flores}, {Westera}, {Whitten},
  {Alcaniz}, {Alonso-Garc{\'\i}a}, {Alencar}, {Alvarez-Cand al}, {Amram},
  {Azanha}, {Barb{\'a}}, {Bernardinelli}, {Borges Fernandes}, {Branco},
  {Brito-Silva}, {Buzzo}, {Caffer}, {Campillay}, {Cano}, {Carvano}, {Castejon},
  {Cid Fernandes}, {Dantas}, {Daflon}, {Damke}, {de la Reza}, {de Melo de
  Azevedo}, {De Paula}, {Diem}, {Donnerstein}, {Dors}, {Dupke}, {Eikenberry},
  {Escudero}, {Faifer}, {Far{\'\i}as}, {Fernandes}, {Fernandes}, {Fontes},
  {Galarza}, {Hirata}, {Katena}, {Gregorio-Hetem},
  {Hern{\'a}ndez-Fern{\'a}ndez}, {Izzo}, {Jaque Arancibia}, {Jatenco-Pereira},
  {Jim{\'e}nez-Teja}, {Kann}, {Krabbe}, {Labayru}, {Lazzaro}, {Lima Neto},
  {Lopes}, {Magalh{\~a}es}, {Makler}, {de Menezes}, {Miralda-Escud{\'e}},
  {Monteiro-Oliveira}, {Montero-Dorta}, {Mu{\~n}oz-Elgueta}, {Nemmen}, {Nilo
  Castell{\'o}n}, {Oliveira}, {Ort{\'\i}z}, {Pattaro}, {Pereira}, {Quint},
  {Riguccini}, {Rocha Pinto}, {Rodrigues}, {Roig}, {Rossi}, {Saha}, {Santos},
  {Schnorr M{\"u}ller}, {Sesto}, {Silva}, {Smith Castelli}, {Teixeira},
  {Telles}, {Thom de Souza}, {Th{\"o}ne}, {Trevisan}, {de Ugarte Postigo},
  {Urrutia-Viscarra}, {Veiga}, {Vika}, {Vitorelli}, {Werle}, {Werner}, \&
  {Zaritsky}}]{2019MNRAS.489..241M}
{Mendes de Oliveira}, C., {Ribeiro}, T., {Schoenell}, W., {et~al.} 2019,
  \mnras, 489, 241

\bibitem[{{M{\'e}ndez-Abreu} {et~al.}(2019){M{\'e}ndez-Abreu}, {de
  Lorenzo-C{\'a}ceres}, {Gadotti}, {Fragkoudi}, {van de Ven},
  {Falc{\'o}n-Barroso}, {Leaman}, {P{\'e}rez}, {Querejeta},
  {S{\'a}nchez-Blazquez}, \& {Seidel}}]{2019MNRAS.482L.118M}
{M{\'e}ndez-Abreu}, J., {de Lorenzo-C{\'a}ceres}, A., {Gadotti}, D.~A.,
  {et~al.} 2019, \mnras, 482, L118

\bibitem[{{M{\'e}ndez-Abreu} {et~al.}(2014){M{\'e}ndez-Abreu}, {Debattista},
  {Corsini}, \& {Aguerri}}]{2014A&A...572A..25M}
{M{\'e}ndez-Abreu}, J., {Debattista}, V.~P., {Corsini}, E.~M., \& {Aguerri},
  J.~A.~L. 2014, \aap, 572, A25

\bibitem[{{M{\'e}ndez-Abreu} {et~al.}(2012){M{\'e}ndez-Abreu},
  {S{\'a}nchez-Janssen}, {Aguerri}, {Corsini}, \&
  {Zarattini}}]{2012ApJ...761L...6M}
{M{\'e}ndez-Abreu}, J., {S{\'a}nchez-Janssen}, R., {Aguerri}, J.~A.~L.,
  {Corsini}, E.~M., \& {Zarattini}, S. 2012, \apjl, 761, L6

\bibitem[{{Men{\'e}ndez-Delmestre} {et~al.}(2007){Men{\'e}ndez-Delmestre},
  {Sheth}, {Schinnerer}, {Jarrett}, \& {Scoville}}]{2007ApJ...657..790M}
{Men{\'e}ndez-Delmestre}, K., {Sheth}, K., {Schinnerer}, E., {Jarrett}, T.~H.,
  \& {Scoville}, N.~Z. 2007, \apj, 657, 790

\bibitem[{{Meurer} {et~al.}(2006){Meurer}, {Hanish}, {Ferguson}, {Knezek},
  {Kilborn}, {Putman}, {Smith}, {Koribalski}, {Meyer}, {Oey}, {Ryan-Weber},
  {Zwaan}, {Heckman}, {Kennicutt}, {Lee}, {Webster}, {Bland -Hawthorn},
  {Dopita}, {Freeman}, {Doyle}, {Drinkwater}, {Staveley-Smith}, \&
  {Werk}}]{2006ApJS..165..307M}
{Meurer}, G.~R., {Hanish}, D.~J., {Ferguson}, H.~C., {et~al.} 2006, \apjs, 165,
  307

\bibitem[{{Minchev} {et~al.}(2011){Minchev}, {Famaey}, {Combes}, {Di Matteo},
  {Mouhcine}, \& {Wozniak}}]{2011A&A...527A.147M}
{Minchev}, I., {Famaey}, B., {Combes}, F., {et~al.} 2011, \aap, 527, A147

\bibitem[{{Minchev} {et~al.}(2012){Minchev}, {Famaey}, {Quillen}, {Di Matteo},
  {Combes}, {Vlaji{\'c}}, {Erwin}, \& {Bland -Hawthorn}}]{2012A&A...548A.126M}
{Minchev}, I., {Famaey}, B., {Quillen}, A.~C., {et~al.} 2012, \aap, 548, A126

\bibitem[{{Mingozzi} {et~al.}(2019){Mingozzi}, {Cresci}, {Venturi}, {Marconi},
  {Mannucci}, {Perna}, {Belfiore}, {Carniani}, {Balmaverde}, {Brusa}, {Cicone},
  {Feruglio}, {Gallazzi}, {Mainieri}, {Maiolino}, {Nagao}, {Nardini}, {Sani},
  {Tozzi}, \& {Zibetti}}]{2019A&A...622A.146M}
{Mingozzi}, M., {Cresci}, G., {Venturi}, G., {et~al.} 2019, \aap, 622, A146

\bibitem[{{Mu{\~n}oz-Mateos} {et~al.}(2015){Mu{\~n}oz-Mateos}, {Sheth},
  {Regan}, {Kim}, {Laine}, {Erroz-Ferrer}, {Gil de Paz}, {Comeron}, {Hinz},
  {Laurikainen}, {Salo}, {Athanassoula}, {Bosma}, {Bouquin}, {Schinnerer},
  {Ho}, {Zaritsky}, {Gadotti}, {Madore}, {Holwerda}, {Men{\'e}ndez-Delmestre},
  {Knapen}, {Meidt}, {Querejeta}, {Mizusawa}, {Seibert}, {Laine}, \&
  {Courtois}}]{2015ApJS..219....3M}
{Mu{\~n}oz-Mateos}, J.~C., {Sheth}, K., {Regan}, M., {et~al.} 2015, \apjs, 219,
  3

\bibitem[{{Nair} \& {Abraham}(2010)}]{2010ApJ...714L.260N}
{Nair}, P.~B. \& {Abraham}, R.~G. 2010, \apjl, 714, L260

\bibitem[{{Neumann} {et~al.}(2020){Neumann}, {Fragkoudi}, {P{\'e}rez},
  {Gadotti}, {Falc{\'o}n-Barroso}, {S{\'a}nchez-Bl{\'a}zquez}, {Bittner},
  {Husemann}, {G{\'o}mez}, {Grand }, {Donohoe-Keyes}, {Kim}, {de
  Lorenzo-C{\'a}ceres}, {Martig}, {M{\'e}ndez-Abreu}, {Pakmor}, {Seidel}, \&
  {van de Ven}}]{2020A&A...637A..56N}
{Neumann}, J., {Fragkoudi}, F., {P{\'e}rez}, I., {et~al.} 2020, \aap, 637, A56

\bibitem[{{Neumann} {et~al.}(2019){Neumann}, {Gadotti}, {Wisotzki}, {Husemann},
  {Busch}, {Combes}, {Croom}, {Davis}, {Gaspari}, {Krumpe}, {P{\'e}rez-Torres},
  {Scharw{\"a}chter}, {Smirnova-Pinchukova}, {Tremblay}, \&
  {Urrutia}}]{2019A&A...627A..26N}
{Neumann}, J., {Gadotti}, D.~A., {Wisotzki}, L., {et~al.} 2019, \aap, 627, A26

\bibitem[{{Oh} {et~al.}(2012){Oh}, {Oh}, \& {Yi}}]{2012ApJS..198....4O}
{Oh}, S., {Oh}, K., \& {Yi}, S.~K. 2012, \apjs, 198, 4

\bibitem[{{Pan} {et~al.}(2015){Pan}, {Kuno}, {Koda}, {Hirota}, {Sorai}, \&
  {Kaneko}}]{2015ApJ...815...59P}
{Pan}, H.-A., {Kuno}, N., {Koda}, J., {et~al.} 2015, \apj, 815, 59

\bibitem[{{P{\'e}rez} {et~al.}(2004){P{\'e}rez}, {Fux}, \&
  {Freeman}}]{2004A&A...424..799P}
{P{\'e}rez}, I., {Fux}, R., \& {Freeman}, K. 2004, \aap, 424, 799

\bibitem[{{P{\'e}rez} {et~al.}(2017){P{\'e}rez}, {Mart{\'\i}nez-Valpuesta},
  {Ruiz-Lara}, {de Lorenzo-Caceres}, {Falc{\'o}n-Barroso}, {Florido},
  {Gonz{\'a}lez Delgado}, {Lyubenova}, {Marino}, {S{\'a}nchez},
  {S{\'a}nchez-Bl{\'a}zquez}, {van de Ven}, \& {Zurita}}]{2017MNRAS.470L.122P}
{P{\'e}rez}, I., {Mart{\'\i}nez-Valpuesta}, I., {Ruiz-Lara}, T., {et~al.} 2017,
  \mnras, 470, L122

\bibitem[{{P{\'e}rez} \&
  {S{\'a}nchez-Bl{\'a}zquez}(2011)}]{2011A&A...529A..64P}
{P{\'e}rez}, I. \& {S{\'a}nchez-Bl{\'a}zquez}, P. 2011, \aap, 529, A64

\bibitem[{{P{\'e}rez} {et~al.}(2007){P{\'e}rez}, {S{\'a}nchez-Bl{\'a}zquez}, \&
  {Zurita}}]{2007A&A...465L...9P}
{P{\'e}rez}, I., {S{\'a}nchez-Bl{\'a}zquez}, P., \& {Zurita}, A. 2007, \aap,
  465, L9

\bibitem[{{P{\'e}rez} {et~al.}(2009){P{\'e}rez}, {S{\'a}nchez-Bl{\'a}zquez}, \&
  {Zurita}}]{2009A&A...495..775P}
{P{\'e}rez}, I., {S{\'a}nchez-Bl{\'a}zquez}, P., \& {Zurita}, A. 2009, \aap,
  495, 775

\bibitem[{{Persic} \& {Salucci}(1991)}]{1991ApJ...368...60P}
{Persic}, M. \& {Salucci}, P. 1991, \apj, 368, 60

\bibitem[{{Petersen} {et~al.}(2016){Petersen}, {Weinberg}, \&
  {Katz}}]{2016MNRAS.463.1952P}
{Petersen}, M.~S., {Weinberg}, M.~D., \& {Katz}, N. 2016, \mnras, 463, 1952

\bibitem[{{Phillips}(1996)}]{1996ASPC...91...44P}
{Phillips}, A.~C. 1996, in Astronomical Society of the Pacific Conference
  Series, Vol.~91, IAU Colloq. 157: Barred Galaxies, ed. R.~{Buta}, D.~A.
  {Crocker}, \& B.~G. {Elmegreen}, 44

\bibitem[{{Popping} {et~al.}(2010){Popping}, {P{\'e}rez}, \&
  {Zurita}}]{2010A&A...521A...8P}
{Popping}, G., {P{\'e}rez}, I., \& {Zurita}, A. 2010, \aap, 521, A8

\bibitem[{{Puxley} {et~al.}(1988){Puxley}, {Hawarden}, \&
  {Mountain}}]{1988MNRAS.231..465P}
{Puxley}, P.~J., {Hawarden}, T.~G., \& {Mountain}, C.~M. 1988, \mnras, 231, 465

\bibitem[{{Querejeta} {et~al.}(2015){Querejeta}, {Meidt}, {Schinnerer},
  {Cisternas}, {Mu{\~n}oz-Mateos}, {Sheth}, {Knapen}, {van de Ven}, {Norris},
  {Peletier}, {Laurikainen}, {Salo}, {Holwerda}, {Athanassoula}, {Bosma},
  {Groves}, {Ho}, {Gadotti}, {Zaritsky}, {Regan}, {Hinz}, {Gil de Paz},
  {Menendez-Delmestre}, {Seibert}, {Mizusawa}, {Kim}, {Erroz-Ferrer}, {Laine},
  \& {Comer{\'o}n}}]{2015ApJS..219....5Q}
{Querejeta}, M., {Meidt}, S.~E., {Schinnerer}, E., {et~al.} 2015, \apjs, 219, 5

\bibitem[{{Rautiainen} \& {Salo}(2000)}]{2000A&A...362..465R}
{Rautiainen}, P. \& {Salo}, H. 2000, \aap, 362, 465

\bibitem[{{Regan} {et~al.}(1999){Regan}, {Sheth}, \&
  {Vogel}}]{1999ApJ...526...97R}
{Regan}, M.~W., {Sheth}, K., \& {Vogel}, S.~N. 1999, \apj, 526, 97

\bibitem[{{Regan} {et~al.}(2006){Regan}, {Thornley}, {Vogel}, {Sheth},
  {Draine}, {Hollenbach}, {Meyer}, {Dale}, {Engelbracht}, {Kennicutt}, {Armus},
  {Buckalew}, {Calzetti}, {Gordon}, {Helou}, {Leitherer}, {Malhotra}, {Murphy},
  {Rieke}, {Rieke}, \& {Smith}}]{2006ApJ...652.1112R}
{Regan}, M.~W., {Thornley}, M.~D., {Vogel}, S.~N., {et~al.} 2006, \apj, 652,
  1112

\bibitem[{{Renaud} {et~al.}(2015){Renaud}, {Bournaud}, {Emsellem}, {Agertz},
  {Athanassoula}, {Combes}, {Elmegreen}, {Kraljic}, {Motte}, \&
  {Teyssier}}]{2015MNRAS.454.3299R}
{Renaud}, F., {Bournaud}, F., {Emsellem}, E., {et~al.} 2015, \mnras, 454, 3299

\bibitem[{{Reynaud} \& {Downes}(1998)}]{1998A&A...337..671R}
{Reynaud}, D. \& {Downes}, D. 1998, \aap, 337, 671

\bibitem[{{Romano} {et~al.}(2008){Romano}, {Mayya}, \&
  {Vorobyov}}]{2008AJ....136.1259R}
{Romano}, R., {Mayya}, Y.~D., \& {Vorobyov}, E.~I. 2008, \aj, 136, 1259

\bibitem[{{Rosales-Ortega}(2011)}]{2011NewA...16..220R}
{Rosales-Ortega}, F.~F. 2011, \na, 16, 220

\bibitem[{{Ryder} \& {Dopita}(1993)}]{1993ApJS...88..415R}
{Ryder}, S.~D. \& {Dopita}, M.~A. 1993, \apjs, 88, 415

\bibitem[{{Sakamoto} {et~al.}(1999){Sakamoto}, {Okumura}, {Ishizuki}, \&
  {Scoville}}]{1999ApJ...525..691S}
{Sakamoto}, K., {Okumura}, S.~K., {Ishizuki}, S., \& {Scoville}, N.~Z. 1999,
  \apj, 525, 691

\bibitem[{{Salo} \& {Laurikainen}(2017)}]{2017ApJ...835..252S}
{Salo}, H. \& {Laurikainen}, E. 2017, \apj, 835, 252

\bibitem[{{Salo} {et~al.}(2010){Salo}, {Laurikainen}, {Buta}, \&
  {Knapen}}]{2010ApJ...715L..56S}
{Salo}, H., {Laurikainen}, E., {Buta}, R., \& {Knapen}, J.~H. 2010, \apjl, 715,
  L56

\bibitem[{{Salo} {et~al.}(2015){Salo}, {Laurikainen}, {Laine}, {Comer{\'o}n},
  {Gadotti}, {Buta}, {Sheth}, {Zaritsky}, {Ho}, {Knapen}, {Athanassoula},
  {Bosma}, {Laine}, {Cisternas}, {Kim}, {Mu{\~n}oz-Mateos}, {Regan}, {Hinz},
  {Gil de Paz}, {Menendez-Delmestre}, {Mizusawa}, {Erroz-Ferrer}, {Meidt}, \&
  {Querejeta}}]{2015ApJS..219....4S}
{Salo}, H., {Laurikainen}, E., {Laine}, J., {et~al.} 2015, \apjs, 219, 4

\bibitem[{{Salo} {et~al.}(1999){Salo}, {Rautiainen}, {Buta}, {Purcell}, {Cobb},
  {Crocker}, \& {Laurikainen}}]{1999AJ....117..792S}
{Salo}, H., {Rautiainen}, P., {Buta}, R., {et~al.} 1999, \aj, 117, 792

\bibitem[{{S{\'a}nchez} {et~al.}(2016){S{\'a}nchez}, {Garc{\'\i}a-Benito},
  {Zibetti}, {Walcher}, {Husemann}, {Mendoza}, {Galbany}, {Falc{\'o}n-Barroso},
  {Mast}, {Aceituno}, {Aguerri}, {Alves}, {Amorim}, {Ascasibar},
  {Barrado-Navascues}, {Barrera-Ballesteros}, {Bekerait{\`e}}, {Bland
  -Hawthorn}, {Cano D{\'\i}az}, {Cid Fernandes}, {Cavichia}, {Cortijo},
  {Dannerbauer}, {Demleitner}, {D{\'\i}az}, {Dettmar}, {de
  Lorenzo-C{\'a}ceres}, {del Olmo}, {Galazzi}, {Garc{\'\i}a-Lorenzo}, {Gil de
  Paz}, {Gonz{\'a}lez Delgado}, {Holmes}, {Igl{\'e}sias-P{\'a}ramo}, {Kehrig},
  {Kelz}, {Kennicutt}, {Kleemann}, {Lacerda}, {L{\'o}pez Fern{\'a}ndez},
  {L{\'o}pez S{\'a}nchez}, {Lyubenova}, {Marino}, {M{\'a}rquez},
  {Mendez-Abreu}, {Moll{\'a}}, {Monreal-Ibero}, {Ortega Minakata},
  {Torres-Papaqui}, {P{\'e}rez}, {Rosales-Ortega}, {Roth},
  {S{\'a}nchez-Bl{\'a}zquez}, {Schilling}, {Spekkens}, {Vale Asari}, {van den
  Bosch}, {van de Ven}, {Vilchez}, {Wild}, {Wisotzki}, {Y{\i}ld{\i}r{\i}m}, \&
  {Ziegler}}]{2016A&A...594A..36S}
{S{\'a}nchez}, S.~F., {Garc{\'\i}a-Benito}, R., {Zibetti}, S., {et~al.} 2016,
  \aap, 594, A36

\bibitem[{{S{\'a}nchez} {et~al.}(2012){S{\'a}nchez}, {Kennicutt}, {Gil de Paz},
  {van de Ven}, {V{\'\i}lchez}, {Wisotzki}, {Walcher}, {Mast}, {Aguerri},
  {Albiol-P{\'e}rez}, {Alonso-Herrero}, {Alves}, {Bakos}, {Bart{\'a}kov{\'a}},
  {Bland-Hawthorn}, {Boselli}, {Bomans}, {Castillo-Morales}, {Cortijo-Ferrero},
  {de Lorenzo-C{\'a}ceres}, {Del Olmo}, {Dettmar}, {D{\'\i}az}, {Ellis},
  {Falc{\'o}n-Barroso}, {Flores}, {Gallazzi}, {Garc{\'\i}a-Lorenzo},
  {Gonz{\'a}lez Delgado}, {Gruel}, {Haines}, {Hao}, {Husemann},
  {Igl{\'e}sias-P{\'a}ramo}, {Jahnke}, {Johnson}, {Jungwiert}, {Kalinova},
  {Kehrig}, {Kupko}, {L{\'o}pez-S{\'a}nchez}, {Lyubenova}, {Marino},
  {M{\'a}rmol-Queralt{\'o}}, {M{\'a}rquez}, {Masegosa}, {Meidt},
  {Mendez-Abreu}, {Monreal-Ibero}, {Montijo}, {Mour{\~a}o}, {Palacios-Navarro},
  {Papaderos}, {Pasquali}, {Peletier}, {P{\'e}rez}, {P{\'e}rez}, {Quirrenbach},
  {Rela{\~n}o}, {Rosales-Ortega}, {Roth}, {Ruiz-Lara},
  {S{\'a}nchez-Bl{\'a}zquez}, {Sengupta}, {Singh}, {Stanishev}, {Trager},
  {Vazdekis}, {Viironen}, {Wild}, {Zibetti}, \&
  {Ziegler}}]{2012A&A...538A...8S}
{S{\'a}nchez}, S.~F., {Kennicutt}, R.~C., {Gil de Paz}, A., {et~al.} 2012,
  \aap, 538, A8

\bibitem[{{S{\'a}nchez-Bl{\'a}zquez} {et~al.}(2011){S{\'a}nchez-Bl{\'a}zquez},
  {Ocvirk}, {Gibson}, {P{\'e}rez}, \& {Peletier}}]{2011MNRAS.415..709S}
{S{\'a}nchez-Bl{\'a}zquez}, P., {Ocvirk}, P., {Gibson}, B.~K., {P{\'e}rez}, I.,
  \& {Peletier}, R.~F. 2011, \mnras, 415, 709

\bibitem[{{S{\'a}nchez-Bl{\'a}zquez} {et~al.}(2014){S{\'a}nchez-Bl{\'a}zquez},
  {Rosales-Ortega}, {M{\'e}ndez-Abreu}, {P{\'e}rez}, {S{\'a}nchez}, {Zibetti},
  {Aguerri}, {Bland-Hawthorn}, {Catal{\'a}n-Torrecilla}, {Cid Fernandes}, {de
  Amorim}, {de Lorenzo-Caceres}, {Falc{\'o}n-Barroso}, {Galazzi}, {Garc{\'\i}a
  Benito}, {Gil de Paz}, {Gonz{\'a}lez Delgado}, {Husemann},
  {Iglesias-P{\'a}ramo}, {Jungwiert}, {Marino}, {M{\'a}rquez}, {Mast},
  {Mendoza}, {Moll{\'a}}, {Papaderos}, {Ruiz-Lara}, {van de Ven}, {Walcher}, \&
  {Wisotzki}}]{2014A&A...570A...6S}
{S{\'a}nchez-Bl{\'a}zquez}, P., {Rosales-Ortega}, F.~F., {M{\'e}ndez-Abreu},
  J., {et~al.} 2014, \aap, 570, A6

\bibitem[{{S{\'a}nchez-Janssen} \& {Gadotti}(2013)}]{2013MNRAS.432L..56S}
{S{\'a}nchez-Janssen}, R. \& {Gadotti}, D.~A. 2013, \mnras, 432, L56

\bibitem[{{Schwarz}(1984)}]{1984MNRAS.209...93S}
{Schwarz}, M.~P. 1984, \mnras, 209, 93

\bibitem[{{Seidel} {et~al.}(2015){Seidel}, {Falc{\'o}n-Barroso},
  {Mart{\'{\i}}nez-Valpuesta}, {D{\'{\i}}az-Garc{\'{\i}}a}, {Laurikainen},
  {Salo}, \& {Knapen}}]{2015MNRAS.451..936S}
{Seidel}, M.~K., {Falc{\'o}n-Barroso}, J., {Mart{\'{\i}}nez-Valpuesta}, I.,
  {et~al.} 2015, \mnras, 451, 936

\bibitem[{{Seidel} {et~al.}(2016){Seidel}, {Falc{\'o}n-Barroso},
  {Mart{\'\i}nez-Valpuesta}, {S{\'a}nchez-Bl{\'a}zquez}, {P{\'e}rez},
  {Peletier}, \& {Vazdekis}}]{2016MNRAS.460.3784S}
{Seidel}, M.~K., {Falc{\'o}n-Barroso}, J., {Mart{\'\i}nez-Valpuesta}, I.,
  {et~al.} 2016, \mnras, 460, 3784

\bibitem[{{Seigar}(2005)}]{2005MNRAS.361L..20S}
{Seigar}, M.~S. 2005, \mnras, 361, L20

\bibitem[{{Seigar} {et~al.}(2005){Seigar}, {Block}, {Puerari}, {Chorney}, \&
  {James}}]{2005MNRAS.359.1065S}
{Seigar}, M.~S., {Block}, D.~L., {Puerari}, I., {Chorney}, N.~E., \& {James},
  P.~A. 2005, \mnras, 359, 1065

\bibitem[{{Seigar} {et~al.}(2006){Seigar}, {Bullock}, {Barth}, \&
  {Ho}}]{2006ApJ...645.1012S}
{Seigar}, M.~S., {Bullock}, J.~S., {Barth}, A.~J., \& {Ho}, L.~C. 2006, \apj,
  645, 1012

\bibitem[{{Sellwood} \& {Wilkinson}(1993)}]{1993RPPh...56..173S}
{Sellwood}, J.~A. \& {Wilkinson}, A. 1993, Reports on Progress in Physics, 56,
  173

\bibitem[{{Shen} \& {Sellwood}(2004)}]{2004ApJ...604..614S}
{Shen}, J. \& {Sellwood}, J.~A. 2004, \apj, 604, 614

\bibitem[{{Sheth} {et~al.}(2013){Sheth}, {Armus}, {Athanassoula}, {Bosma},
  {Gadotti}, {Munoz-Mateos}, {Hinz}, {Regan}, {Laurikainen}, {Jarrett},
  {Zaritsky}, {Menendez-Delmestre}, {Madore}, {Elmegreen}, {Knapen}, {Salo},
  {Schinnerer}, {Kim}, {Ho}, {Elmegreen}, {Buta}, {Cisternas}, {Laine},
  {Comeron}, {Donovan Meyer}, {D'Onghia}, \& {Salim}}]{2013sptz.prop10043S}
{Sheth}, K., {Armus}, L., {Athanassoula}, E., {et~al.} 2013, {Not Dead Yet!
  Completing Spitzer's Legacy with Early Type Galaxies}, Spitzer Proposal

\bibitem[{{Sheth} {et~al.}(2008){Sheth}, {Elmegreen}, {Elmegreen}, {Capak},
  {Abraham}, {Athanassoula}, {Ellis}, {Mobasher}, {Salvato}, {Schinnerer},
  {Scoville}, {Spalsbury}, {Strubbe}, {Carollo}, {Rich}, \&
  {West}}]{2008ApJ...675.1141S}
{Sheth}, K., {Elmegreen}, D.~M., {Elmegreen}, B.~G., {et~al.} 2008, \apj, 675,
  1141

\bibitem[{{Sheth} {et~al.}(2010){Sheth}, {Regan}, {Hinz}, {Gil de Paz},
  {Men{\'e}ndez-Delmestre}, {Mu{\~n}oz-Mateos}, {Seibert}, {Kim},
  {Laurikainen}, {Salo}, {Gadotti}, {Laine}, {Mizusawa}, {Armus},
  {Athanassoula}, {Bosma}, {Buta}, {Capak}, {Jarrett}, {Elmegreen},
  {Elmegreen}, {Knapen}, {Koda}, {Helou}, {Ho}, {Madore}, {Masters},
  {Mobasher}, {Ogle}, {Peng}, {Schinnerer}, {Surace}, {Zaritsky},
  {Comer{\'o}n}, {de Swardt}, {Meidt}, {Kasliwal}, \&
  {Aravena}}]{2010PASP..122.1397S}
{Sheth}, K., {Regan}, M., {Hinz}, J.~L., {et~al.} 2010, \pasp, 122, 1397

\bibitem[{{Sheth} \& {S$^4$G Team}(2014)}]{2014AAS...22320502S}
{Sheth}, K. \& {S$^4$G Team}. 2014, in American Astronomical Society Meeting
  Abstracts, Vol. 223, American Astronomical Society Meeting Abstracts \#223,
  205.02

\bibitem[{{Sheth} {et~al.}(2002){Sheth}, {Vogel}, {Regan}, {Teuben}, {Harris},
  \& {Thornley}}]{2002AJ....124.2581S}
{Sheth}, K., {Vogel}, S.~N., {Regan}, M.~W., {et~al.} 2002, \aj, 124, 2581

\bibitem[{{Sheth} {et~al.}(2005){Sheth}, {Vogel}, {Regan}, {Thornley}, \&
  {Teuben}}]{2005ApJ...632..217S}
{Sheth}, K., {Vogel}, S.~N., {Regan}, M.~W., {Thornley}, M.~D., \& {Teuben},
  P.~J. 2005, \apj, 632, 217

\bibitem[{{Shlosman} {et~al.}(1989){Shlosman}, {Frank}, \&
  {Begelman}}]{1989Natur.338...45S}
{Shlosman}, I., {Frank}, J., \& {Begelman}, M.~C. 1989, \nat, 338, 45

\bibitem[{{Simkin} {et~al.}(1980){Simkin}, {Su}, \&
  {Schwarz}}]{1980ApJ...237..404S}
{Simkin}, S.~M., {Su}, H.~J., \& {Schwarz}, M.~P. 1980, \apj, 237, 404

\bibitem[{{Toomre}(1964)}]{1964ApJ...139.1217T}
{Toomre}, A. 1964, \apj, 139, 1217

\bibitem[{{Torres-Flores} {et~al.}(2014){Torres-Flores}, {Amram}, {Mendes de
  Oliveira}, {Plana}, {Balkowski}, {Marcelin}, \&
  {Olave-Rojas}}]{2014MNRAS.442.2188T}
{Torres-Flores}, S., {Amram}, P., {Mendes de Oliveira}, C., {et~al.} 2014,
  \mnras, 442, 2188

\bibitem[{{van Zee}(2000)}]{2000AJ....119.2757V}
{van Zee}, L. 2000, \aj, 119, 2757

\bibitem[{{Vanhala} \& {Cameron}(1998)}]{1998ApJ...508..291V}
{Vanhala}, H. A.~T. \& {Cameron}, A.~G.~W. 1998, \apj, 508, 291

\bibitem[{{Vera} {et~al.}(2016){Vera}, {Alonso}, \&
  {Coldwell}}]{2016A&A...595A..63V}
{Vera}, M., {Alonso}, S., \& {Coldwell}, G. 2016, \aap, 595, A63

\bibitem[{{Verley} {et~al.}(2007{\natexlab{a}}){Verley}, {Combes},
  {Verdes-Montenegro}, {Bergond}, \& {Leon}}]{2007A&A...474...43V}
{Verley}, S., {Combes}, F., {Verdes-Montenegro}, L., {Bergond}, G., \& {Leon},
  S. 2007{\natexlab{a}}, \aap, 474, 43

\bibitem[{{Verley} {et~al.}(2007{\natexlab{b}}){Verley}, {Leon},
  {Verdes-Montenegro}, {Combes}, {Sabater}, {Sulentic}, {Bergond}, {Espada},
  {Garc{\'{\i}}a}, {Lisenfeld}, \& {Odewahn}}]{2007A&A...472..121V}
{Verley}, S., {Leon}, S., {Verdes-Montenegro}, L., {et~al.} 2007{\natexlab{b}},
  \aap, 472, 121

\bibitem[{{V{\'e}ron-Cetty} \& {V{\'e}ron}(2010)}]{2010A&A...518A..10V}
{V{\'e}ron-Cetty}, M.-P. \& {V{\'e}ron}, P. 2010, \aap, 518, A10

\bibitem[{{Villa-Vargas} {et~al.}(2010){Villa-Vargas}, {Shlosman}, \&
  {Heller}}]{2010ApJ...719.1470V}
{Villa-Vargas}, J., {Shlosman}, I., \& {Heller}, C. 2010, \apj, 719, 1470

\bibitem[{{Wada}(2004)}]{2004cbhg.symp..186W}
{Wada}, K. 2004, Coevolution of Black Holes and Galaxies, 186

\bibitem[{{Wada} \& {Habe}(1992)}]{1992MNRAS.258...82W}
{Wada}, K. \& {Habe}, A. 1992, \mnras, 258, 82

\bibitem[{{Walcher} {et~al.}(2014){Walcher}, {Wisotzki}, {Bekerait{\'e}},
  {Husemann}, {Iglesias-P{\'a}ramo}, {Backsmann}, {Barrera Ballesteros},
  {Catal{\'a}n-Torrecilla}, {Cortijo}, {del Olmo}, {Garcia Lorenzo},
  {Falc{\'o}n-Barroso}, {Jilkova}, {Kalinova}, {Mast}, {Marino},
  {M{\'e}ndez-Abreu}, {Pasquali}, {S{\'a}nchez}, {Trager}, {Zibetti},
  {Aguerri}, {Alves}, {Bland -Hawthorn}, {Boselli}, {Castillo Morales}, {Cid
  Fernandes}, {Flores}, {Galbany}, {Gallazzi}, {Garc{\'\i}a-Benito}, {Gil de
  Paz}, {Gonz{\'a}lez-Delgado}, {Jahnke}, {Jungwiert}, {Kehrig}, {Lyubenova},
  {M{\'a}rquez Perez}, {Masegosa}, {Monreal Ibero}, {P{\'e}rez}, {Quirrenbach},
  {Rosales-Ortega}, {Roth}, {Sanchez-Blazquez}, {Spekkens}, {Tundo}, {van de
  Ven}, {Verheijen}, {Vilchez}, \& {Ziegler}}]{2014A&A...569A...1W}
{Walcher}, C.~J., {Wisotzki}, L., {Bekerait{\'e}}, S., {et~al.} 2014, \aap,
  569, A1

\bibitem[{{Wang} {et~al.}(2020){Wang}, {Athanassoula}, {Yu}, {Wolf}, {Shao},
  {Gao}, \& {Randriamampand ry}}]{2020ApJ...893...19W}
{Wang}, J., {Athanassoula}, E., {Yu}, S.-Y., {et~al.} 2020, \apj, 893, 19

\bibitem[{{Wang} {et~al.}(2012){Wang}, {Kauffmann}, {Overzier}, {Tacconi},
  {Kong}, {Saintonge}, {Catinella}, {Schiminovich}, {Moran}, \&
  {Johnson}}]{2012MNRAS.423.3486W}
{Wang}, J., {Kauffmann}, G., {Overzier}, R., {et~al.} 2012, \mnras, 423, 3486

\bibitem[{{Weilbacher} {et~al.}(2012){Weilbacher}, {Streicher}, {Urrutia},
  {Jarno}, {P{\'e}contal-Rousset}, {Bacon}, \&
  {B{\"o}hm}}]{2012SPIE.8451E..0BW}
{Weilbacher}, P.~M., {Streicher}, O., {Urrutia}, T., {et~al.} 2012, in Society
  of Photo-Optical Instrumentation Engineers (SPIE) Conference Series, Vol.
  8451, \procspie, 84510B

\bibitem[{{Weilbacher} {et~al.}(2014){Weilbacher}, {Streicher}, {Urrutia},
  {P{\'e}contal-Rousset}, {Jarno}, \& {Bacon}}]{2014ASPC..485..451W}
{Weilbacher}, P.~M., {Streicher}, O., {Urrutia}, T., {et~al.} 2014, in
  Astronomical Society of the Pacific Conference Series, Vol. 485, Astronomical
  Data Analysis Software and Systems XXIII, ed. N.~{Manset} \& P.~{Forshay},
  451

\bibitem[{{Whyte} {et~al.}(2002){Whyte}, {Abraham}, {Merrifield}, {Eskridge},
  {Frogel}, \& {Pogge}}]{2002MNRAS.336.1281W}
{Whyte}, L.~F., {Abraham}, R.~G., {Merrifield}, M.~R., {et~al.} 2002, \mnras,
  336, 1281

\bibitem[{{Willett} {et~al.}(2013){Willett}, {Lintott}, {Bamford}, {Masters},
  {Simmons}, {Casteels}, {Edmondson}, {Fortson}, {Kaviraj}, {Keel}, {Melvin},
  {Nichol}, {Raddick}, {Schawinski}, {Simpson}, {Skibba}, {Smith}, \&
  {Thomas}}]{2013MNRAS.435.2835W}
{Willett}, K.~W., {Lintott}, C.~J., {Bamford}, S.~P., {et~al.} 2013, \mnras,
  435, 2835

\bibitem[{{Yi} {et~al.}(2011){Yi}, {Lee}, {Sheen}, {Jeong}, {Suh}, \&
  {Oh}}]{2011ApJS..195...22Y}
{Yi}, S.~K., {Lee}, J., {Sheen}, Y.-K., {et~al.} 2011, \apjs, 195, 22

\bibitem[{{Yi} {et~al.}(2005){Yi}, {Yoon}, {Kaviraj}, {Deharveng}, {Rich},
  {Salim}, {Boselli}, {Lee}, {Ree}, {Sohn}, {Rey}, {Lee}, {Rhee}, {Bianchi},
  {Byun}, {Donas}, {Friedman}, {Heckman}, {Jelinsky}, {Madore}, {Malina},
  {Martin}, {Milliard}, {Morrissey}, {Neff}, {Schiminovich}, {Siegmund},
  {Small}, {Szalay}, {Jee}, {Kim}, {Barlow}, {Forster}, {Welsh}, \&
  {Wyder}}]{2005ApJ...619L.111Y}
{Yi}, S.~K., {Yoon}, S.~J., {Kaviraj}, S., {et~al.} 2005, \apjl, 619, L111

\bibitem[{{Young} {et~al.}(1996){Young}, {Allen}, {Kenney}, {Lesser}, \&
  {Rownd}}]{1996AJ....112.1903Y}
{Young}, J.~S., {Allen}, L., {Kenney}, J. D.~P., {Lesser}, A., \& {Rownd}, B.
  1996, \aj, 112, 1903

\bibitem[{{Zurita} {et~al.}(2020{\natexlab{a}}){Zurita}, {Florido}, {Bresolin},
  {P{\'e}rez}, \& {P{\'e}rez-Montero}}]{2020MNRAS.tmp.2303Z}
{Zurita}, A., {Florido}, E., {Bresolin}, F., {P{\'e}rez}, I., \&
  {P{\'e}rez-Montero}, E. 2020{\natexlab{a}}, \mnras
  [\eprint[arXiv]{2007.12292}]

\bibitem[{{Zurita} {et~al.}(2020{\natexlab{b}}){Zurita}, {Florido}, {Bresolin},
  {P{\'e}rez-Montero}, \& {P{\'e}rez}}]{2020arXiv200712289Z}
{Zurita}, A., {Florido}, E., {Bresolin}, F., {P{\'e}rez-Montero}, E., \&
  {P{\'e}rez}, I. 2020{\natexlab{b}}, arXiv e-prints, arXiv:2007.12289

\bibitem[{{Zurita} \& {P{\'e}rez}(2008)}]{2008A&A...485....5Z}
{Zurita}, A. \& {P{\'e}rez}, I. 2008, \aap, 485, 5

\bibitem[{{Zurita} {et~al.}(2004){Zurita}, {Rela{\~n}o}, {Beckman}, \&
  {Knapen}}]{2004A&A...413...73Z}
{Zurita}, A., {Rela{\~n}o}, M., {Beckman}, J.~E., \& {Knapen}, J.~H. 2004,
  \aap, 413, 73

\bibitem[{{Zurita} {et~al.}(2001){Zurita}, {Rozas}, \&
  {Beckman}}]{2001Ap&SS.276..491Z}
{Zurita}, A., {Rozas}, M., \& {Beckman}, J.~E. 2001, \apss, 276, 491

\end{thebibliography}
%
%
\begin{appendix}
%
%
%
\clearpage
\onecolumn
%
%
\section{Tabulated SF classifications and sources of H$\alpha$ imaging}\label{SF_class_sources}
%
%
The following data are listed in Table~\ref{table_SF_class_sources} for all the galaxies in our sample:
%
%
\begin{itemize}
\item Column 1: Galaxy identification.
\item Column 2: Total stellar mass ($M_{\ast}$) from \citet[][]{2015ApJS..219....3M}.
\item Column 3: Atomic H{\sc\,i} gas content ($M_{\rm HI}$) from HyperLEDA (see Eq.~\ref{gasfrac}).
\item Column 4: Revised Hubble stage ($T$) from \citet[][]{2015ApJS..217...32B}.
\item Column 5: Bar torque parameter ($Q_{\rm b}$) from \citet[][]{2016A&A...587A.160D}.
\item Column 6: Bar SF class (A, B, C, N, U) -- and subclasses a/b -- as described in Table~\ref{sf_system}, based on GALEX FUV imaging.
\item Column 7: Determination of active (rA) and passive (rP) inner rings (including uncertain cases rU), 
as described in Sect.~\ref{class_met}, using GALEX FUV imaging.
\item Column 8: As in column 6 (bar SF class), but determined from continuum-subtracted H$\alpha$ images.
\item Column 9: As in column 7 (inner rings SF activity), but determined from continuum-subtracted H$\alpha$ images.
\item Column 10: Flagging of GALEX UV images from All-Sky Imaging Survey (AIS) (Yes/No).
\item Column 11: Literature sources of H$\alpha$ images. These are taken from 
(1) \citet[][]{1993ApJS...88..415R}, 
(2) \citet[][]{1996RMxAA..32...89G}, 
(3) \citet[][]{1996AJ....112.1903Y}, 
(4) \citet[][]{1999A&A...345...59L}, 
(5) \citet[][]{1999AJ....118..730H}, 
(6) \citet[][]{2000AJ....119.2757V}, 
(7) \citet[][]{2001ApJS..135..125K}, 
(8) \citet[][]{2001ApJ...559..878H}, 
(9) \citet[][]{2003A&A...400..451G}, 
(10) \citet[][]{2003ApJS..147...29G}, 
(11) \citet[][]{2003PASP..115..928K},
(12) \citet[][]{2004A&A...414...23J}, 
(13) \citet[][]{2004A&A...426.1135K}, 
(14) \citet[][]{2004AJ....128.2170H}, 
(15) \citet[][]{2006ApJS..162...97K}, 
(16) \citet[][]{2006ApJS..165..307M}, 
(17) \citet[][]{2006Ap.....49..287K}, 
(18) \citet[][]{2008MNRAS.388..500E}, 
(19) \citet[][]{2008AJ....136.1259R}, 
(20) \citet[][]{2009ApJ...703..517D},
(21) \citet[][]{2012A&A...538A...8S}, 
(22) \citet[][]{2014arXiv1401.8123G}, 
(23) \citet[][]{2014MNRAS.442.2188T}, 
(24) \citet[][]{2014A&A...569A...1W}, 
(25) \citet[][]{2015A&A...579A.102B}, 
(26) \citet[][]{2015Ap.....58..453K}, 
(27) \citet[][]{2016A&A...594A..36S}, 
(28) \citet[][]{2018ApJ...855..107G}, 
(29) \citet[][]{2018A&A...611A..28G}, 
(30) \citet[][]{2019A&A...622A.176C}, 
(31) \citet[][]{2019MNRAS.484.5009E}, 
(32) \citet[][]{2019MNRAS.489..241M}, 
(33) NASA/IPAC Extragalactic Database (NED) linking images from Palomar/Las Campanas Atlas of 
Nearby Galaxies \citep[https://ha-atlas.obs.carnegiescience.edu/, see][]{2003ApJS..147...29G}, 
(34) ESO archive, based on observations made at the European Southern Observatory using the Very Large Telescope under programs:
\begin{itemize}
\item (34.1) \href{http://archive.eso.org/wdb/wdb/eso/sched_rep_arc/query?progid=60.A-9313(A)}{60.A-9313(A)}, \citet[][]{2015A&A...584A..90G}.
\item (34.2) \href{http://archive.eso.org/wdb/wdb/eso/sched_rep_arc/query?progid=096.D-0263(A)}{096.D-0263(A)}, \citet[][]{2018MNRAS.473.1359L}.
\item (34.3) \href{http://archive.eso.org/wdb/wdb/eso/sched_rep_arc/query?progid=095.D-0172(A)}{095.D-0172(A)}, \citet[][]{2018A&A...613A..35K}.
\item (34.4) \href{http://archive.eso.org/wdb/wdb/eso/sched_rep_arc/query?progid=296.B-5054(A)}{296.B-5054(A)}, \citet[][]{2019A&A...627A.136I}.
\item (34.5) \href{http://archive.eso.org/wdb/wdb/eso/sched_rep_arc/query?progid=60.A-9319(A)}{60.A-9319(A)}\footnote{\label{FOOT1}The authors of 
the proposal have published no paper.}.
\item (34.6) \href{http://archive.eso.org/wdb/wdb/eso/sched_rep_arc/query?progid=097.B-0640(A)}{097.B-0640(A)}, \citet[][]{2019MNRAS.482..506G}.
\item (34.7) \href{http://archive.eso.org/wdb/wdb/eso/sched_rep_arc/query?progid=098.A-0364(A)}{098.A-0364(A)}\footref{FOOT1}.
\item (34.8) \href{http://archive.eso.org/wdb/wdb/eso/sched_rep_arc/query?progid=1100.B-0651(A)}{1100.B-0651(A)}, \citet[][]{2019ApJ...887...80K}.
\item (34.9) \href{http://archive.eso.org/wdb/wdb/eso/sched_rep_arc/query?progid=1100.B-0651(B)}{1100.B-0651(B)}, \citet[][]{2019ApJ...887...80K}.
\item (34.10) \href{http://archive.eso.org/wdb/wdb/eso/sched_rep_arc/query?progid=1100.B-0651(C)}{1100.B-0651(C)}, \citet[][]{2019ApJ...887...80K}.
\item (34.11) \href{http://archive.eso.org/wdb/wdb/eso/sched_rep_arc/query?progid=094.B-0321(A)}{094.B-0321(A)}, \citet[][]{2019A&A...622A.146M}.
\item (34.12) \href{http://archive.eso.org/wdb/wdb/eso/sched_rep_arc/query?progid=0100.A-0607(A)}{0100.A-0607(A)}, \citet[][]{2019A&A...631A.114B}.
\item (34.13) \href{http://archive.eso.org/wdb/wdb/eso/sched_rep_arc/query?progid=0101.A-0282(A)}{0101.A-0282(A)}, \citet[][]{2019A&A...631A.114B}.
\item (34.14) \href{http://archive.eso.org/wdb/wdb/eso/sched_rep_arc/query?progid=60.A-9100(A)}{60.A-9100(A)}\footnote{MUSE commissioning.}.
\item (34.15) \href{http://archive.eso.org/wdb/wdb/eso/sched_rep_arc/query?progid=0101.D-0748(A)}{0101.D-0748(A)}, \citet[][]{2020AJ....159..167L}.
\item (34.16) \href{http://archive.eso.org/wdb/wdb/eso/sched_rep_arc/query?progid=097.D-0408(A)}{097.D-0408(A)}, \citet[][]{2020AJ....159..167L}.
\item (34.17) \href{http://archive.eso.org/wdb/wdb/eso/sched_rep_arc/query?progid=0104.B-0668(A)}{0104.B-0668(A)}\footref{FOOT1}.
\item (34.18) \href{http://archive.eso.org/wdb/wdb/eso/sched_rep_arc/query?progid=0103.D-0440(A)}{0103.D-0440(A)}, \citet[][]{2020MNRAS.495..992L}.
\item (34.19) \href{http://archive.eso.org/wdb/wdb/eso/sched_rep_arc/query?progid=096.B-0449(A)}{096.B-0449(A)}\footref{FOOT1}.
\item (34.20) \href{http://archive.eso.org/wdb/wdb/eso/sched_rep_arc/query?progid=0103.B-0582(A)}{0103.B-0582(A)}\footref{FOOT1}.
\end{itemize}
\end{itemize}
%
\onecolumn
\begin{tiny}
\begin{longtable}{| c| c| c| c| c| c| c| c| c| c| c|}
\hline
\multicolumn{1}{|c}{\centering Galaxy} & \multicolumn{1}{|c}{\centering log$_{10}$ $M_{\ast}$} & \multicolumn{1}{|c}{\centering log$_{10}$ $M_{\rm HI}$} & \multicolumn{1}{|c}{\centering $T$} &
\multicolumn{1}{|c}{\centering $Q_{\rm b}$} & \multicolumn{1}{|c}{\centering SF class (FUV)} & \multicolumn{1}{|c}{\centering Ring SF (FUV)} &
\multicolumn{1}{|c}{\centering SF class (H$\alpha$)} & \multicolumn{1}{|c|}{\centering Ring SF (H$\alpha$)} & \multicolumn{1}{|c|}{\centering AIS (FUV)} & \multicolumn{1}{|c|}{\centering Source H$\alpha$} \tabularnewline
\multicolumn{1}{|c}{\centering } & \multicolumn{1}{|c}{\centering ($M_{\odot}$)} & \multicolumn{1}{|c}{\centering ($M_{\odot}$)} & \multicolumn{1}{|c}{\centering } &
\multicolumn{1}{|c}{\centering } & \multicolumn{1}{|c}{\centering (A,B,C,N,U)} & \multicolumn{1}{|c}{\centering (rA,rP)} &
\multicolumn{1}{|c}{\centering (A,B,C,N,U)} & \multicolumn{1}{|c|}{\centering (rA,rP)} & \multicolumn{1}{|c|}{\centering (Yes/No)} & \multicolumn{1}{|c|}{\centering } \tabularnewline
\multicolumn{1}{|c}{\centering } & \multicolumn{1}{|c}{\centering } & \multicolumn{1}{|c}{\centering } & \multicolumn{1}{|c}{\centering } &
\multicolumn{1}{|c}{\centering } & \multicolumn{1}{|c}{\centering +a/b} & \multicolumn{1}{|c}{\centering } &
\multicolumn{1}{|c}{\centering +a/b} & \multicolumn{1}{|c|}{\centering } & \multicolumn{1}{|c|}{\centering } & \multicolumn{1}{|c|}{\centering } \tabularnewline
\hline
\hline
ESO013-016 & 9.56 & 9.11 & 6 & 0.52 & Ca & rA & - & - & Y & - \\
ESO026-001 & 9.37 & 8.88 & 5 & 0.58 & A & - & - & - & Y & - \\
ESO027-001 & 10.00 & 9.40 & 3 & 0.25 & A & - & - & - & Y & - \\
ESO048-017 & 9.20 & 9.11 & 9 & - & Ca & - & - & - & Y & - \\
ESO079-005 & 9.30 & 9.46 & 9 & - & Cb & - & - & - & Y & - \\
\end{longtable}
\end{tiny}

\twocolumn
%
%
\clearpage
\onecolumn
%
%
%
\section{Statistical trends are unaffected by the presence of AGN or the depth of the UV imaging}\label{app_AIS_Agn}
%
%
%
Here we test possible dependences of the classifications performed in Sect~\ref{class_met} on the depth of the FUV imaging. 
We also check whether the presence of AGN, which are known to be responsible for photoionization in the central regions of galaxies, 
can be responsible for the nuclear FUV and H$\alpha$ emission (class A) and thus affect our statistics.

\begin{figure*}[!b]
\centering
\includegraphics[width=0.49\textwidth]{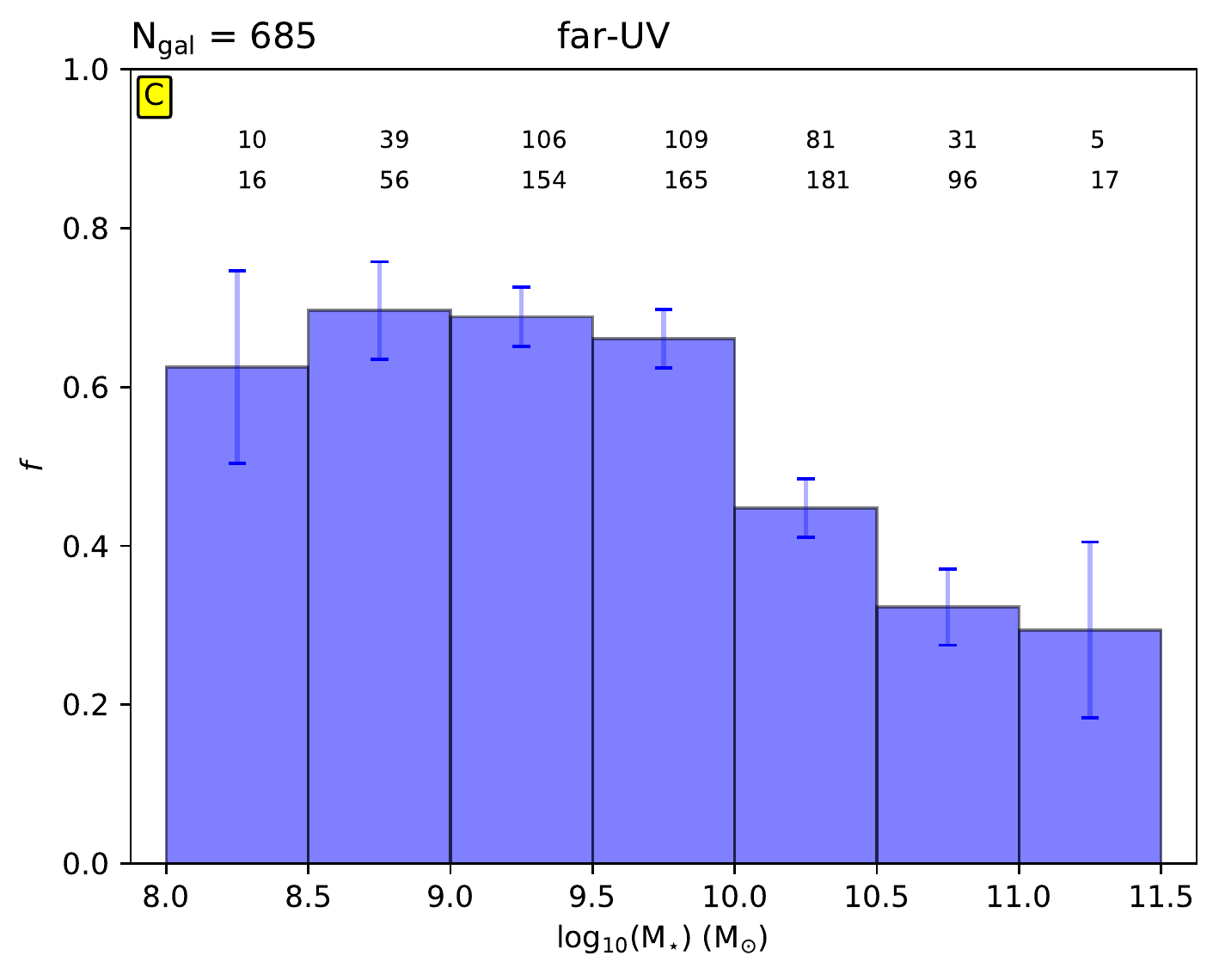}
\includegraphics[width=0.49\textwidth]{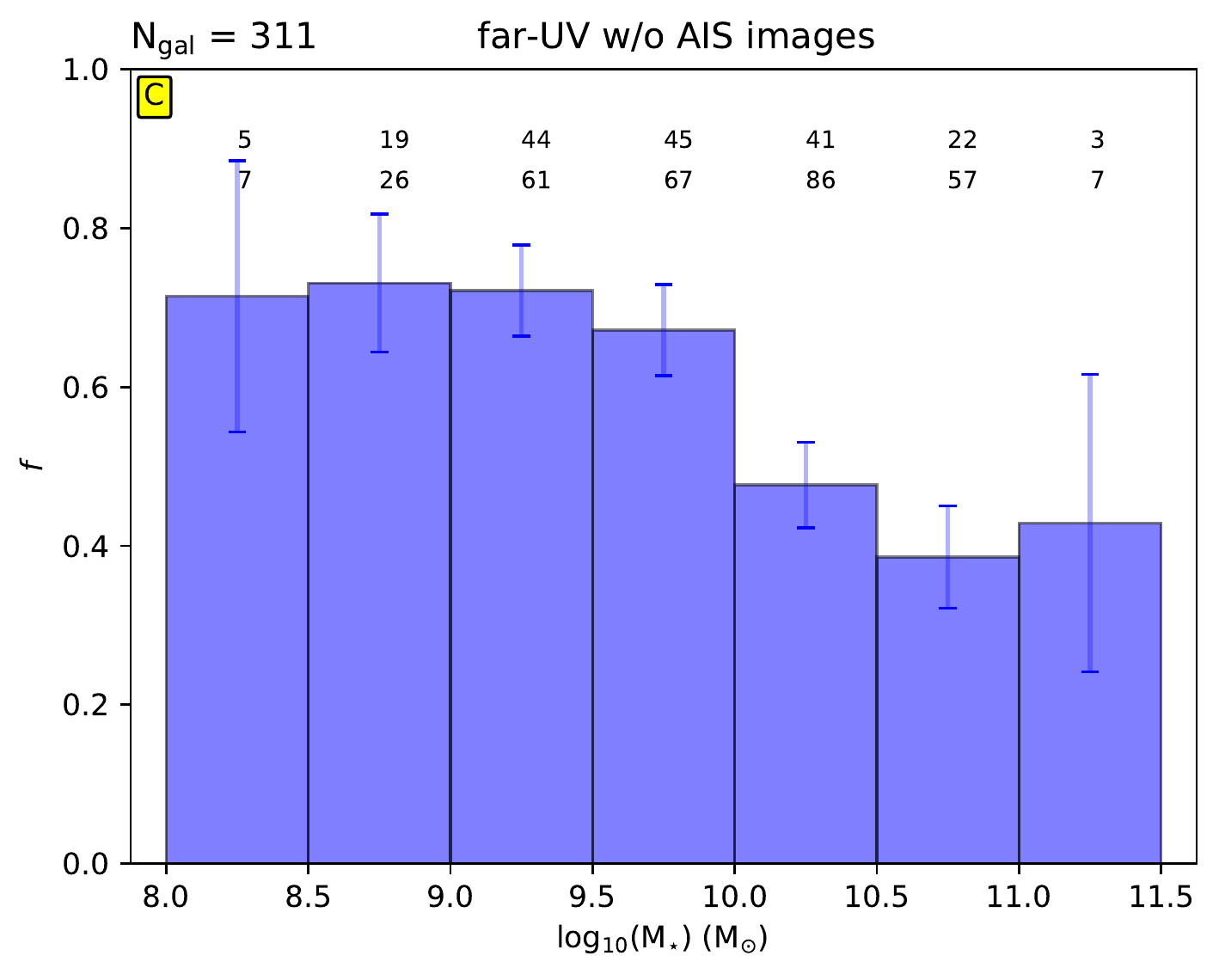}
\caption{
Fraction of galaxies classified as C (SF along bar), as seen in FUV images, as a function of the decimal logarithm of the total stellar mass. 
 \emph{Left panel}: Sample that includes images from the GALEX All-Sky Imaging Survey (AIS, with fairly short exposure times); 
 \emph{Right panel}: Only  galaxies from the FUV sample with exposure times of 1000 seconds or more (i.e., excluding AIS). 
The numbers in each of the mass bins are shown above the bars for each $M_{\ast}$-bin (of width 0.5 dex).
}
\label{AIS-NOAIS}
\end{figure*}

A number of galaxies in the FUV sample belong to the GALEX All-Sky Imaging Survey (AIS); they  had low exposure times 
(on the order of 100 seconds),  and thus the detection of SF along the bar (category C) can be compromised by the depth of the images. 
In Fig.~\ref{AIS-NOAIS} we show the frequency of galaxies of class C versus the total stellar mass of the host galaxy, including and excluding the 
AIS images (the latter decreases the sample size by almost $50$\%). We confirm that the statistical trends presented in 
Fig.~\ref{MSTAR} are not affected by the depth of the FUV images: 
as shown in Sect.~\ref{mstar_SF}, the less massive the galaxy is, the more common class C is. 

\begin{figure*}[!b]
\includegraphics[width=0.49\textwidth]{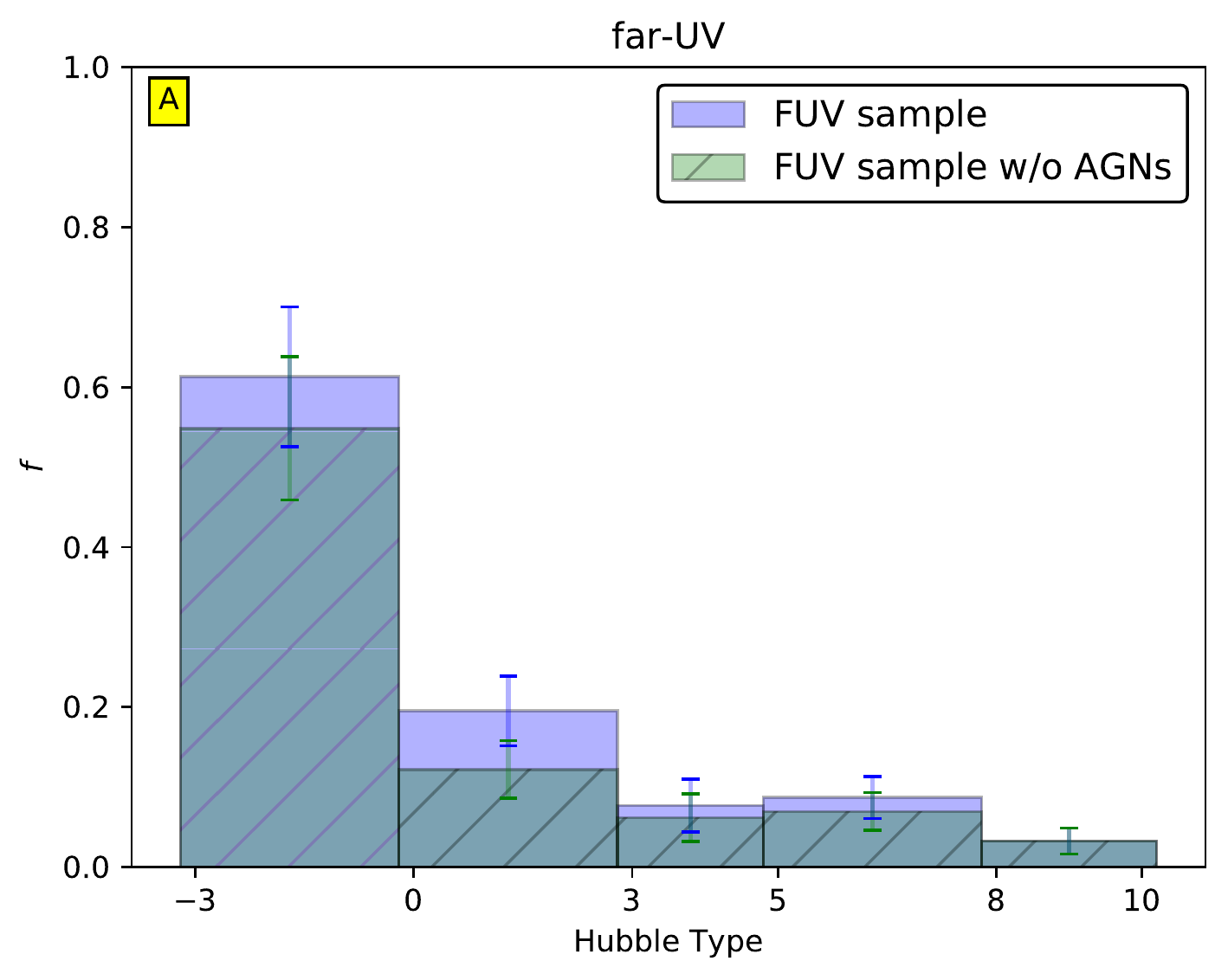}
\includegraphics[width=0.49\textwidth]{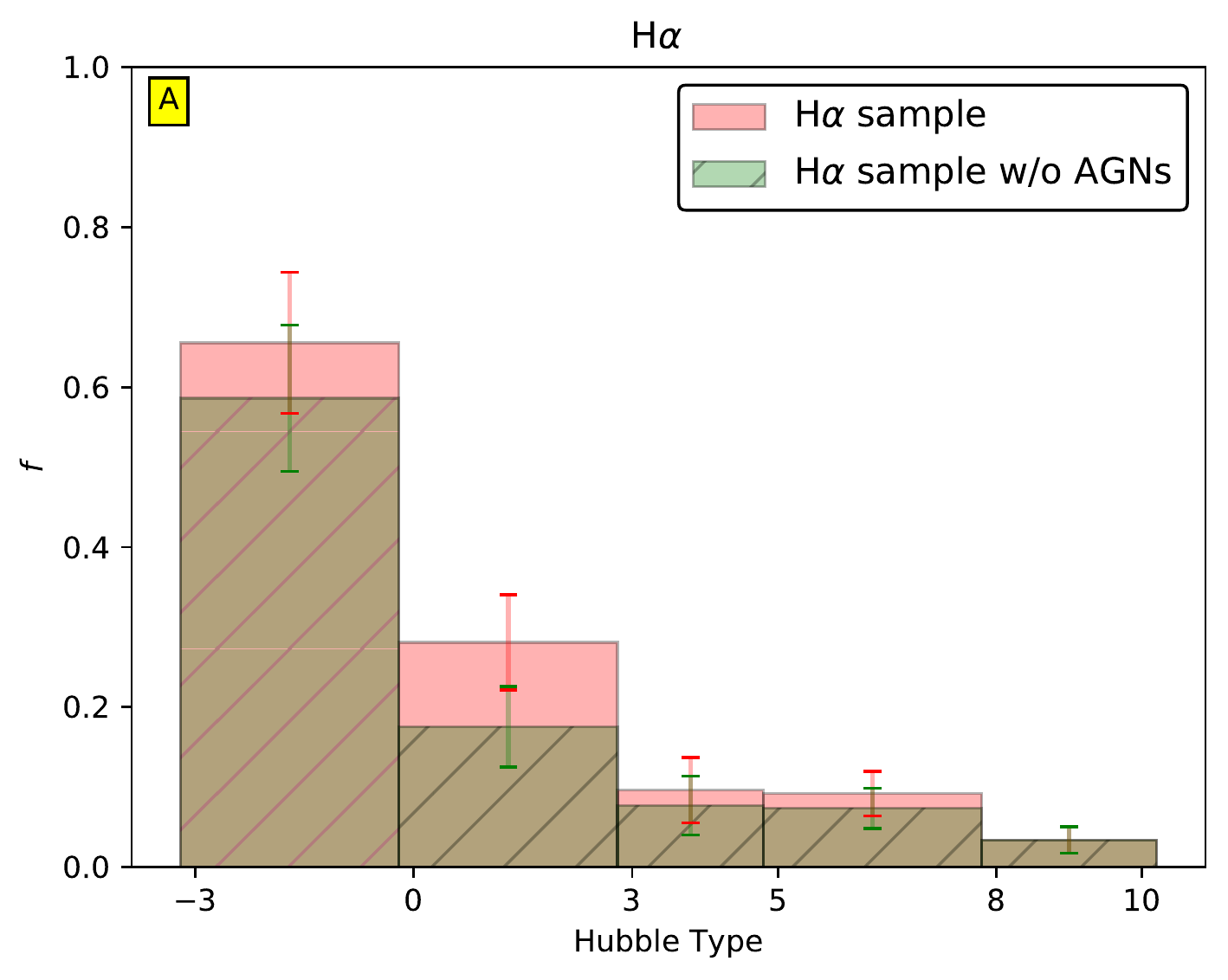}
\caption{Fraction of galaxies of SF class A (star formation only in the central regions) 
as a function of the revised Hubble stage for the FUV sample (\emph{left panel}) and for the H$\alpha$ sample (\emph{right panel}), 
including and excluding those galaxies that have an AGN according to \citet[][]{2010A&A...518A..10V}.}
\label{AGN-NOAGN}
\end{figure*}

The second possible shortcoming is that the H$\alpha$ and FUV emission detected in 
the center of some galaxies may be due to AGN photoionization of surrounding gas, and not to SF. 
This is relevant for the assignment of class A in our classification system. 
In Fig. \ref{AGN-NOAGN} we show the fraction of class A galaxies with and without AGN \citep[according to][]{2010A&A...518A..10V} 
as a function of the Hubble type, obtaining similar statistical trends (differences are smaller than the binomial counting error bars per $T$-bin). 
In conclusion, we  verified that AGN are not a major source of uncertainty in our statistical analysis. 
%
%
%
%
%
%
%
%
\clearpage
\onecolumn
%
%
\section{Fraction of active inner rings}\label{inner_rings_appendix}
\begin{figure*}[!h]
\centering
\includegraphics[width=0.49\textwidth]{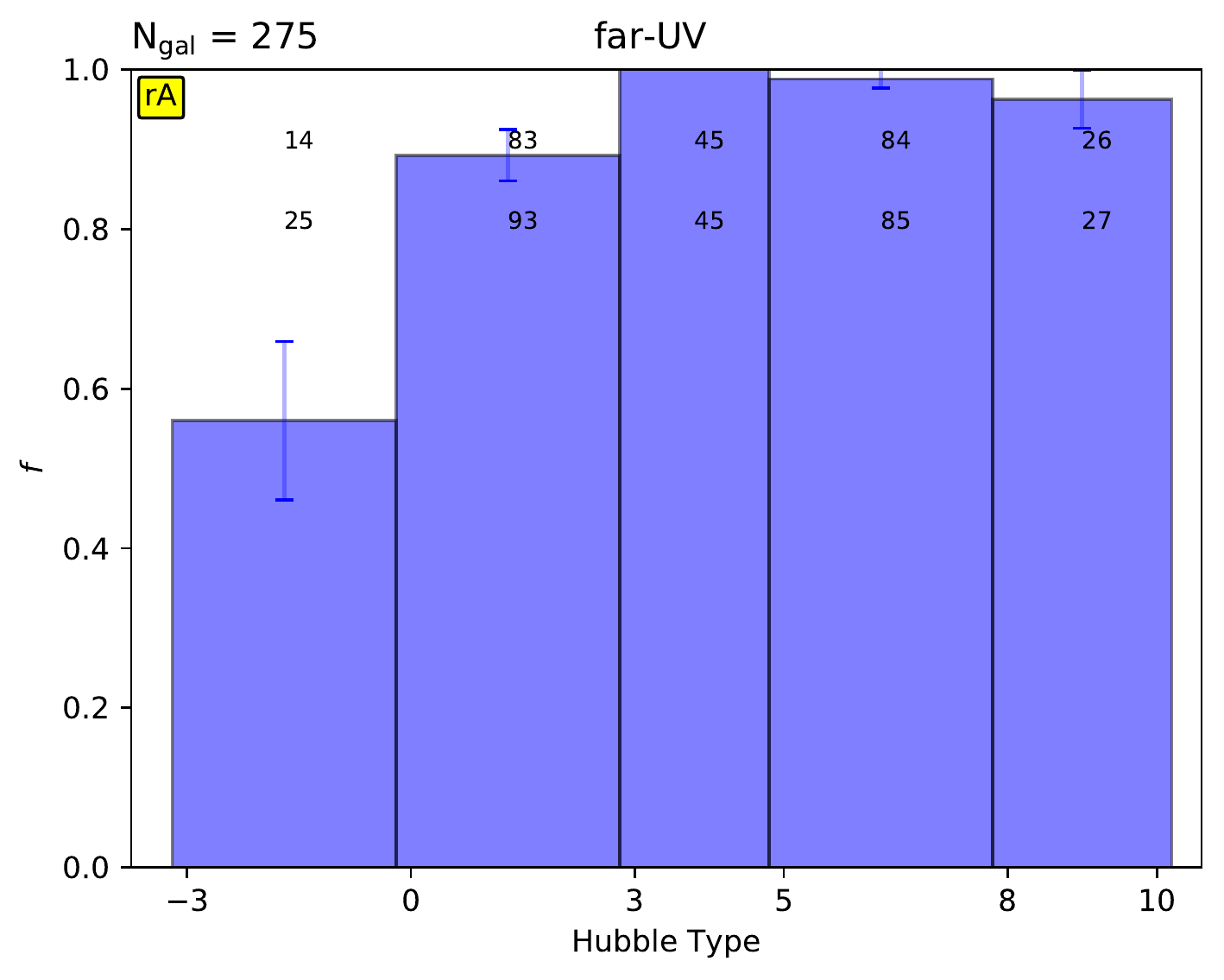}
\includegraphics[width=0.49\textwidth]{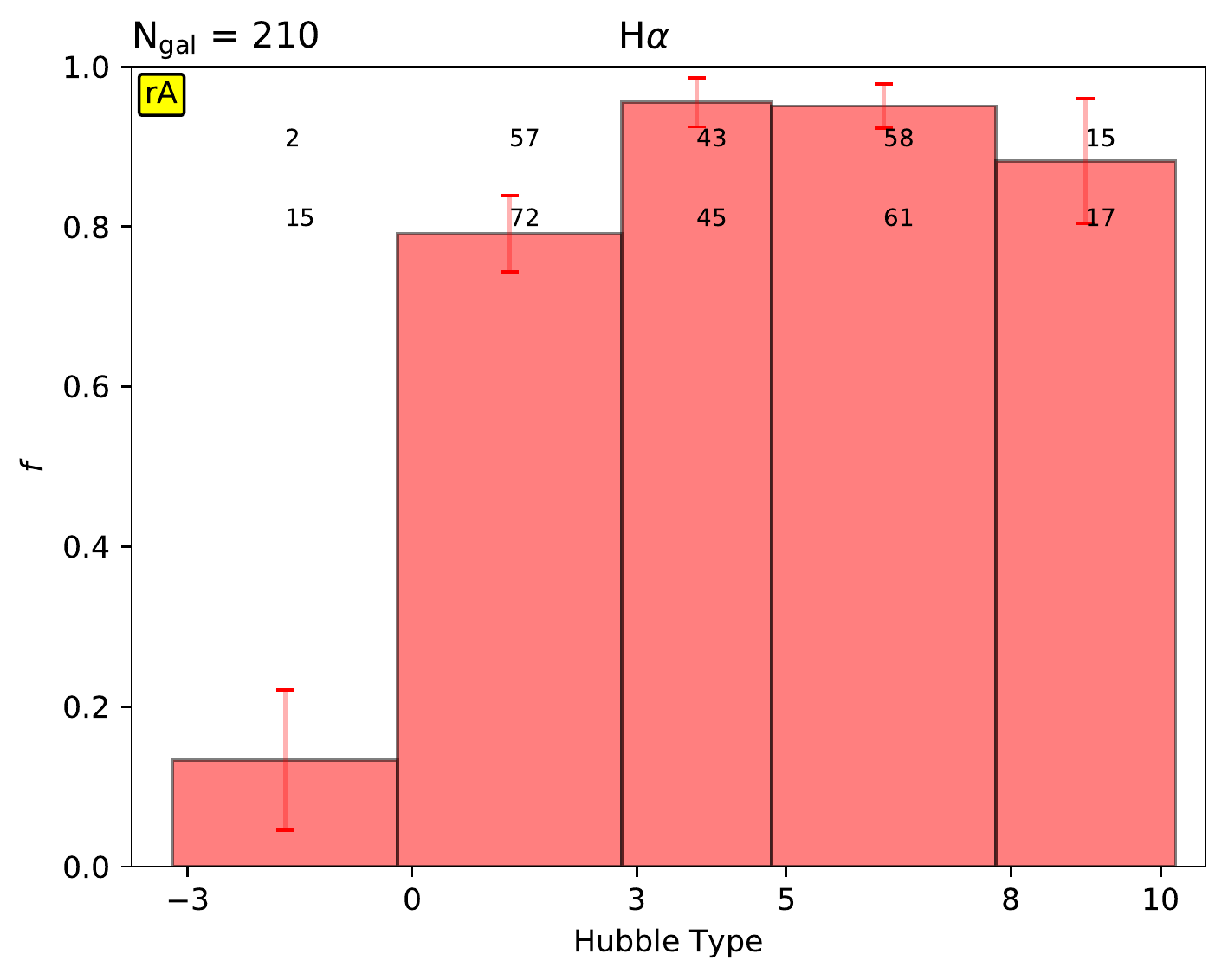}
\caption{
Fraction of inner rings that host SF as a function of the morphological type of the galaxy, 
identified based on the flux in GALEX FUV (left) and continuum-subtracted H$\alpha$ (right) imaging.
}
\label{appendixrings}
\end{figure*}

\begin{figure*}[!h]
\centering
\includegraphics[width=0.49\textwidth]{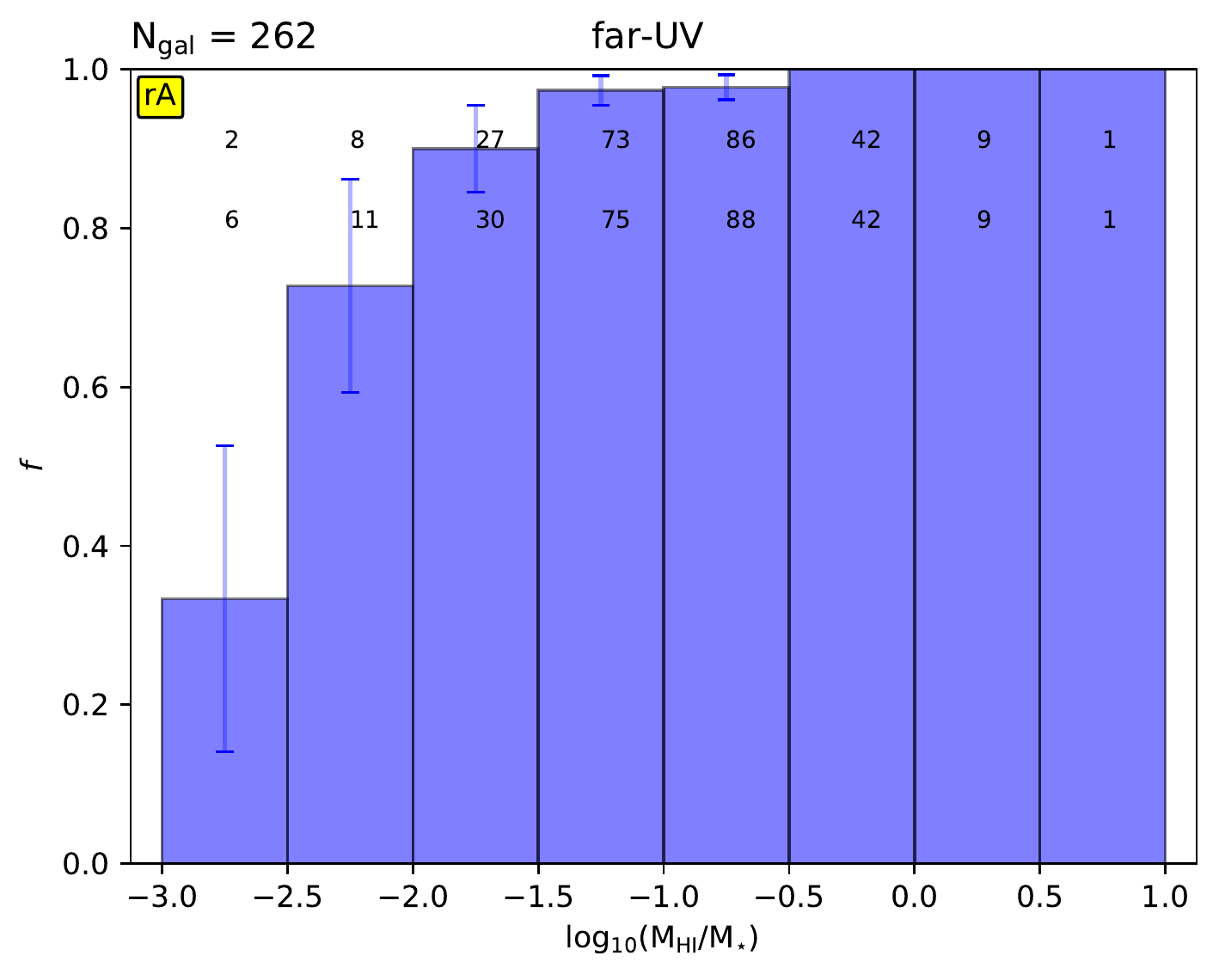}
\includegraphics[width=0.49\textwidth]{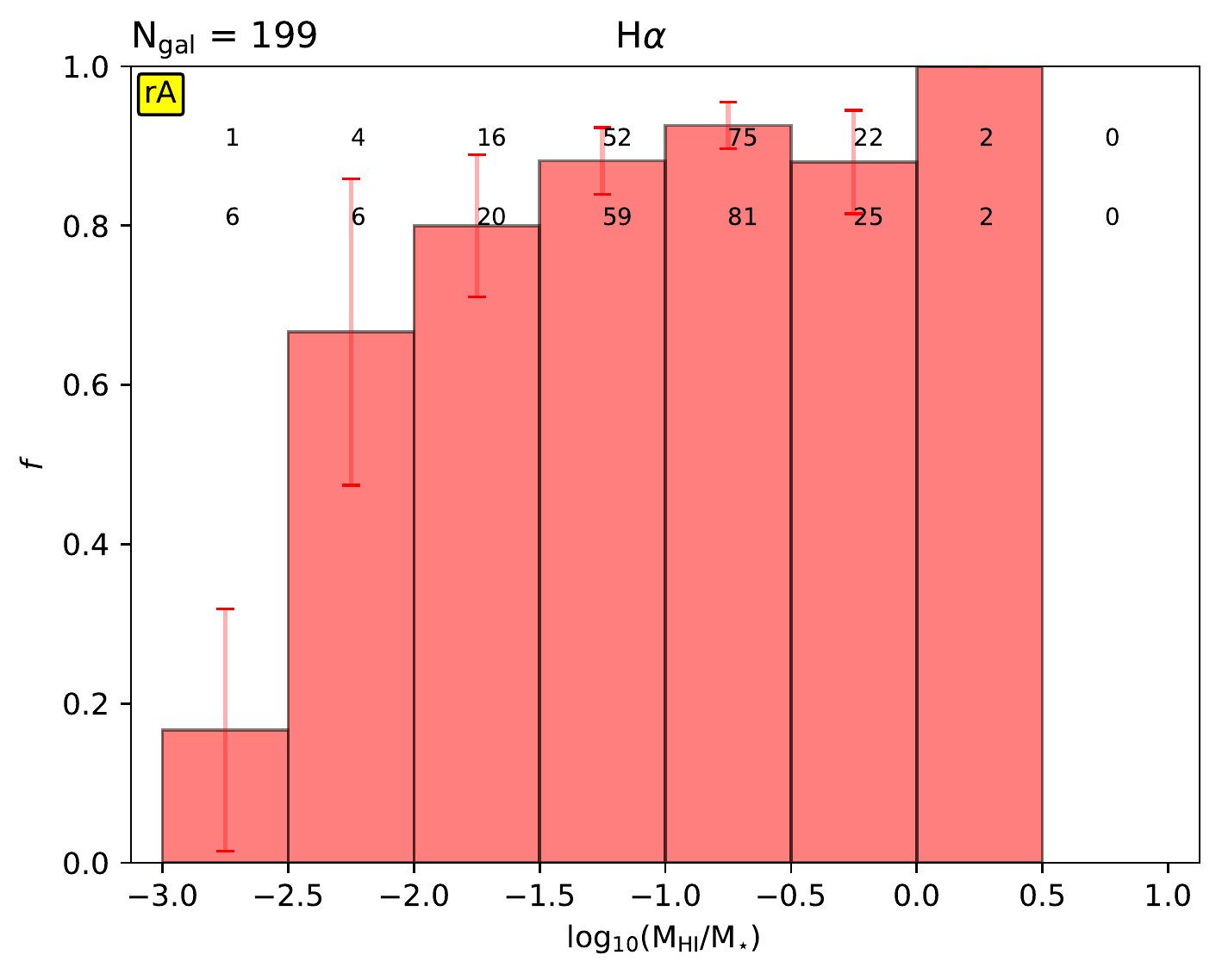}
\caption{
As in Fig.~\ref{appendixrings}, but as a function of the H{\sc\,i} gas content normalized by the total stellar mass.
}
\label{appendixrings2}
\end{figure*}

%
%
%
In this section we study the fraction of inner rings that have undergone recent SF, 
based on the classifications from Sect.~\ref{class_met}, using both GALEX FUV and continuum-subtracted H$\alpha$ imaging. 
We only probe those inner (pseudo)rings classified by \citet[][]{2015ApJS..217...32B} in the S$^4$G. 
We confirm that passive rings are mostly hosted by early-type galaxies (Fig.~\ref{appendixrings}), 
mainly lenticular galaxies in which the fraction of active rings is $\lesssim 50 \%$. 
Naturally, this is a consequence of passive rings being harboured by 
galaxies with low relative amounts of H{\sc\,i} gas (Fig.~\ref{appendixrings2}). 
%
%
%
%
%
%
%
\clearpage
\onecolumn
%
%
\section{Average bars as a function of the total stellar mass}\label{mass_average_bar}
\begin{figure*}[!h]
\centering
\includegraphics[width=0.99\textwidth]{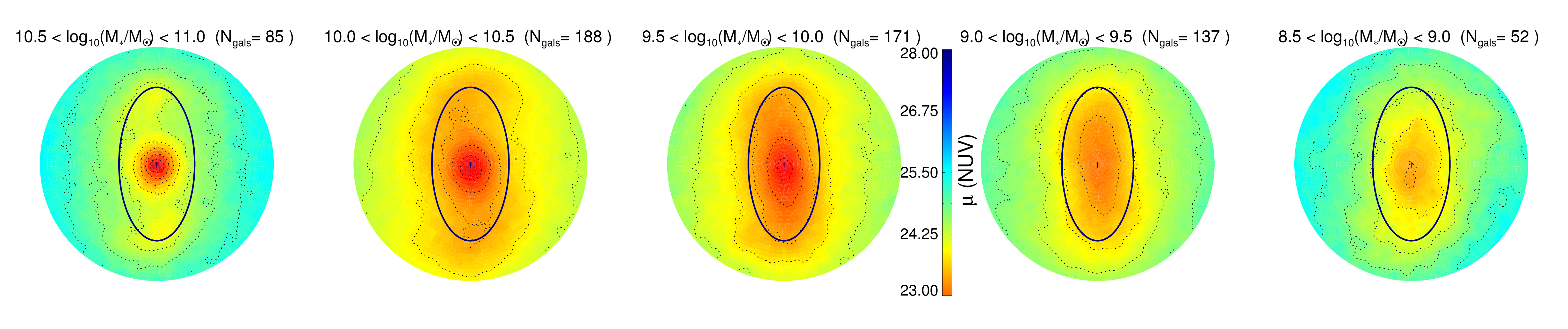}

\includegraphics[width=0.99\textwidth]{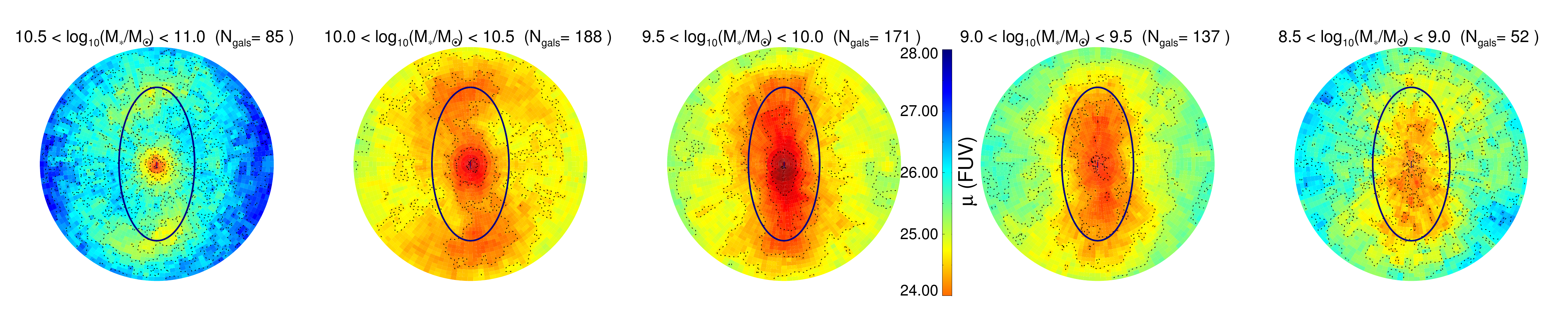}
\caption{
As in Fig.~\ref{Fig_ttype_bars_NUV}, but binning the sample as a function of the total stellar mass of the host galaxy.
}
\label{Fig_mass_bars_NUV}
\end{figure*}

\begin{figure*}[!h]
\centering
\includegraphics[width=0.49\textwidth]{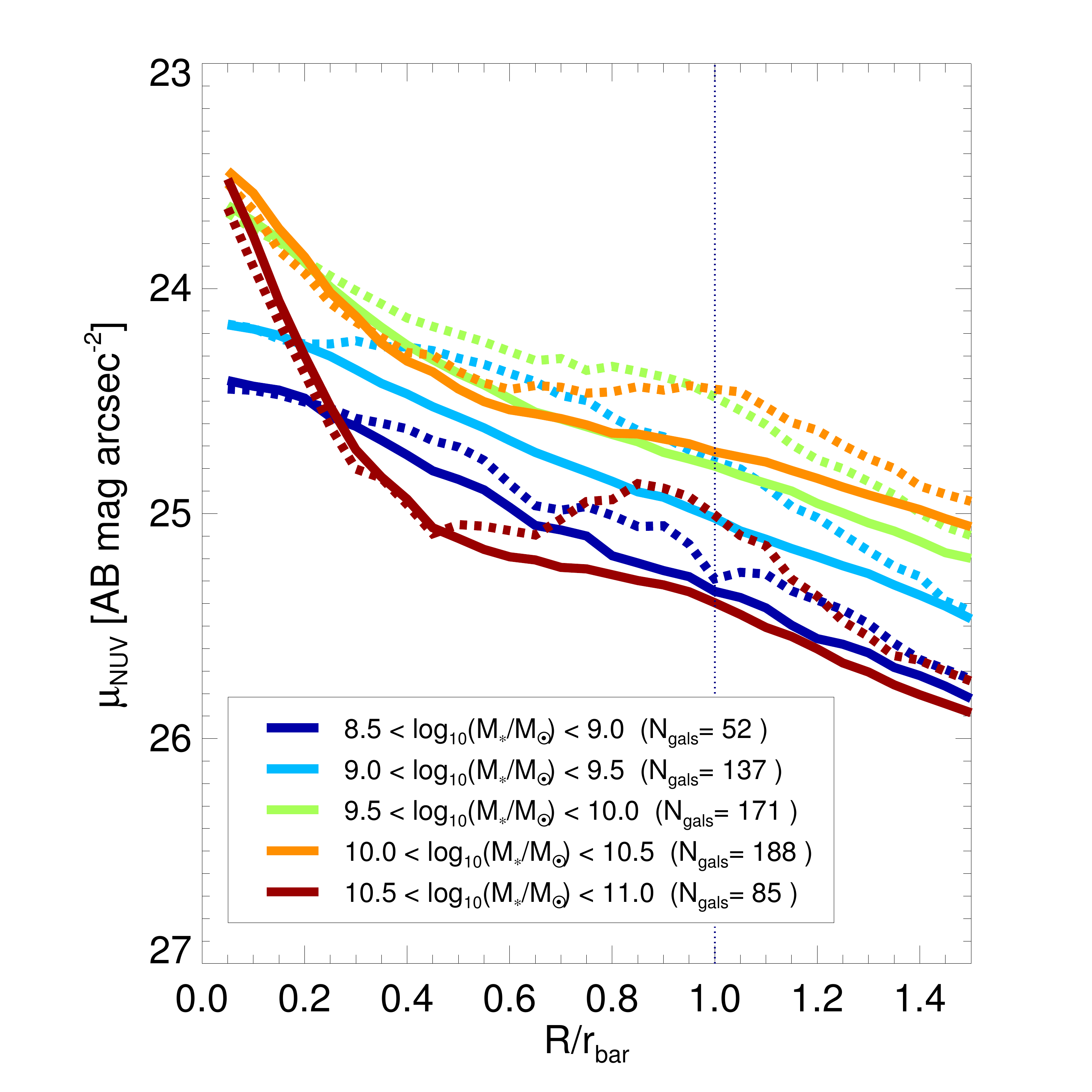}
\includegraphics[width=0.49\textwidth]{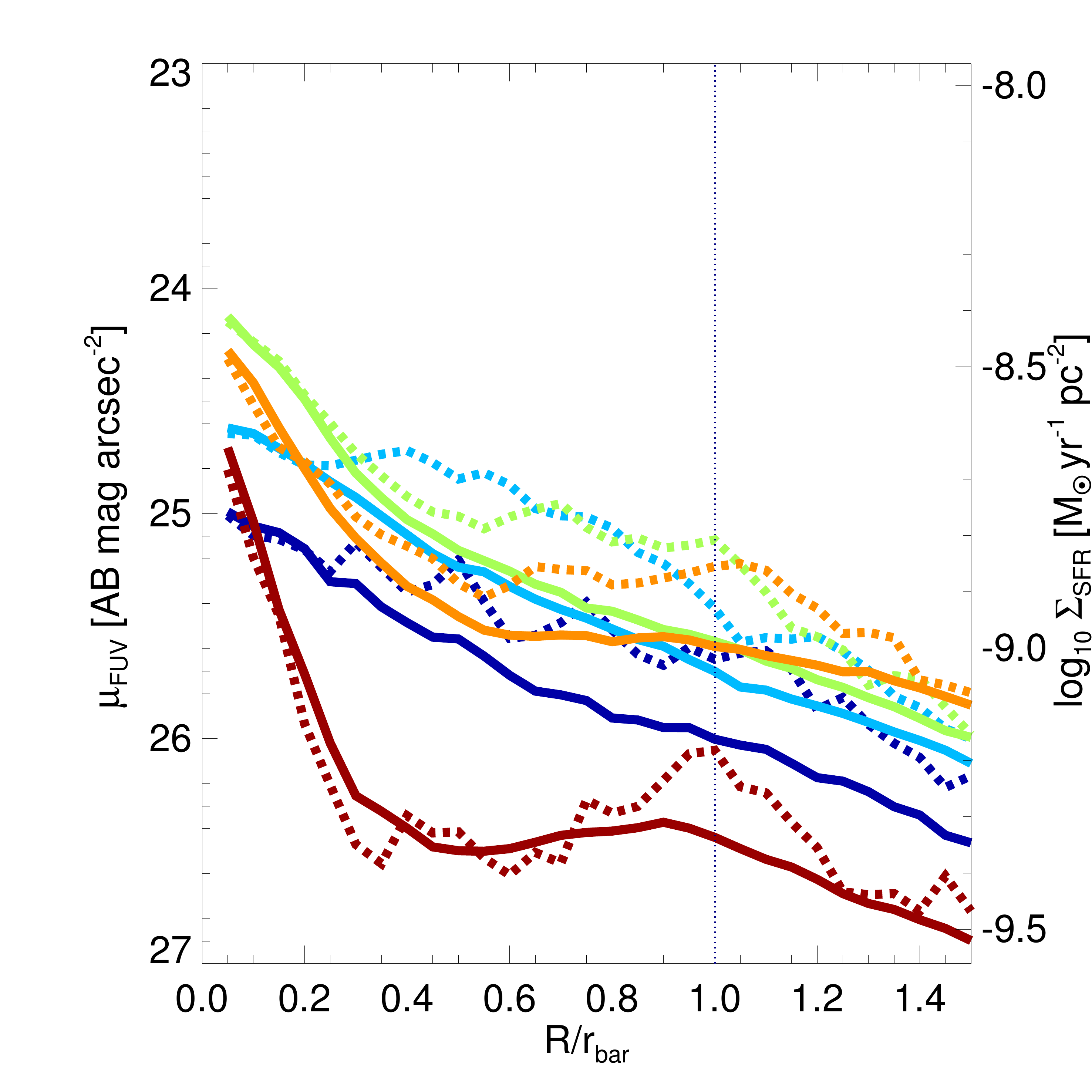}
\caption{
As in Fig.~\ref{Fig_ttype_bars_NUV_1D}, but binning  galaxies with respect to the total stellar mass. 
Azimuthally averaged mean FUV and NUV density profiles are derived from the bar stacks in Fig.~\ref{Fig_mass_bars_NUV}.
}
\label{Fig_mass_bars_NUV_1D}
\end{figure*}

%
%
We characterize the spatial distribution of SF in bars by stacking GALEX NUV and FUV images (see Sect.~\ref{bar_uv_stack}) 
after binning the parent sample as a function of the total stellar mass (Figs.~\ref{Fig_mass_bars_NUV} and \ref{Fig_mass_bars_NUV_1D}).
%
%
%
%
%
%
%
\clearpage
\onecolumn
%
%
\section{Comparative analysis of mean NUV luminosity profiles for  non-barred and weakly or strongly barred galaxies (complementary figures)}\label{bars_NUV}
\begin{figure*}[!h]
\centering
\includegraphics[width=0.49\textwidth]{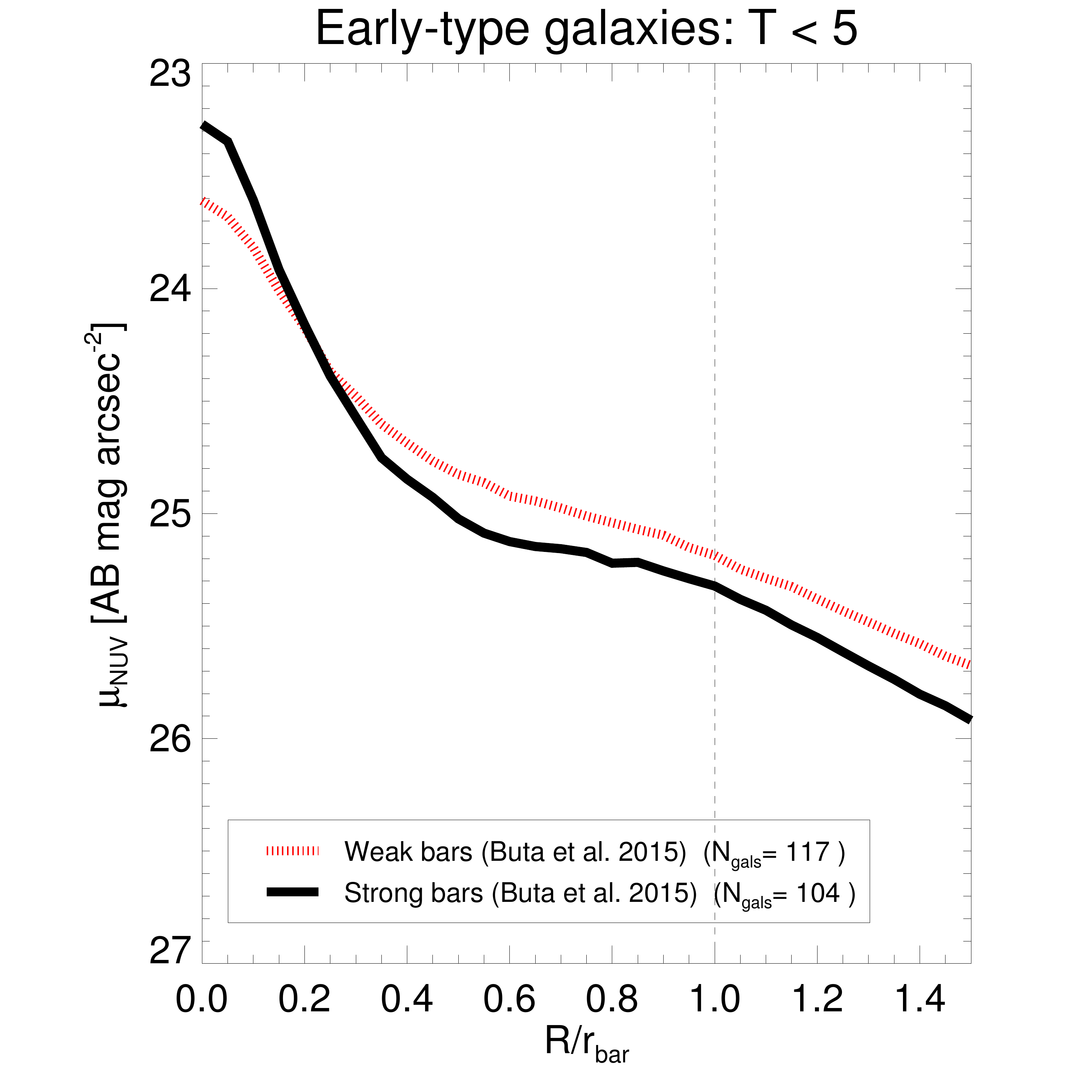}
\includegraphics[width=0.49\textwidth]{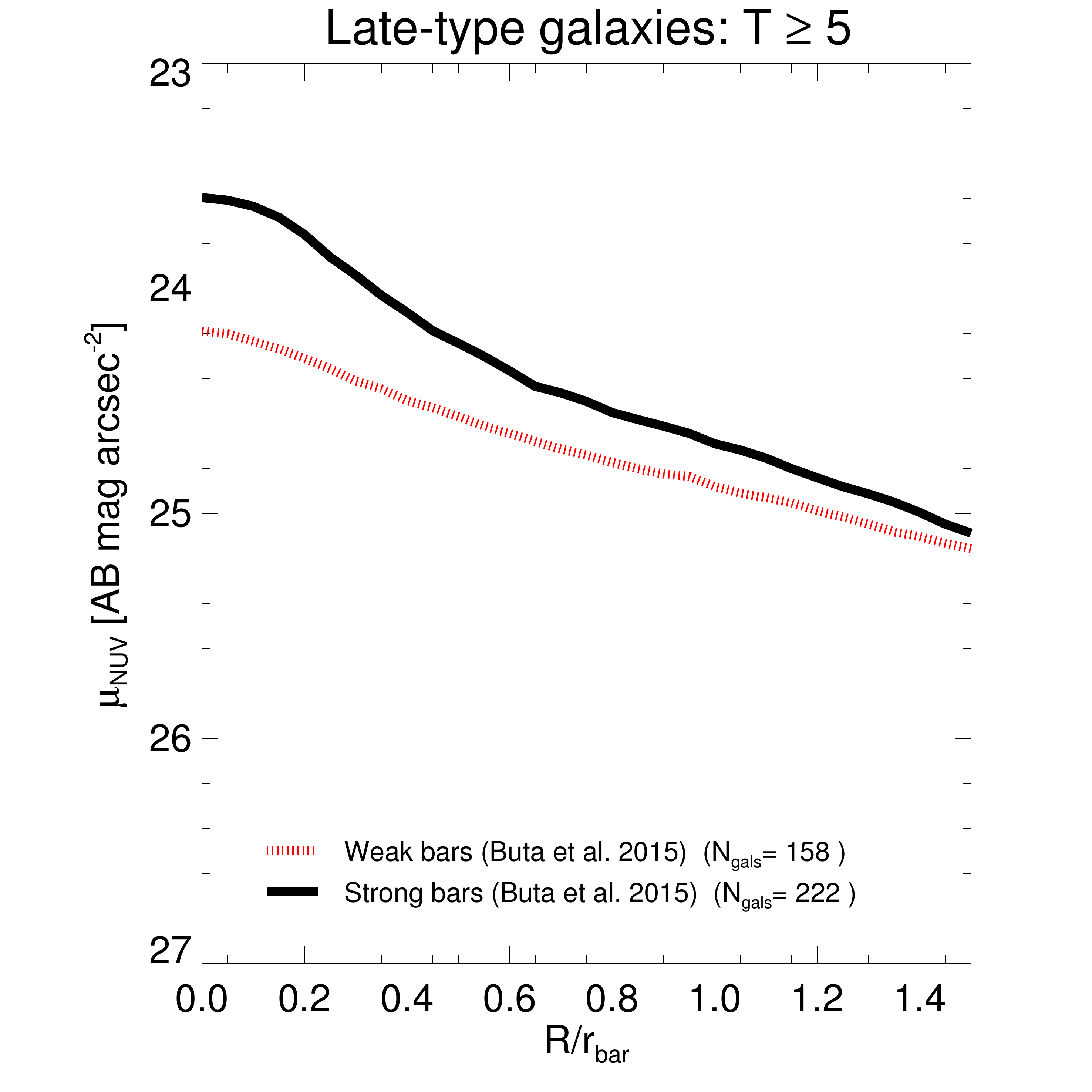}
\caption{
As in Fig.~\ref{Fig_family_bars_UV_1D}, but using NUV imaging.
}
\label{Fig_family_bars_NUV_1D}
\end{figure*}

\begin{figure}[!h]
\centering
\includegraphics[width=0.49\textwidth]{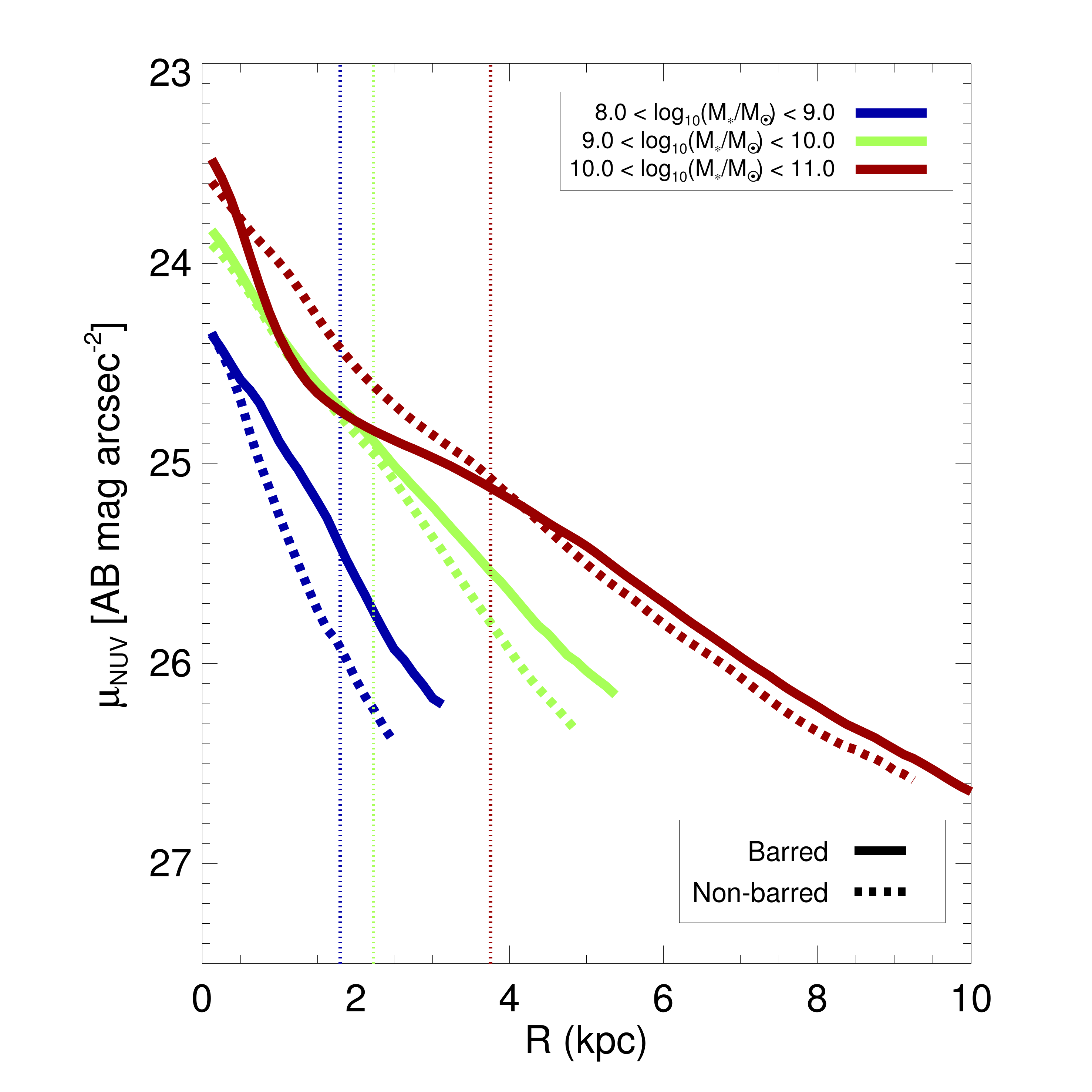}
\includegraphics[width=0.5\textwidth]{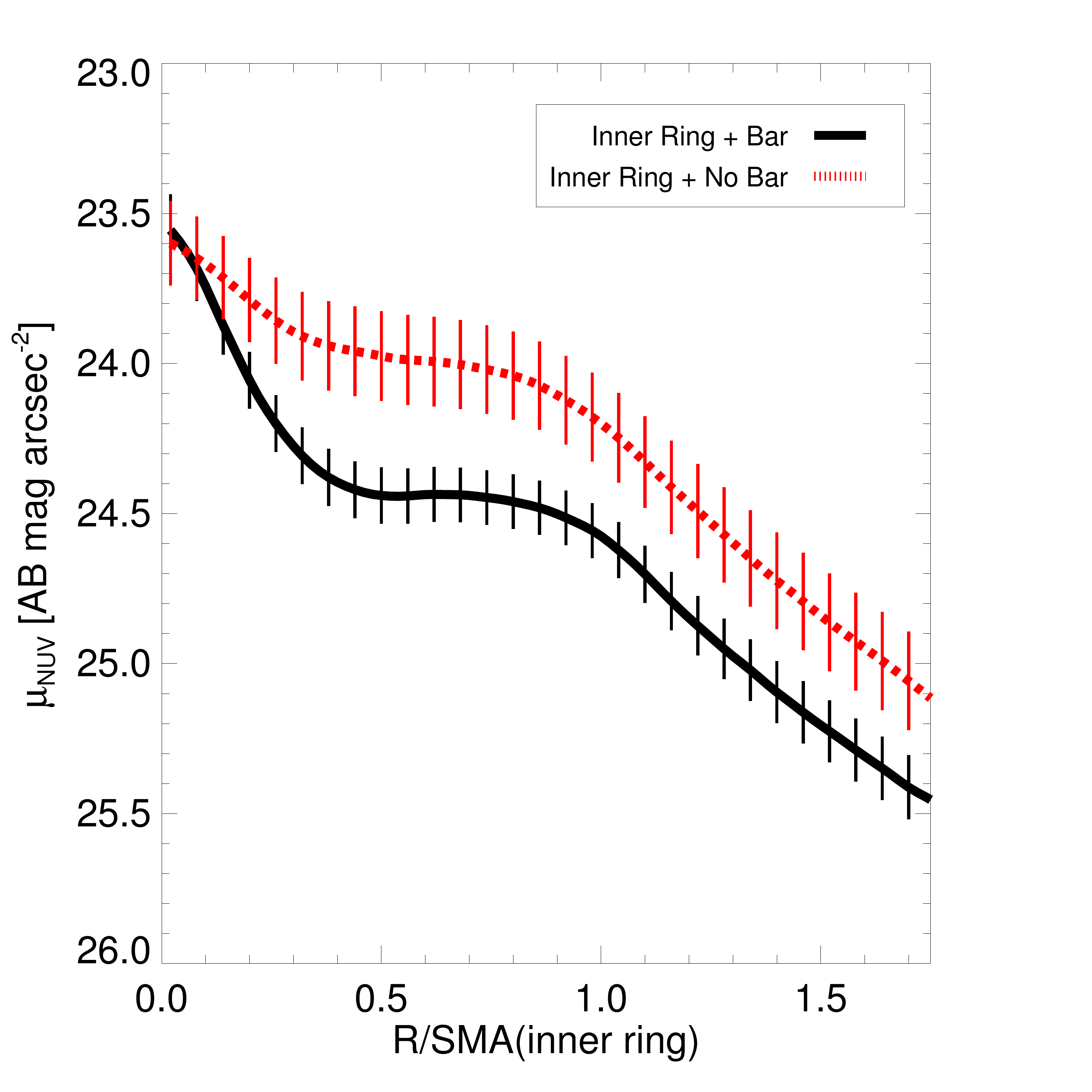}
\caption{
As in Fig.~\ref{Fig_mass_bars_NUV_FUV_1Dstack_bars_separated} (left) and Fig.~\ref{Fig_inner_rings_SMA} (right), 
but using NUV imaging.
}
\label{Fig_mass_bars_NUV_1Dstack_bars_separated}
\end{figure}

%
%
By averaging GALEX NUV images, we confirm the differences in SF between weakly or strongly barred and  non-barred galaxies 
(Fig.~\ref{Fig_family_bars_NUV_1D} and left panel of Fig.~\ref{Fig_mass_bars_NUV_1Dstack_bars_separated}) 
that were reported using FUV in Sects.~\ref{diffs_strong_weak}~and~\ref{1-Dstacks}, for different $T$- and $M_{\ast}$-bins. 
We also study average radial distribution of SF in inner-ringed galaxies using NUV 
(right panel of Fig.~\ref{Fig_mass_bars_NUV_1Dstack_bars_separated}), and confirm that 
the shapes of the profiles for barred and non-barred galaxies hosting inner rings are very similar (as reported in Sect.~\ref{1-Dstacks} using FUV).
%
%
%
%
%
%
\end{appendix}
%
%
\end{document}